\documentclass[11pt]{article}
\usepackage{jcappub,natbib,float,caption}

\usepackage{graphicx,amsmath,amssymb,color,bm}
\usepackage[dvipsnames]{xcolor}
\usepackage[caption=false]{subfig}
\usepackage{arydshln}
\usepackage{rotating}
\usepackage{cleveref}

\def\be{\begin{equation}}
\def\ee{\end{equation}}
\def\bea{\begin{eqnarray}}
\def\eea{\end{eqnarray}}
\newcommand{\vs}{\nonumber\\}
\def\bg#1\eg{\begin{gather}#1\end{gather}}

\def\Mpch{\,h^{-1}\,{\rm Mpc}}

\newcommand{\refeq}[1]{Eq.~(\ref{eq:#1})}

\newcommand{\refsec}[1]{Sec.~\ref{sec:#1}}          
\newcommand{\refapp}[1]{App.~\ref{app:#1}}

\renewcommand{\v}[1]{\bm{#1}}
\def\P{\mathcal{P}}

\DeclareMathOperator{\tr}{tr}
\DeclareMathOperator{\TF}{TF}

\newcommand{\vx}{\v{x}}

\newcommand{\vk}{\v{k}}

\newcommand{\vp}{\v{p}}

\def\lapl{\nabla^2}

\newcommand{\<}{\langle}
\renewcommand{\>}{\rangle}

\renewcommand{\d}{\delta}
\def\dK{\delta^{\rm K}}

\newcommand{\nhat}{\hat{n}}
\newcommand{\vnhat}{\v{\hat{n}}}
\newcommand{\vhat}[1]{\v{\hat{#1}}}

\def\Mpch{h^{-1}\,\text{Mpc}}

\newcommand{\Om}{\Omega_m}

\def\Del{\mathcal{D}}

\def\cH{\mathcal{H}}

\renewcommand{\vec}{\bm}
\newcommand*{\p}  {\partial}

\newcommand*{\df}  {\delta}

\newcommand*{\non}  {\nonumber}
\newcommand*{\lb}  {\left(}
\newcommand*{\rb}  {\right)}

\newcommand*{\la}  {\left\langle}
\newcommand*{\ra}  {\right\rangle}
\newcommand*{\eps}  {\epsilon}

\newcommand{\ba}{\[\begin{aligned}}
\newcommand{\ea}{\end{aligned}\]}

\newcommand{\eq}[1]{\begin{align}#1\end{align}}
\newcommand{\eeq}[1]{\begin{equation}#1\end{equation}}

\newcommand{\comment}[1]{}

\title{An EFT description of galaxy intrinsic alignments}

\author{Zvonimir Vlah$^{[a]}$,}
\author{Nora Elisa Chisari$^{[b]}$,}
\author{Fabian Schmidt$^{[c]}$}

\affiliation[a]{Theory Department, CERN, 1 Esplanade des Particules, CH-1211 Gen\` eve 23, Switzerland.}
\affiliation[b]{Institute for Theoretical Physics, Utrecht University, Princetonplein 5, 3584 CE Utrecht, The Netherlands.}
\affiliation[c]{Max-Planck-Institut f\"ur Astrophysik, Karl-Schwarzschild-Str. 1, 85741 Garching, Germany.}

\emailAdd{zvonimir.vlah@cern.ch}
\emailAdd{n.e.chisari@uu.nl}
\emailAdd{fabians@mpa-garching.mpg.de}

\abstract{
  We present a general perturbative effective field theory (EFT) description of
galaxy shape correlations, which are commonly known as intrinsic alignments.
This rigorous approach extends current analytical modelling strategies in that it only relies on the equivalence principle. 
  We present our results in terms of 
  three-dimensional statistics for two- and three-point functions of both galaxy shapes and number counts. 
  In case of the two-point function, we recover the well-known linear alignment result at leading order, but also present the full 
  next-to-leading order expressions. In case of the three-point function we present leading order results for all the auto-
  and cross-correlations of galaxy shapes and densities. We use a spherical tensor basis to decompose the tensor 
  perturbations in different helicity modes, which allows us to make use of isotropy and parity properties in the 
  correlators. Combined with the results on projection presented in a forthcoming companion paper, our framework is directly applicable to accounting for intrinsic alignment contamination in 
  weak lensing surveys, and to extracting cosmological information from intrinsic alignments. 
}

\begin{document}

\maketitle
\flushbottom

\section{Introduction}
\label{sec:intro}

Intrinsic galaxy alignments are the correlations between intrinsic galaxy shapes and the large-scale structure, i.e. before the effect of gravitational lensing on the observed shapes \cite[see][for reviews]{Troxel15,Joachimi15}. 
These are known to exist between luminous red galaxies at low redshift \cite{Hirata07,okumura/jing:2009_I, okumura/jing:2009_II,Singh15,Singh16}, and are considered a contaminant for cosmology with weak gravitational 
lensing \cite{Heavens00,Hirata04,Mandelbaum06}. Despite these alignments being measured to high significance in current galaxy surveys, there have been  relatively few efforts on 
exploring possible models for this signal. Historically, the literature has adopted the ``linear tidal alignment'' model (LA) \cite{Catelan01,Blazek11} and 
only recently have alternatives been considered. The LA model relies on assuming that projected intrinsic galaxy shapes are proportional to 
the projected tidal field of the large-scale structure. While this model seems to accurately describe alignments for red galaxy pairs separated by large distances 
($\gtrsim 10\Mpch$), recent measurements of the alignment signal show an excess above the linear prediction for smaller separations \cite{Singh15,Johnston19}. 
In contrast, the alignment correlations of blue galaxies remain poorly constrained, with only upper limits currently available \cite{Mandelbaum06,Heymans13,Johnston19}, 
and proposed models rely on the impact of tidal torques from the large-sale structure on the direction of the galactic angular momenta \cite{Heavens00,Crittenden01,Hirata04,Schafer17}. 

There are two main motivations for improving on the modelling of intrinsic galaxy alignments. First, obtaining unbiased parameter constraints from weak lensing surveys requires accurate 
modelling of all potential contributions to observed correlations of galaxy shapes, which includes 
the intrinsic alignment signal \cite{Joachimi10}.
With the advent of the next generation of weak lensing surveys such as {\it Euclid} \cite{Euclid}, the Large Synoptic Survey Telescope (LSST) \cite{Ivezic} and {\it WFIRST} \cite{WFIRST}, 
requirements on the accuracy of the alignment model are becoming more strict \cite{Krause16}. 
Notably, the most important contribution to consider is the correlation between the density field responsible for the lensing 
of galaxy shapes, which at the same time sources the alignments of physically nearby galaxies. This is typically known as the ``GI'' contamination \cite{Hirata04}. 
Second, it has been shown that intrinsic alignments 
encode information from phenomena such as primordial non-Gaussianity and gravitational waves from inflation in complementary manner to other probes 
of the large-scale structure \cite{Chisari14,schmidt/pajer/zaldarriaga,schmidt/chisari/dvorkin,Chisari16b}. Accurate extraction of this information requires a reliable alignment model.

Beyond the linear predictions, intrinsic alignment modelling has been extended using tools like standard perturbation theory (SPT) \cite{Blazek15,Blazek17,Schmitz18}, 
the halo model \cite{Schneider10,Joachimi13} and cosmological hydrodynamical simulations 
\cite{Kiessling15,Codis15,Chisari15,Chisari16,Velliscig15a,Velliscig15b,Tenneti14,Tenneti15a,Tenneti15b,Tenneti16,Tenneti17,Chisari17,Hilbert17}. 
These methods target different questions regarding modelling of galaxy intrinsic alignments. Analytical predictions like those provided by SPT aim to be fast and extend 
the LA model to smaller scales. They are designed to work in regimes where density perturbations are small, but suffer from lack of convergence at small scales 
\cite{bernardeau/etal:2001,carlson/white/padmanabhan:2009, scoccimarro/frieman/1996}. The halo model approach aims for a phenomenological prediction of 
the nonlinear behaviour of galaxy alignments within halos. Cosmological simulations can serve as a validation tool for all these models and help constrain their 
free parameters in as much as the simulated galaxy population is representative of the observed one.

In this work, we pursue an approach closer to the Eulerian perturbation theory, and describe how a complete bias-like expansion up to a given order 
in perturbations can be constructed to describe the statistics of any {\it tensorial} field in the large-scale structure using the effective field theory (EFT) 
of the large-scale structure \cite{Baumann12}. This method is based on the description of the matter density distribution as an effective fluid on large scales. 
Small-scale physics is integrated out and encapsulated in a set of free parameters that can be fit to either simulations or observations. This systematic approach 
has previously been applied to {\it scalar} fields representing number counts of galaxies and halos (see \cite{biasreview} for a review). 
Since it also includes contributions that involve higher spatial derivatives of the density and tidal fields, it consistently solves the issue of unphysical 
dependence of SPT predictions on very small-scale modes which are not correctly described by perturbation theory.

This EFT of tracer shapes allows us to extend the modelling of intrinsic alignments from linear scales to the quasi-linear regime. In our approach, 
we model three-dimensional galaxy shapes within the EFT by identifying all terms that correspond to gravitational observables at a given order 
and that satisfy the symmetries of a trace-free tensor. We then employ this expansion to consistently compute auto-correlations of intrinsic galaxy shapes, 
and the cross-correlation between intrinsic shapes and a set of scalar biased tracers, such as a generic galaxy population. Our work, however, can be 
straightforwardly extended to applications that are of interest to weak lensing surveys, such as the modelling of cross-correlations between intrinsic shapes 
and the matter density field (this includes the case of the lensing by the cosmic microwave  background \cite{Troxel14,Hall14,Chisari15c,2016MNRAS.461.4343L}). In addition, 
it is also possible to model the selection effects induced by intrinsic alignments in scalar biased tracers observed by galaxy redshift surveys 
\cite{hirata:2009,krause/hirata:2011,Martens18} along the lines of \cite{pkgs}. Finally, the EFT prescription for scalar biased tracers can also be applied 
to intrinsic correlations of galaxy sizes, as well as other galaxy properties. Size correlations have recently been detected in current surveys \cite{Joachimi15b} 
and can pose a challenging contaminant in the attempt to constrain cosmological magnification via observed galaxy size perturbations \cite{Joachimi15b,Ciarlariello15,Ciarlariello16}.

This work is organised as follows. Section~\ref{sec:bias} gives a general description of the expansion of three-dimensional galaxy shapes in terms 
of local gravitational observables. 
In section~\ref{sec:ssn}, we introduce the spherical tensor basis which is used to decompose the galaxy shape tensors.
We then proceed to compute three-dimensional statistics (both two- and three-point functions) in section~\ref{sec:Pk3D}. 
In section~\ref{sec:res}, we present explicit one-loop power spectrum and tree-level bispectrum results and give estimates of the contribution of each 
term in our intrinsic alignment framework. These contributions are scaled by unknown bias coefficients which should be constrained from simulations, observations, or marginalised over in cosmological analyses. 
In section~\ref{sec:rsd} we outline briefly additional physical/observational effects that can affect the observables. In particular we mention the galaxy shape projection on the sky and redshift space distortions.
These are the topic of an upcoming companion paper \cite{vlah/chisari/schmidt2019}.
We conclude in section~\ref{sec:concl}. Most of the detailed calculations are delegated to the appendices. 
In appendix~\ref{app:SVT} we perform the decomposition of the tensor fields in spherical tensor components that allows us to isolate the relevant 
correlator contributions to the two- and three-point functions. 
The detailed tensor field perturbative expansion is performed in appendix~\ref{App:PT_shear_field} and one-loop power spectrum and tree-level bispectrum results are obtained. Some renormalisation concepts concerning the tensor field of biased tracers are addressed in appendix~\ref{app:reno}.

For numerical results, we assume a Euclidean $\Lambda$CDM cosmology with $\Omega_m=0.295,~\Omega_b=0.047,~n_s=0.968,~\sigma_8=0.835$ and $h=0.688$. We consider the case of adiabatic Gaussian perturbations and General Relativity.  The case of primordial non-Gaussianity is straightforward to include following Ref.~\cite{schmidt/chisari/dvorkin}. Furthermore, in perturbative calculations, we make the usual approximation of setting the $n$-th order growth factor $D^{(n)}(\tau)$ to $[D(\tau)]^n$. Notation conventions and a list of most important quantities used in the paper are given in the table~\ref{tab:notation}.

\begin{table*}
\centering
\begin{tabular}{l|l}
\hline
\hline
$f(\vk) \equiv \int d^3 \vx\, f(\vx) e^{-i\vk\cdot\vx}$  & Fourier transform conventions\\[3pt]
$\int_{\v{p}} \equiv \int \frac{d^3 \v{p}}{(2\pi)^3}$ & Momentum integral \\[3pt]
$\mathcal F (\vec p_1, \vec p_1) \df(\vec p_2) \df(\vec p_2)$
 & Integration over repeated momenta \\
$\  \equiv \int_{\vec p_1} \int_{\vec p_2} \mathcal F (\vec p_1, \vec p_2) \df(\vec p_1) \df(\vec p_2)$ &  \\[3pt]
$\dK_{ij}$ & Kronecker symbol \\[3pt]
$\tilde \df^{\rm K}_{ij} = N_0^{-1} \dK_{ij}$ & Normalized Kronecker symbol (\refeq{Y_basis}) \\[3pt]
$\df^{\rm D}(\vx)$ & Dirac delta function in real space \\[3pt]
$\df^{\rm D}_{\vec k-\vec k'} \equiv (2\pi)^3 \d^{\rm D}(\vec k - \vec k')$ & Dirac delta function in Fourier space \\[3pt]
$\Del_{ij} \equiv \p_i\p_j/\nabla^2 - (1/3)\dK_{ij}  $ & Shear derivative \\[3pt]
$\vk_{1\cdots n} \equiv \vk_1 + \cdots + \vk_n$ & Sum notation
($\vk_{1n} \equiv \vk_{1\cdots n}$ in appendix~\ref{App:PT_shear_field}) \\[3pt]
$\vec Y_{ij}^{(m)}(\vx)$ & Rank two spherical tensor basis \\[3pt]
$\< O(\vk_1) \cdots O(\vk_n)\>'\,$
 & $n$-point correlator without momentum conservation\footnotemark\\[2pt]
\hline
$\d_{\rm m}$ & Fractional matter density perturbation \\[3pt]
$\df_L$ & Linear-order fractional matter density perturbation \\[3pt]
$\Theta_{ij} \equiv \Del_{ij} \d_{\rm m}$ & Scaled tidal field \\[3pt]
$\Pi^{[1]}_{ij} \equiv \Del_{ij} \d_{\rm m} + (1/3)\dK_{ij} \d_{\rm m}$ &
Scaled Hessian of gravitational potential  \\[3pt]
$v^i$ & Matter velocity field \\[3pt]
\hline
$\d_{\rm n}$ & Fractional galaxy number density perturbation \\[3pt]
$\d_{\rm s}$ & Fractional galaxy size density perturbation \\
& \quad (trace of shape tensor) \\[3pt]
$S_{ij}$ & 3D galaxy shape tensor \\[3pt]
$g_{ij}$ & Trace-free part of 3D galaxy shape tensor \\[3pt]
$\gamma_{ij}$ & Projected shape on the sky \\
& \quad (trace-free part of projected shape tensor) \\[3pt]
\hline
\end{tabular}
\caption{List of notation and most important quantities used in this paper.
  Fields in Fourier space are understood to be integrated over repeated
  momentum variables.
\label{tab:notation}}
\end{table*}

\section{General bias expansion for shapes}
\label{sec:bias}

Galaxy shapes, more specifically ellipticities, are a spin-2 field on the
sky, and their description is analogous to polarization; specifically,
shapes refer to the trace-free part of the second-moment tensor
of the projected image of the galaxy. Unlike the effect
of gravitational lensing on shapes, which is a projection effect, intrinsic
galaxy shape correlations have to be described in three-dimensional
space, since they arise from physical interactions \cite{schmidt/chisari/dvorkin}. 
Our goal therefore will be to first describe the statistics of the three-dimensional galaxy shape tensor, 
and then to project it on the sky to obtain the observed spin-2 shape field. 
Note that the projection mixes the trace- and trace-free parts of the three-dimensional shape tensor, 
so even though we are mostly interested in the trace-free part after projection,
we need to specify the full three-dimensional shape tensor to begin with.
As we will describe, the three-dimensional shapes of galaxies in their rest frame can be described through 
a well-defined ``bias'' expansion in terms of local gravitational observables. Hence, throughout this paper, all indices refer
to three-dimensional spatial indices. 

\footnotetext{Explicitly, we write 
\eeq{ \la O(\vk_1) \cdots O(\vk_n)\ra = (2\pi)^3 \d^{\rm D}(\vk_{1\cdots n}) \la O(\vk_1) \cdots O(\vk_n)\ra' \,,}
where the Dirac delta function ensures the total momentum conservation, and is a consequence of the statistical translation invariance.}

\subsection{Definitions and preliminary considerations}

The fractional number density perturbation, $\d_{\rm n}$, of galaxies at a space-time position $(\vx,\tau)$ is defined as
\be
\d_{\rm n}(\vx,\tau) \equiv \frac{n_{\rm g}(\vx,\tau)}{\< n_{\rm g}(\tau)\>} - 1\,.
\label{eq:deltan}
\ee
The EFT of large-scale structure provides an effective description for this field which we summarize in the next section. 
The EFT allows the prediction of galaxy count statistics for a given cosmology, i.e. ``galaxy clustering'', up to a number a free bias parameters that need to be determined from simulations or observations.

As in the case of galaxy clustering, we should disentangle intrinsic physical effects (from the point of view of an observer comoving with a galaxy) 
from projection effects which deal with the mapping from the rest frame of the galaxy to the coordinates of the observer. In the context of galaxy shapes, 
this means that we describe physical alignment effects in terms of the three-dimensional galaxy shape in its rest frame, \emph{not} in terms of the shape projected on the sky plane.
We will come back to the effect of projection in section~\ref{sec:rsd}, and in more detail in the accompanying paper \cite{vlah/chisari/schmidt2019}.

We assume that, for each galaxy $\alpha$, its light distribution is described by the symmetric second-moment tensor 
$I_{ij}(\vx_\alpha)$.\footnote{We adopt the notation in which the vectorial quantities in the invariant form are represented as
\eeq{
\vec V = V_i \vec e^i,
}
where $\vec e^i$ are the basis vectors and $V_i$ are the vector components. 
In analogy to this notation, invariant form of the tensorial quantities can be represented as 
\eeq{
\vec T = T_{ij} \vec e^i \otimes \vec e^j,
}
where the basis is given as the Cartesian product of the two basis vectors $\vec e^i$, 
and $T_{ij}$ are the usual tensor components in the given basis. 
}
This is akin to modelling the 3D light distribution of each galaxy with an ellipsoid, where three degrees of freedom represent the lengths of ellipsoid semiaxes,
and the residual three parameters describe the relative orientation of the ellipsoid (three Euler angles) at the position $\vx_\alpha$.
We can then formally define a tensor field of intrinsic galaxy shapes as a function of comoving location,
\eeq{
  I_{ij}(\vx) = \sum_\alpha I_{ij}(\vx_\alpha) \d_D(\vx-\vx_\alpha)\, ,
  \label{eq:Iijdef}
}
where the sum runs over all galaxies in the survey. The trace of the field returns a scalar field which describes the number-weighted galaxy size, i.e.
\eeq{
\tr[I_{ij}(\vx)] = \overline{s^2} (1+\d_{\rm s}(\vx)),
}
where $\overline{s^2}$ is defined through $\< I_{ij} \> = \dK_{ij}\, \overline{s^2}/3$ and $\d_{\rm s}(\vx)$ is the corresponding 
size fluctuation field.\footnote{An alternative definition for the tensor field of interest could be
\eeq{
I_{ij}(\vx) = \sum_\alpha \frac{I_{ij}(\vx_\alpha)}{{\tr \< I_{\alpha, ij}\>}} \d_D(\vx-\vx_\alpha)\,,
  \label{eq:Iijdef2}
}
where we use the estimated flux normalization for each individual object.
With this choice, the trace of the shape field $I_{ij}$ yields the galaxy number density, i.e. $\tr[ I_{ij}(\vx) ] = \< n_{\rm g}\>(1+\d_{\rm n}(\vx))$. 
Our EFT expansion will apply to either definition, as well as any other definition in terms of physical observables. However, the values of the bias parameters will depend on the chosen definition.}
More generally, we define the shape fluctuation field
\be
S_{ij}(\vx) =  \frac{ I_{ij}(\vx) - \< I_{ij}(\vx) \>}{\tr \< I_{ij} \> } = \frac{I_{ij}(\vx)}{\tr \< I_{ij}\> } - \frac{1}{3} \dK_{ij}
= g_{ij}(\vx) + \frac13 \d_{\rm s}(\vx) \dK_{ij}\,,
\label{eq:Sij}
\ee
where we have introduced the trace-free tensor $g_{ij}$, which describes galaxy shape perturbations.
As was already mentioned, different normalizations of the field $S_{ij}$ (i.e. $I_{ij}$) might be more 
convenient in different situations; for example, one might introduce a number density weighting in \refeq{Iijdef2}. 
However all these different definitions do not change the bias expansion which we introduce below, which is agnostic to the
precise definition of the galaxy shape field. We will only need the fact that
$g_{ij}(\vx)$ is a symmetric trace-free tensor field. We return to comment on the possible effects of this choice 
later on when we discuss the bias expansion in Section \ref{ss:biasexp}.  

\subsection{Review of bias expansion for scalar tracers}

Any scalar tracer in the Universe can be connected to local gravitational observables by means of a bias expansion,
\eeq{
  \d_{\rm n}(\vx,\tau) = \sum_O b_O^{(\rm n)} [O]
  \label{eq:dn}
}
where $[O]$ are renormalised operators constructed out of the density field, gravitational potential and other perturbations, and the $b_O^{(\rm n)}$ 
are the corresponding (renormalised) bias parameters. We will define and
perform the renormalisation procedure when we turn to statistics in
\refsec{Pk3D}. If the physical processes that determine the properties of the tracers are {\it local}, then the need to satisfy 
the equivalence principle suggests that the operators in the bias expansion are those corresponding to the density and the tidal field. This expansion 
allows us in principle to predict any given statistics of the biased tracers, and to connect them to the underlying cosmological model. 
Note further that the bias expansion is not unique: at any given order in perturbations, the complete set of operators forms the basis of a vector space, 
and any other basis can serve as an equivalent choice. 

Reference \cite{MSZ} showed that one such complete basis of the general galaxy bias expansion (at lowest order in derivatives) consists of all scalar combinations of a set of operators, $\vec \Pi^{[n]}$. 
The operators $\vec \Pi^{[n]}$ are second-rank tensors and thus the components always carry two indices, i.e. $\Pi^{[n]}_{ij}$. 
These tensors are defined recursively starting from 
\be
\Pi^{[1]}_{ij}(\vx,\tau) = \frac{2}{3\Om\cH^2} \partial_{x,i}\partial_{x,j}\Phi(\vx,\tau)
= \frac13 \dK_{ij}\, \d_{\rm m} + \Theta_{ij}\,,
\label{eq:hatPi}
\ee
which is proportional to the Hessian of the gravitational potential and contains the leading gravitational 
observables at a given spacetime position $\vx,\tau$: the matter density perturbation $\d_{\rm m}$ and scaled tidal field $\Theta_{ij}$. 
Note the superscript $[1]$, to be distinguished from $(1)$, refers to the fact that $\vec \Pi^{[1]}$ \emph{starts} at first order in perturbation theory, 
but contains higher order terms as well. Thus, $\Theta_{ij}$ is the scaled fully nonlinear tidal field, while $\Theta_{ij}^{(1)}$ will stand for its linear-order component. 
To avoid any confusion, we introduce the label ``generation''  for $[n]$ and reserve the label ``order'' for the  perturbative expansion $(n)$.  
For example, $\lb \vec \Pi^{[1]} \rb^{(3)}$ is then an operator of the first generation and third order. We can define the higher generation tensors $\vec \Pi^{[n]}$ by convective time derivatives:
\be
\vec \Pi^{[n]} = \frac{1}{(n-1)!} \left[(\cH f)^{-1}\frac{D}{D\tau} \vec \Pi^{[n-1]} - (n-1) \vec \Pi^{[n-1]}\right]\,,
\label{eq:Pindef}
\ee
where
\be
\frac{D}{D\tau} \equiv \frac{\partial}{\partial\tau} + v^i \frac{\partial}{\partial x^i}
\label{eq:DDtau}
\ee
is the convective (or Lagrangian) time derivative, and $f=d\ln D/d\ln a$ is the logarithmic growth rate. Including all operators at a given order constructed out of the $\vec \Pi^{[n]}$ then consistently incorporates the effect of time evolution on the final tracer field \cite{MSZ} (see also Section~2 of \cite{biasreview}). 

For example, the set of operators that contribute to the expansion for a scalar field up to cubic order are\footnote{When 
dealing with notation of the trace and trace-free bias operators constructed from the $\vec \Pi$ tensor fields we 
adopt the shorthand notation, e.g. $\tr\big[ \big( \Pi^{[1]} \big)^2 \big]$, instead of writing the tensor elements 
explicitly as $\tr\big[ \Pi_{ij}^{[1]}  \Pi_{jk}^{[1]}  \big]$. 
}
\eq{
{\rm (1)} \qquad &\tr\big[ \Pi^{[1]} \big], \non\\
{\rm (2)} \qquad &\tr\big[ \big( \Pi^{[1]} \big)^2 \big],~~ \Big( \tr\big[ \Pi^{[1]} \big] \Big)^2, \non\\
{\rm (3)} \qquad &\tr\big[ \big( \Pi^{[1]} \big)^3 \big],~~\tr\big[ \big( \Pi^{[1]} \big)^2 \big] \tr\big[ \Pi^{[1]} \big],
~~ \Big( \tr\big[ \Pi^{[1]} \big] \Big)^3,~~~\tr\big[ \Pi^{[1]} \Pi^{[2]} \big].
\label{eq:EulBasis_tr}
}
Notice that this expansion is not only applicable to galaxy densities, but in general to any scalar field that satisfies the local bias assumption. 
In particular, it can be also applied to other properties of galaxies such as their sizes, in which case
\be
  \d_{\rm s} (\vx,\tau) = \sum_O b_O^{({\rm s})} [O]\,.
\label{eq:ds}
\ee
Equations (\ref{eq:dn}) and (\ref{eq:ds}) assume a deterministic relation between the biased tracer field 
and the local gravitational operators. Beyond this set of operators, we should also consider the contributions 
from higher derivative operators and stochasticity attributed to the initial 
conditions on very small scales, which we cover in section \ref{app:higherderiv}.

Equivalently to Eq. (\ref{eq:dn}), we can construct observables from the Fourier-transformed
tracer overdensity field:
\eq{
\df_{\rm n} (\vec k) &= \sum_{n=1}^\infty   (2\pi)^3 \df^{\rm D}_{\vec k - \vec p_{1\cdots n}}K_{\rm n}^{(n)} (\vec p_1 \ldots , \vec p_n) \df_L(\vec p_1) \ldots  \df_L(\vec p_n).
\label{eq:deltag}
}
where $K_{\rm n}^{(n)}$ are bias-dependent symmetrised kernels 
that will be introduced in section \ref{subsec:PT_Fourer}.

Similarly to the case of biased tracers, the EFT \cite{Baumann12, carrasco/hertzberg/senatore} 
provides a standard expansion for the matter field, i.e. what is needed to make theoretical predictions of gravitational lensing (e.g \cite{Foreman+:2016, Modi++:2017}). Up to third order, this is given by
\eeq{
\df_{\rm m} (\vec x) = \df^{(1)} + \df^{(2)} + \df^{(3)} + 2\pi c_s^2 R_{\rm nl}^2 \nabla^2 \df^{(1)},
}
where $c_s^2$ is the effective sound speed of the evolved matter density field, corresponding 
to the leading EFT counterterm in the matter density field,
and $R_{\rm nl}$ is the characteristic scale of the nonlinear regime.
It is again often convenient to work with the Fourier transformed field, 
\eeq{
\df_{\rm m} (\vec k) =
\sum_{n=1}^\infty   (2\pi)^3 \df^{\rm D}_{\vec k - \vec p_{1\cdots n}}F^{(n)} (\vec p_1 \ldots , \vec p_n) \df_L(\vec p_1) \ldots  \df_L(\vec p_n) + \mbox{counterterms},
\label{eq:delta_pt}
}
where $F^{(n)}$ are the symmetrised SPT kernels \cite{bernardeau/etal:2001} and $\vec p_{1\cdots n} \equiv \vec p_1 + \ldots + \vec p_n$.

\subsection{Bias expansion of shapes}
\label{ss:biasexp}

For galaxy shapes, we now want to perform a similar deterministic expansion of the intrinsic three-dimensional shape at lowest order in derivatives,
\eq{
  g_{ij} &= \sum_O b_O^{({\rm g})} [O_{ij}]\,.
  \label{eq:gij}
}
and we need to determine which operators $O_{ij}$ need to be included at a given order in perturbation theory. 
The basic building block for this expansion is still the set of tensor fields $\vec \Pi^{[n]}$ of all generations $n\geq 1$. 
Rather than considering scalar combinations, we now need to take into account all trace-free tensor combinations.
We denote the trace-free operator associated with a generic tensor operator $T_{ij}$ by 
\eq{
  \TF[T]_{ij} &\equiv T_{ij}-\frac{1}{3}\dK_{ij}\dK_{kl}T^{kl}\,.
}
For example, $g_{ij} = \TF[S]_{ij}$. 
Notice that in order to be consistent under renormalisation, trace and trace-free parts in general are multiplied 
by different bias parameters (see appendix~\ref{app:reno} for further discussion), which is why we perform the bias expansion for $g_{ij}$ here, not $S_{ij}$. 

By taking all trace-free combinations of the $\vec \Pi^{[n]}$, we obtain the following list of operators up to third order:
\bea
{\rm 1^{st}} \ && \ \TF[\Pi^{[1]}]_{ij} \label{eq:EulBasis_TF} \\[3pt] 
{\rm 2^{nd}} \ && \ \TF[\Pi^{[2]}]_{ij}\,,\  \TF[(\Pi^{[1]})^2]_{ij}\,,\  \TF[\Pi^{[1]}]_{ij} \tr[\Pi^{[1]}] \nonumber\\[3pt] 
{\rm 3^{rd}} \ && \ \TF[\Pi^{[3]}]_{ij}\,,\  
\TF[\Pi^{[1]} \Pi^{[2]}]_{ij}\,,\  \TF[\Pi^{[2]}]_{ij} \tr[\Pi^{[1]}]\,,   
\nonumber\\[3pt] 
&& \  \TF[(\Pi^{[1]})^3]_{ij}\,,\  \TF[(\Pi^{[1]})^2]_{ij} \tr[\Pi^{[1]}]\,,\  
\TF[\Pi^{[1]}]_{ij} (\tr[\Pi^{[1]}])^2\,,\  
\TF[\Pi^{[1]}]_{ij} \tr[(\Pi^{[1]})^2]\,.
\nonumber
\eea
The number of operators is higher than in the scalar case: we have one
more second order and three more third-order operators. 
As argued by \cite{MSZ}, we do not need to include $\tr[\Pi^{[n]}]$ with $n>1$ in either scalar or trace-free tensor basis, since those operators can always be re-expressed 
in terms of lower order operators that we have already included.  
This \emph{does not} hold for $\TF[\Pi^{[n]}]$ however: the term $\TF[\Pi^{[2]}]$ needs to be included at second order, since
\eeq{
\Pi^{[2]}_{ij} \propto \frac{\partial_i\partial_j}{\nabla^2} \left[ \d_{\rm m}^2 - \frac32 (\Theta_{kl})^2 \right]
}
cannot be expressed as a linear combination of the other second order bias operators.  
This is an important difference to the case of galaxy bias, where $\vec \Pi^{[2]}$ only appears at third order (via the nonlocal operator variously called $O_{\rm td}, \Gamma_3$ and others).
$\TF[\Pi^{[2]}]$ corresponds to the term $t_{ij}$ recently also included in \cite{Blazek17,Schmitz18}. More generally, in the expansion of galaxy shapes at $n$-th order we have terms 
up to including the $n$-th generation, whereas we only obtain terms up to $(n-1)$-th generation in the bias expansion of galaxy number counts (and other scalar observables).

We emphasise again that, while the bias expansion for 3D shapes is independent of how the shape field is precisely defined, 
the values of the bias parameters $b_O^{(\rm g)}$ will, in general, differ depending on this definition. 
For example, when galaxy number weighting is included, one obtains a contribution to one of the second-order bias coefficients given by
\eeq{
b_{\tr[ \Pi^{[1]}]\TF[ \Pi^{[1]}]} \supset b_1 b_{\TF[ \Pi^{[1]}]} \,,
}
where $b_1$ is the linear bias corresponding to the number counts of the galaxy sample whose shape field 
is being measured.\footnote{Consideration of the galaxy number weighting of the shape fluctuation field $S_{ij}$ 
can be viewed as a reinterpretation of Eq.~\eqref{eq:Sij}. 
We could thus decompose the $S_{ij}$ field as 
\eeq{
S_{ij}(\vx) =  \lb 1 + \d_{\rm n}(\vx) \rb \lb \tilde g_{ij}(\vx) + \frac13 \tilde \d_{\rm s}(\vx) \dK_{ij} \rb, 
}
where we have newly defined shape $\tilde g_{ij}$ and size $\tilde \d_{\rm s}$ fields.
The connection to the earlier definition in Eq.~\eqref{eq:Sij} is simply given by the multiplicative, $(1 + \d_{\rm n})$, number weighting factor
\eeq{
 g_{ij}(\vx) =  \lb 1 + \d_{\rm n}(\vx) \rb  \tilde g_{ij}(\vx), ~~~~ \d_{\rm s}(\vx) =  \lb 1 + \d_{\rm n}(\vx) \rb \tilde \d_{\rm s}(\vx).
}
At linear order we still have $b^{\rm g}_1 = \tilde b^{\rm g}_1$ and $b^{\rm s}_1 = \tilde b^{\rm s}_1$, while at second order we obtain terms proportional to $b^{\rm n}_1 \tilde b^{\rm g}_1$ and $b^{\rm n}_1 \tilde b^{\rm s}_1$. However, in either definition there also exists an independent operator ${\tr[ \Pi^{[1]}]\TF[ \Pi^{[1]}]}$,
 as part of the expansion of $g_{ij}(\vx)$ and also $\tilde g_{ij}(\vx)$, whose free coefficient can absorb the difference and thus renders the two expansions formally equivalent.
}

This dependence of the values of bias parameters on the definition of the tracer is of course not unique to shapes: weighted 
galaxy number counts (e.g., by luminosity) also lead to different bias parameters than unweighted counts.
The central point is that, when leaving all bias parameters free, the EFT expansion is able to describe any physical tracer, 
regardless of its selection or weighting as long as these are based on physical observables.

\subsection{Selection effects}

So far, we have assumed that the intrinsic, three-dimensional shape of galaxies (as measured through its surface brightness) 
does not involve any preferred directions apart from those introduced by the large-scale tidal field which is encoded in the 
operators in \refeq{EulBasis_TF}. Then, the statistics of projected galaxy shapes can be derived from \refeq{gij} by projecting 
the three-dimensional shape tensor $g_{ij}$ to a tensor $\gamma_{ij}$ on the sky. For the in-depth treatment of these projection 
effects we refer the reader to the accompanying paper \cite{vlah/chisari/schmidt2019}.

In reality, both observed number counts and shapes can also depend on the orientation of the galaxy with respect to the line of sight. 
For example, the detection probability of luminous red galaxies is higher when they are aligned along the line of sight \cite{hirata:2009,Krause11,Martens18}. 
Similarly, edge-on disks could be detected more easily due to higher signal-to-noise. However, in this case dust absorption within the galaxy could also play a role. 
In order to include such effects, one has to allow for the line of sight to appear in the shape expansion. Ref.~\cite{pkgs} recently 
provided the complete enumeration of these terms for the galaxy density.
Here, we generalise their result to shapes, i.e. to a trace-free three-dimensional tensor instead of a scalar tracer. We obtain the following relevant subset of new terms up to second order:
\bea
{\rm 1^{st}} \ && \ -  \label{eq:selection} \\[3pt] 
{\rm 2^{nd}} \ && \ \TF[ \nhat^k\nhat^l \Pi^{[1]}_{ik} \Pi^{[1]}_{jl} ]\,,\non\
\nhat^k\nhat^l \Pi^{[1]}_{kl} \TF[\Pi^{[1]}]_{ij} \\[3pt] 
{\rm 3^{rd}} \ && \  \TF[ \nhat^k\nhat^l \Pi^{[1]}_{ik} \Pi^{[2]}_{jl} ]\,,\non\
     \nhat^k\nhat^l \Pi^{[1]}_{kl} \TF[\Pi^{[2]}]_{ij}\,,\  
     \nhat^k\nhat^l \Pi^{[2]}_{kl} \TF[\Pi^{[1]}]_{ij}
\,.
\eea
Here, we have not included terms which are proportional to $\nhat^i \nhat^j$. This is because such contributions disappear after projection onto the sky, 
which is the observationally relevant case we focus on in this paper. Further, we have neglected writing the cubic contributions that are of order $(\Pi^{[1]})^3$, 
as  they disappear in the power spectrum after renormalisation.

Note that there is no additional contribution at linear order (and lowest order in derivatives). This is because we do not have any additional indices to contract with $\nhat^k \nhat^l$, so a linear 
term involving $\hat{\v{n}}$ has to be proportional to $\nhat^i\nhat^j$. The same holds for all higher-order terms of the form $\TF[\Pi^{[n]}]_{ij}$. At second order, 
there are two contributions, the second of which involves $\nhat^k\nhat^l \Pi^{[1]}_{kl} \equiv \Pi^{[1]}_\parallel$, which is the leading selection effect that appears 
in the bias expansion of the galaxy density (e.g., \cite{pkgs}). Thus, the second contribution could arise due to the selection effect for aligned galaxies mentioned above, 
coupled with the fact that the shear field is weighted by the galaxy number density. At third order, terms of similar structure appear but now involving  $\vec \Pi^{[2]}$ 
(there is no selection contribution involving $\vec \Pi^{[3]}$ at this order following the above reasoning, while we have not written any of the $(\Pi^{[1]})^3$ contributions).

\subsection{Higher derivatives}
\label{app:higherderiv}

So far, we have worked to lowest order in spatial derivatives.  Essentially,
this assumes that galaxy shapes are perfectly local functions of density
and tidal field along the fluid trajectory.  This is clearly an approximation.  
To go beyond it, we promote the bias parameters in \refeq{gij}
to spatial functionals of the operators.  Then, expanding the operators
inside these functionals in a formal Taylor series, one can show that
one obtains spatial derivatives acting on the operators $O_{ij}$, 
where for each derivative we obtain one power of $R_*$, the typical spatial
extent of the kernel. For example, in the case where dark matter halos are treated 
as a biased tracer, it is often assumed that the typical scale associated with the derivative 
expansion is approximately the halo Lagrangian radius, i.e. $R_* \sim R_L$.
The leading higher derivative term in \refeq{gij} is
\be
R_*^2 \nabla^2 \TF[\Pi^{[1]}]_{ij}\,.
\label{eq:hderivleading}
\ee
At higher order, we obtain terms such as
\be
R_*^2 \nabla^2 \TF[(\Pi^{[1]})^2]_{ij}\,,\  
R_*^2 \TF[ \partial_k \Pi^{[1]} \partial^k \Pi^{[1]}]_{ij}\,, 
\ee
and many others; these terms increase rapidly in number at higher order.  Fortunately,
as long as $R_*$ is of order the halo Lagrangian radius or smaller, 
these terms are fairly small. For example, \refeq{hderivleading} is suppressed by $(R_*k)^2$. In that case, it is sufficient to
keep only \refeq{hderivleading} for the 1-loop power spectrum.  
If the scale $R_*$ is larger than the spatial length scale (nonlinear scale)
controlling the perturbative expansion, then one can correspondingly include
higher-order derivative terms. This might be the case for very massive
halos \cite{Fujita2016, lazeyras/schmidt, abidi/baldauf}, for example, or due to radiative-transfer effects \cite{Pritchard:2006ng,Coles:2007be,Pontzen:2014ena,Schmidt:2017lqe,McQuinn:2018zwa,cabass/schmidt}.

\subsection{Stochasticity}
\label{app:stoch}

The small-scale modes that we integrate over to define the bias
parameters $b_O$ in \refeq{gij} (which are really \emph{responses} of
galaxy shapes to long-wavelength perturbations) lead to stochasticity
or scatter around the mean relations.  In analogy with stochastic
contributions to galaxy bias \cite{biasreview}, \refeq{gij} is generalized to
\eeq{
g_{ij}(\vx,\tau) = \sum_O \left[b_O^{({\rm g})}(\tau) + \eps_O(\vx,\tau) \right] O_{ij}(\vx,\tau) + \eps_{ij}(\vx,\tau)\,.
\label{eq:gij2}
}
The fields $\eps_{ij},\,\eps_O$ are uncorrelated with the $O_{ij}$, 
have vanishing expectation value and moreover are completely described by 
their one-point distributions: $ \< \eps_O(\vx) \eps_{O'}(\vx)\>'$
etc.  While the $\eps_O$ are scalar fields, $\eps_{ij}$ is a trace-free
tensor in keeping with the symmetry properties of $g_{ij}$.  In fact,
$\eps_{ij}$ after projection describes the leading order \emph{shape noise} contribution
to the shape power spectrum
\eeq{
\<\eps_{ij}(\vk) \eps_{kl}(\vk') \>' = \left( \dK_{ik}\dK_{jl} + \dK_{il}\dK_{jk} -  \frac{2}{3} \dK_{ij} \dK_{kl}\right) P^{\rm g}_\eps\,,
\label{eq:stoch_power}
}
where $P^{\rm g}_\eps$ is a white-noise (constant) power spectrum on large scales, 
with corrections toward smaller scales scaling as $k^2 R_*^2$. 
Using the requirement that $\eps_{ij}$ inherits the tensorial properties of the $g_{ij}$ field,
it can be easily shown that this is indeed the only trace-free tensorial structure that we 
can form that does not depend on the direction of the $\vk$ mode (see appendix~\ref{App:SVT_PS}). Since the formation process of tracers is local in real space, the noise cannot depend on $\vec k$ in the limit that $k\to 0$.

Beyond linear order, the higher-order stochastic contributions up to third order are given by
\bea
{\rm 1^{st}} \ && \ \eps_{ij} \label{eq:stochBasis} \\[3pt] 
{\rm 2^{nd}} 
\ && \ \eps^\d_{ij} \tr[ \Pi^{[1]}]\, ,  \non\\
\ && \ \eps_{ \Pi^{[1]}} \TF[\Pi^{[1]}]_{ij}\, , \non\\
{\rm 3^{rd}} 
\ &&  \ \eps^{\d^2}_{ij}  \Big( \tr\big[  \Pi^{[1]} \big] \Big)^2, ~~ \eps^{K^2}_{ij} \, \tr\big[ \big(  \Pi^{[1]} \big)^2 \big]\, , \non\\
\ && \ \eps_{ \Pi^{[2]}}\TF[\Pi^{[2]}]_{ij}\,,\  \eps_{[ \Pi^{[1]}]^2} \TF[(\Pi^{[1]})^2]_{ij}\,,\  \eps_{ \Pi^{[1]} \Pi^{[1]}}\TF[\Pi^{[1]}]_{ij} \tr[ \Pi^{[1]}] \,.
\non
\eea
The second-order contributions on the second line in particular will become
relevant in the three-point function (bispectrum) of shapes, sizes and number counts.

\section{The shape, size and number count fields}
\label{sec:ssn}

We are now ready to obtain expressions for the galaxy shape, size,
and number count fields. We proceed in two steps: first, we transform
all operators to Fourier space. Second, we decompose the tensors into
irreducible spherical tensor components, which greatly simplifies
both the loop integrals and sky projections.

\subsection{Perturbation expansion in Fourier space}
\label{subsec:PT_Fourer}

We use perturbation theory to compute the behaviour of the tracer overdensity and shape fields in the mildly nonlinear regime. 
Given that our goal is to investigate one-loop power spectra and tree-level bispectra, 
it is required to obtain the explicit expressions for the field up to the third order. 
Thus in order to proceed we need to compute the perturbative contributions from each of the bias terms in Eq.~\eqref{eq:EulBasis_tr} and Eq.~\eqref{eq:EulBasis_TF}.
Since, for any operator $\vec{O}^{\rm TF}$ in the list \refeq{EulBasis_TF} there
is an operator $\vec{O}$ such that
\be
O_{ij}^{\rm TF} = \TF[O]_{ij} - \frac13 \dK_{ij} \tr[O],
\ee
we can perform the field expansion in terms of the operators $\vec{I}$
and then take the trace and trace-free components at the end. The latter
will precisely yield the operators in  \refeq{EulBasis_TF}, while the
former will lead to \refeq{EulBasis_tr} with some trivial degeneracies.

Starting with the leading term $\Pi^{[1]}$ in Eq.\eqref{eq:hatPi}, in Fourier space we have
\eeq{
\Pi^{[1]}_{ij}(\vec k) =  \frac{k_ik_j}{k^2} \df_{\rm m}(\vec k),
}
where the perturbative expansion (up to the EFT counterterms) of the matter field $\df_{\rm m}$ is given in Eq.~\eqref{eq:delta_pt}.
Writing the explicit form of these higher order bias operators is straightforward though tedious. For this reason
we refer the interested reader to appendix~\ref{App:PT_shear_field} for the detailed derivation. 

\begin{table}[t!]
\centering
\begin{tabular}{ c c  c  c | c  } 
\hline
 Order  & Gen. & Family & Operator & Kernel ~$K^{(n,q,p)}_{ij}$ \\ \hline \hline
\textbf{(1)} & [1] & 1 & $\Pi^{[1]}_{ij}$  & $\frac{p_ip_j}{p^2} $ \\  \hline
\textbf{(2)} & [1] & 1 & $\Pi^{[1]}_{ij}$ & $\frac{\vec p_{12,i} \vec p_{12,j}} { p_{12}^2 } F^{(2)} (\vec p_1, \vec p_2)$ \\  \cdashline{2-4}
       & [2] & 1 & $\Pi^{[2]}_{ij}$ 
            &$\pi^{(2,2)}_{ij} (\vec p_1,\vec p_2) $\\
       &  & 2 & $\Big[ \big( \Pi^{[1]} \big)^2 \Big]_{ij}$  
         & $\frac{1}{2} \frac{\vec p_1 . \vec p_2}{p_1^2 p_2^2} (p_{1i}p_{2j} + p_{2i}p_{1j})$ \\
       &  & 3 & $\Pi^{[1]}_{ij} \tr \big[ \Pi^{[1]} \big] $  
         & $ \frac{1}{2} \lb \frac{p_{1i}p_{1j} }{p_1^2} + \frac{p_{2i}p_{2j} }{p_2^2}  \rb $ \\ \hline         
\textbf{(3)} & [1] & 1 & $\Pi^{[1]}_{ij}$ & $\frac{\vec p_{123,i} \vec p_{123,j}} { p_{123}^2 } F^{(3)} (\vec p_1, \vec p_2, \vec p_3)$ \\  \cdashline{2-4}
       & [2] & 1 & $\Pi^{[2]}_{ij}$ 
         & $\pi^{(2,3)}_{ij} (\vec p_1, \vec p_2, \vec p_3)$ \\ 
       &  & 2 & $\Big[ \big( \Pi^{[1]} \big)^2 \Big]_{ij}$  
         & $\Bigg[ \frac{ (\vec p_{12}. \vec p_{3} ) ~ \vec p_{12,\{ i}  p_{3,j\}}}{3 p_{12}^2 p_3^2} F^{(2)} (\vec p_1, \vec p_2) \Bigg]_{\rm sim}$  \\
       &  & 3 & $\Pi^{[1]}_{ij} \tr \big[ \Pi^{[1]} \big] $  
       & $\Bigg[ \frac{1}{3}\bigg( \frac{\vec p_{12,i} \vec p_{12,j}} { p_{12}^2 }+\frac{ p_{3,i} p_{3,j}}{p_3^2}  \bigg) 
                 F^{(2)} (\vec p_1, \vec p_2) \Bigg]_{\rm sim}$ \\ \cdashline{2-4}    
       & [3] & 1 & $\Pi^{[3]}_{ij}$ 
         & $\pi^{(3,3)}_{ij} (\vec p_1, \vec p_2, \vec p_3)$ \\ 
       &  & 2 & $\Big[ \big( \Pi^{[1]} \big)^3 \Big]_{ij}$  
         & $\bigg[ \frac{ (\vec p_1 . \vec p_2)(\vec p_2.  \vec p_3)}{ 3 p_1^2 p_2^2 p_3^2} ( \vec p_{1,i} \vec p_{3,j} ) \bigg]_{\rm sim} $ \\
       &  & 3 & $\Big[ \Pi^{[1]} \Pi^{[2]} \Big]_{ij} $  
           & $\bigg[ \frac{\vec p_{\{1,i} \vec p_{1,m}}{p_1^2} \pi^{(2,2)}_{mj\}} (\vec p_2,\vec p_3) \bigg]_{\rm sim}$ \\
        &  & 4 & $\Pi^{[2]}_{ij} \tr \big[ \Pi^{[1]} \big] $  
           & $\Big[ \tfrac{1}{3} \pi^{(2,2)}_{ij} (\vec p_1,\vec p_2) \Big]_{\rm sim}$ \\
       &  & 5 & $\big[ \big( \Pi^{[1]} \big)^2 \big]_{ij}\tr\big[ \Pi^{[1]} \big] $  
         & $\bigg[ \frac{(\vec p_1.\vec p_2) p_{1,\{ i} p_{2,j\}} }{6 p_1^2 p_2^2} \bigg]_{\rm sim}$  \\    
       &  & 6 & $\Pi^{[1]}_{ij} \Big(\tr\big[ \Pi^{[1]} \big] \Big)^2$  
         & $\bigg[ \frac{ p_{1,i} p_{1,j} }{ 3 p_1^2 }\bigg]_{\rm sim}$ \\    
       &  & 7 & $ \Pi^{[1]}_{ij} \tr\big[ \big( \Pi^{[1]} \big)^2 \big] $  
         & $\bigg[ \frac{ p_{1,i} p_{1,j} (\vec p_2 . \vec p_3)^2 }{3 p_1^2p_2^2p_3^2 } \bigg]_{\rm sim}$ \\ \hline    
\end{tabular}
\captionof{table}{
Table of bias kernels $K^{(n,q,p)}_{ij}$ up to the third order in PT. 
All the operators are categorised according to the PT order $n$ and operator generation $q$ they belong to. 
Within each order $n$ and generation $q$, we assign an additional labelling number, that we call ``family number'' $p$. 
Thus, every operator is uniquely identified by the set of integers $(n,q,p)$. Note that the family number is arbitrarily given 
by ordering in this table for convenience in referring to these operators in the following section. 
Importantly, at fixed generation and family number $(q,p)$, operators of different order have the same bias coefficient. 
The label ``sim'' indicates that kernels are to be symmetrised in all $\vec p_i$ momenta.
}
\label{tb:Bias_kernels}
\end{table}

Combining all these results we can write down the full perturbative expansion for the bias 
expansion of the number density $\df_{\rm n}$, size $\df_{\rm s}$ and shape $g_{ij}$ fields at each order in PT. 
Collecting all the terms in Eq.~\eqref{eq:EulBasis_TF} and using the expansions detailed 
above we get the contributions that we summarise in table \ref{tb:Bias_kernels}.

Finally, the biased tracer observables of both density and shape fields can now be
obtained by taking trace and trace-free components of these operators.
We can thus write 
\eq{
\hspace{-2cm} a \in \{ {\rm n},\, {\rm s} \}: ~~  \d_{a}(\vk,\tau) &= \sum_O b_O^{(a)}(\tau) \tr [O_{ij}](\vk,\tau)\,, \non\\
  g_{ij}(\vk,\tau) &= \sum_O b_O^{({\rm g})}(\tau) \TF [O_{ij}](\vk,\tau)\,,
  \label{eq:d_and_gij}
}
where by the set of operators $O_{ij}$ we mean the operators in table~\ref{tb:Bias_kernels}, in addition to the derivative and stochastic operators.
Writing the scalar field bias expansion in this form generates several degenerate terms.
This is expected, since taking the trace of all the operators in \refeq{EulBasis_TF} leads to a subset of linearly independent operators. These degeneracy relations between the trace of 
the operators in table~\ref{tb:Bias_kernels} are explicitly given in appendix~\ref{App:degeneracy}. Once these degeneracies are removed, we recover known results on the bias expansion of the galaxy density (see \cite{biasreview} for a review).

Similarly to the case of scalar fields in Eq.~\eqref{eq:deltag}, it is convenient to work with the Fourier transformed
tensorial field expansion when describing the shape field. We define the expansion of the Fourier transformed symmetric 
rank two tensor field given in Eq. (\ref{eq:Sij}) as
\eeq{
S_{ij}(\vk) = \sum_{n=1}^\infty (2\pi)^3 \d^{\rm D}(\vk-\vp_{1\cdots n}) \left[K^{(n)}_{ij}(\vp_1,\cdots\vp_n) \right] \d_L(\vp_1) \cdots \d_L(\vp_n)\,.
\label{eq:SijPT}
}
where $K^{(n)}_{ij}$ are bias-dependent symmetrised kernels. Notice that we do not indicate explicitly that it refers to the shape field, since we will only deal with this biased field in the following. 
The expressions given in Eq.~\eqref{eq:d_and_gij} can now be obtained by taking either trace or trace-free components of these bias expansion representation of the $\vec S(\vk)$ tensor. 
We also note that the scalar bias kernel $K^{(n)}$ used in case of the scalar field in Eq.~\eqref{eq:deltag},
can be expressed just as the trace of $K^{(n)}_{ij}$, once the degeneracies mentioned above are removed.
Within each order and generation we assign an additional labelling number to each operator $O_{ij}$, which we call family number, 
in order to distinguish all the operators and biases just by their order, generation, and family numbers. 
Note that the family number is arbitrarily given by the ordering given in table~\ref{tb:Bias_kernels} and carries no physical meaning.
For the kernels in Eq.~\eqref{eq:SijPT} we can write an explicit expression in terms of operators given in table~\ref{tb:Bias_kernels} 
(in addition to the derivative and stochastic operators). 
We can thus write
\eq{
\label{eq:bias_kernel}
K^{(n)}_{ij}(\vp_1,\cdots\vp_n) &=  \sum_{q,p} \bigg( c_{q,p}^{(\rm s)}  \tfrac{1}{3} \dK_{ij} \tr \left[ \vec K^{(n,q,p)} (\vp_1,\cdots \vp_n) \right] \\
&\hspace{5cm} + c_{q,p}^{(\rm g)} \TF \left[ \vec K^{(n,q,p)} (\vp_1,\cdots \vp_n) \right]_{ij} \bigg), \non
}
where we have now inserted the bare bias parameters, as they are yet to be renormalised. 
Note that at each PT order $n$ we have contributions of generations such that $q \leq n$, and indices 
`$\rm s$' and `$\rm g$' are respectively labels for bias coefficients of size density and shapes (or equivalently trace and trace-free parts).
Notice that these bias coefficients depend only on the generation $q$ and family $p$ numbers but not the PT order $n$.
One might naively expect that for the field of same tracers the same bias coefficients could potentially be used for 
both trace and trace-free part. However, the renormalisation shows that we are in general not allowed to make this assumption; $c_{q,p}^{(\rm s)}$ and $c_{q,p}^{(\rm g)}$ are different in general. 
For more extensive discussion on this technical but important point we refer the reader to appendix~\ref{app:reno}.
In order to familiarise ourselves with the notation introduced here, it is useful to explicitly write the kernels for 
e.g. the size density field and compare it to the expansion used in previous work. Note this expansion is the same in the case of the number count field, but with different bias coefficients.
We can write down the first and second order kernels as
\eq{
\df_{\rm s}^{(1)} (\vec k) &= \df^{\rm D}_{\vec k - \vp} \tr[K^{(1)}_{ij}](\vp) \df_L(\vp) = \lb c_{1,1}^{\rm s} + c_{R_*}^{\rm s} R_*^2 k^2 \rb \df_L(\vec k), \\
\df_{\rm s}^{(2)} (\vec k) &= \df^{\rm D}_{\vec k - \vp_1 - \vp_2} \tr[K^{(2)}_{ij}](\vp_1, \vp_2)  \df_L(\vp_1)  \df_L(\vp_2) \non\\
&= \df^{\rm D}_{\vec k - \vp_1 - \vp_2} \bigg( 
c_{1,1}^{\rm s} F^{(2)}(\vp_1,\vp_2)  + c_{2,1}^{\rm s} \pi^{(2,2)}_{ij}(\vec p_1,\vec p_2) 
+ c_{2,2}^{\rm s} \frac{(\vp_1 . \vp_2)^2}{p_1^2 p_2^2} + c_{2,3}^{\rm s} \bigg) \df_L(\vp_1)  \df_L(\vp_2). \non\\
&= \df^{\rm D}_{\vec k - \vp_1 - \vp_2} \bigg( 
\tilde c_{1,1}^{\rm s} F^{(2)}(\vec p_1,\vec p_2) 
+ \tilde c_{2,2}^{\rm s} \frac{(\vp_1 . \vp_2)^2}{p_1^2 p_2^2} + \tilde c_{2,3}^{\rm s} \bigg) \df_L(\vp_1)  \df_L(\vp_2). \non
}
Here we have used the shorthand notation for Dirac delta and integration over repeated momenta described in Table I.
In the first line, we have included the leading higher-derivative term as well,
since it can be conveniently combined with the linear bias term.
As we shall discuss in section \ref{sec:bias_ren}, when the trace is taken in the expression 
for $\df_{\rm s}^{(2)}$, the three operators appearing after the second equality exhibit a degeneracy (see Eq. \eqref{eq:trace_second_order}),
and thus one of them can be reabsorbed in the other two operators.  This gives us the last line in the expression for $\df_{\rm s}^{(2)}$. 
We can compare this expression to some of the usual bias expansions and nomenclature used in the literature. 
For example, in ref. \cite{mcdonald/roy:2009} the biased tracer field at second order is expressed in terms 
of second order dark matter field $\df^{(2)}$, square of linear density $[\df_L]^2$ and scaled tidal 
operator\footnote{Note that ref. \cite{mcdonald/roy:2009} uses the $s_{ij}$ label for the tidal field,
instead of $\Theta_{ij}$ used here.} $[\Theta^{(1)}]^2$.
In this basis we can write the above size density field as
\eq{
\df_{\rm s}^{(1)} (\vec k) &= b_1^{\rm s} \lb 1 - \frac{1}{2} b^{\rm s}_{R_*} R^2_* k^2 \rb \df_L(\vec k), \\
\df_{\rm s}^{(2)} (\vec k) &= b_1^{\rm s} \df_{\rm m}^{(2)}(\vec k) + \frac12 b_2^{\rm s} [\df_L]^2(\vec k) + \frac12 b_{\Theta^2}^{\rm s} [\Theta^{(1)}]^2(\vec k) \non\\
&= \df^{\rm D}_{\vec k - \vp_1 - \vp_2} \bigg( 
b_1^{\rm s} F^{(2)}(\vp_1,\vp_2) + \frac12 b_2^{\rm s} + \frac12 b_{\Theta^2}^{\rm s} \lb \frac{(\vp_1 . \vp_2)^2}{p_1^2 p_2^2} -\frac{1}{3} \rb\bigg) \df_L(\vp_1)  \df_L(\vp_2). \non
}
We see that in this second order example it is easy to identify the relation of the bias parameters in the two bases.
These are related by the simple linear transformation
(see also Appendix~C of \cite{biasreview}) 
\eeq{
b_1^{\rm s} = \tilde c_{1,1}^{\rm s},~~~ b^{\rm s}_{R_*} = -2  \tilde c_{R_*}^{\rm s}/{\tilde c_{1,1}^{\rm s}},~~~ 
b_2^{\rm s} = 2 \tilde c_{2,3}^{\rm s} , ~~{\rm and}~~ b_{\Theta^2}^{\rm s} = 2 \tilde c_{2,2}^{\rm s} - \frac{2}{3}  \tilde c_{2,3}^{\rm s}. 
}
Of course, similar relations would be obtained for higher generation bias parameters from the higher order terms in the field expansion.

To reiterate, we see that at the second order in PT, 
we have nominally four bias parameters appearing in both shape and size (or number count) fields. 
These, we label $c_{1,1}$, $c_{2,1}$, $c_{2,2}$ and $c_{2,3}$. 
Linear bias $c_{1,1}$ is the first generation bias parameter and since it is the only one in any of the fields (number, size or shape), we 
can drop the second label and simply write $c_1 \equiv c_{1,1}$. The rest of these parameters carry the common generation index in first place, 
while the second index indicates the fact that there are three operators in this generation. 
As we have seen above, when the trace of these operators is taken, operators are degenerate, and we can redefine
the parameters in order to reduce the operator basis, making it consistent with Eq.~\eqref{eq:EulBasis_tr}. 
However, this will not be the case when trace-free (shape) components are considered.
In this case, no such degeneracy appears at the level of the field, and all of the parameters and operators listed above need to be retained.
Degeneracies as well as renormalisation are further discussed in section \ref{sec:bias_ren} and again, in more detail, in appendix~\ref{App:PT_shear_field}.

\subsection{Irreducible spherical tensor decomposition}

So far we have been discussing the expansion and parametrisation of biased tracers, describing their number densities, sizes and shapes. 
Eventually we are interested in computing statistical correlators, such as power spectra and bispectra between these tracer fields. 
On such correlators and statistical quantities we moreover impose some symmetry properties, like spatial homogeneity, isotropy and parity invariance,
which motivates us to use a specific decomposition of our biased fields in order to optimally utilise these properties.

In this section we particularly focus on isotropy and the consequences it has on the decomposition of the tensor (rank two) fields we have 
been discussing in the previous section. Given this setting, it is useful to decompose our symmetric tensors in the basis tensors that 
have simple transformation properties under the irreducible representations of the SO(3) rotation group. Such tensors we call spherical 
tensors \cite{sakurai2011} of multipole order $\ell=2$. They transform under rotation as
\eeq{
\vec Y_{ij}^{(m)}(\vhat k') = \sum_{m'= - 2}^2  \mathcal D^2_{mm'} (\phi,\theta,\psi) \vec Y_{ij}^{(m')}(\vhat k),
\label{eq:Y_rotation}
}
where $\mathcal D^{2}$ is the rank five Wigner matrix.
We can now decompose any tensor field $T_{ij}(\vec k)$ into these basis functions in terms of helicity $m=0,\pm1,\pm2$ defined with respect to the wavevector $\vk$:
\eq{
\label{eq:SVT_decomp}
T_{ij} (\vec k) 
&= \sum_{\ell = 0,2} \sum_{m=-\ell}^\ell T^{(m)}_\ell (\vec k) \left(\vec{Y}^{(m)}_\ell(\vhat k)\right)_{ij} \\
&= \frac{1}{3}T^{(0)}_0 (\vec k) \dK_{ij} Y^{(0)}  + \sum_{m=-2}^2T^{(m)}_2 (\vec k) \vec Y^{(m)}_{ij}(\vhat k). \non
}
Here, $\vec Y^{(0)}_{0, ij} = Y^{(0)} \dK_{ij}$ is the single helicity-0 mode, and $Y^{(0)}=1$ is defined for symmetry. 
Often, it will be convenient to drop the tensor indices $\ell$ on the basis vectors; 
hence, we choose the boldface symbol $\vec{Y}$ in order to emphasise that we are dealing with a basis  $\ell=2$ tensor, while for the $\ell=0$ component we will use the explicit form given above.
$T^{(m)}_\ell$ are the spherical tensor components transforming under rotation analogously to the basis tensors given in Eq.~\eqref{eq:Y_rotation}.
Spherical tensor components can be directly obtained from $T_{ij}$ by projections using the basis vectors:
\eq{
T^{(0)}_0 (\vec k) = {\rm tr}[ \vec T (\vec k) ] , ~~~  T^{(m)}_2 (\vec k) =  \vec Y^{(m)*}_2(\vhat k) . \vec  T (\vec k) . 
\label{eq:SVT_decomp_inv}
}
It is important to stress that this decomposition is relying on symmetry properties only and no dynamical properties of the field are used thus far.
In other words, the decomposition is equally valid in cases when PT/EFT methods are used to describe the field, as we do in this paper, 
or if one would use e.g. N-body or hydro simulations to describe these fields. In this sense, this decomposition allows us to 
modularize and separate the properties of statistical symmetry of our system from the explicit dynamic characteristics. 

In order to construct the basis tensors explicitly, we first define an orthonormal basis of 3D Euclidean space through
\eeq{
\vec e_1= \frac{\vec k \times \vnhat}{|\vec k \times \vnhat |},~~ \vec e_2 =\vhat k \times \vec e_1~~ \vec e_k = \vhat k\,,
\label{eq:basis}
}
where $\vnhat$ is chosen to be an arbitrary, non-collinear direction to $\vhat k$.
From this, the helicity basis can be constructed as
\be
\vec e^0 = \vec e_k,~~
\vec e^\pm = \mp \tfrac{1}{\sqrt 2} \lb \vec e_1 \mp i \vec e_2 \rb\,.
\ee
Using this basis, the tensor basis functions are defined as
\eq{
& \vec Y^{(0)}_{ij} = N_0 \Big( \hat k_i \hat k_j - \tfrac{1}{3} \dK_{ij} \Big), 
\qquad \vec Y^{(\pm 1)}_{ij} = N_1 \Big( \hat k_j e_i^{\pm} + \hat k_i e_j^{\pm} \Big),
\qquad \vec Y^{(\pm 2)}_{ij} = N_2 e_i^{\pm} e_j^{\pm},
\label{eq:Y_basis}
}
where $N_{0,1,2}=\left\{ \sqrt{\tfrac{3}{2}}, \sqrt{\frac{1}{2}}, 1\right\}$ are normalisation constants so that the basis is orthonormal. Notice that $\vec{Y}^{(m)}_{ij}$ are trace-free.

More details and properties of the helicity basis are given in \refapp{SVT}.
Most importantly, at the power spectrum level, the only nonvanishing contributions involve two fields of the same helicity. 
This is due to statistical isotropy (the absence of preferred directions). 
In addition, parity guarantees that the power spectra of opposite helicity will be the same. 
This fact significantly reduces the number of terms we have to consider in the following.
Similar reductions occur also for higher order statistics.  

\section{Three-dimensional correlation statistics}
\label{sec:Pk3D}

We are now ready to derive the statistics of scalar (size, number) and tensor (shape) fields. Our results will be the two-point functions at 1-loop (next-to-leading) order as well as the three-point functions at leading order.

\subsection{Two-point functions}

There are three types of two-point functions to consider: the auto-correlations of scalar and trace-free tensor fields, and their cross-correlation. We write these as
\eq{
\label{eq:contributions_ss_sg_gg}
  \< \d_a(\vk) \d_b(\vk') \>' &= (2\pi)^3 \df^{\rm D}_{\vec k+\vec k'}P^{ab}(k)\,,\quad\mbox{where}\quad
  a,\,b \in \{{\rm n},\, {\rm s} \} \\
\< \d_a(\vk) g_{ij}(\vk') \>' &= (2\pi)^3 \df^{\rm D}_{\vec k+\vec k'}P^{a{\rm g}}_{ij}(k) \vs
\< g_{ij}(\vk) g_{kl}(\vk') \>' &= (2\pi)^3 \df^{\rm D}_{\vec k+\vec k'}P^{\rm gg}_{ijkl}(k) \,. \non
}
We reiterate that, in our notation, the different fields are denoted as
\eq{
 {\rm n}\::&\  \mbox{fractional galaxy number density perturbation} \\
 {\rm s}\::&\  \mbox{fractional galaxy size perturbation} \vs
  g_{ij}\  ({\rm g})\::&\  \mbox{trace-free galaxy shape perturbation.} \non
  }
In order to obtain expressions for these spectra we can consider the power spectrum of the 
full 3D shape field and then project out the trace and trace-free part.   
We define the power spectrum of the full 3D shape field as 
\eeq{
\la S^{\alpha}_{ij} (\vec k)  S^{\beta}_{kl} (\vec k') \ra = (2\pi)^3 \df^{\rm D}_{\vec k+\vec k'} P^{\alpha\beta}_{ijkl} (k),
}
where ($\alpha$,$\beta$) are the labels of different bias samples (e.g. different mass halos, LRG's, etc). 
However, we will drop these indices here for simplicity and they can be easily reinstated if needed.
Using the decomposition into spherical tensor 
basis given in Eq.~\eqref{eq:SVT_decomp} we can introduce the power spectra of the spherical tensor field 
components $S^{(m)}_2$. Given that the tensors are invariant quantities and the basis spherical tensors 
transform according to Eq.\eqref{eq:Y_rotation}, the components $S^{(m)}_2$ also transform as spherical tensors.
Statistical isotropy and homogeneity then give for the component power spectra 
\eq{
\la S_\ell^{(m)}(\vec k) ~S_{\ell'}^{(m')} (\vec k') \ra = (2\pi)^3 \dK_{mm'} \df^{\rm D}_{\vec k + \vec k'} P^{(m)}_{\ell \ell'} (k),
}
where $\ell$ and $\ell'$ can take values 0 or 2 representing the trace or trace-free components (or alternatively the total angular momentum),
and $m$ and $m'$ are helicity components. We see that as a consequence of statistical isotropy different helicity components do not correlate.
Given that we started from the real shear tensor it follows that $S_\ell^{(m)*}(\vec k) = (-1)^\ell S_\ell^{(m)}(-\vec k)$ which together with the 
parity invariance gives $P^{(m)}_{\ell \ell'} (k) = P^{(-m)}_{\ell \ell'} (k)$.  Combining the above results, the shear-shear tensor 
power spectrum can be decomposed into six independent scalar contributions, i.e. we have
\eeq{
P_{ijkl} 
= \frac{1}{9} \dK_{ij} \dK_{kl}  P^{(0)}_{00}  + \frac{1}{3}  \dK_{\{ij} \vec Y^{(0)}_{kl\}} P^{(0)}_{02} 
+ \vec Y^{(0)}_{ij} \vec Y^{(0)}_{kl}  P^{(0)}_{22} + \sum_{q=1}^2 (-1)^q \vec Y^{(q)}_{\{ij} \vec Y^{(-q)}_{kl\}} P^{(q)}_{22}, 
\label{eq:decomp_ps_main}
}
where $ \dK_{\{ij} \vec Y^{(0)}_{kl\}} = \vec Y^{(0)}_{ij} \dK_{kl} + \dK_{ij} \vec Y^{(0)}_{kl}$, and in the last term we also symmetrize over $(q)$ and $(-q)$ contributions, 
i.e.  $\vec Y^{(q)}_{\{ij} \vec Y^{(-q)}_{kl\}} = \vec Y^{(q)}_{ij} \vec Y^{(-q)}_{kl} + \vec Y^{(-q)}_{ij} \vec Y^{(q)}_{kl}$ (notice that we do not divide by 2).
The first term here is the auto-correlation of the trace components, while the next term describes the cross-correlation of the trace component of one field with the trace-free component of the other. If one is correlating different tracers, then this term separates into two contributions $\propto P^{\alpha\beta}_{02},\ P^{\alpha\beta}_{20}$. 
Finally, the remaining terms correspond to the auto-correlation of the trace-free parts, which consists of three distinct helicity components.
We have thus reduced the, naively, $6\times 6$ components of $P_{ijlm}(\vec{k})$
to 5 (6 in case of different tracers being cross-correlated, as $P^{(0)}_{02} \neq P^{(0)}_{20}$) functions of $k$ only. 
For a more detailed derivation of this decomposition we refer the reader to appendix~\ref{App:SVT_PS}.
It is worth stressing again that the decomposition above is valid in the fully nonlinear case and does not rely on any PT considerations.

We are interested in the auto- and cross-correlators of galaxies and shapes that can be obtained 
by taking the trace or trace-free components, i.e. $\df_{\rm s} = { \rm tr} \big[ \vec S \big] $ 
and $g_{ij} =  { \rm TF} \big[ \vec S \big]_{ij}$. 
Combining the expressions in Eq. \eqref{eq:contributions_ss_sg_gg} and \eqref{eq:decomp_ps_main}
we have
\eq{
\label{eq:spectra_all}
P^{ab}(k) & = P^{ab(0)}_{00} (k),\quad\mbox{where}\quad  a,\,b \in \{ {\rm n},\, {\rm s} \}  \\
P^{a{\rm g}}_{ij} (\vec k) & = \vec Y^{(0)}_{ij}(\vhat k) P^{a{\rm g}(0)}_{02}(k), \non\\
P^{\rm gg}_{ijkl} (\vec k) & = \vec Y^{(0)}_{ij}(\vhat k) \vec Y^{(0)}_{kl}(\vhat k) P^{{\rm gg}(0)}_{22}(k) + \sum_{q=1}^2 (-1)^q  \vec Y^{(q)}_{\{ij}(\vhat k) \vec Y^{(-q)}_{kl\}}(\vhat k) P^{{\rm gg}(q)}_{22}(k)\,. \non
}
The helicity $m =\pm 1$ and $m=\pm 2$ contributions only appear in the shape-shape auto-spectra. 
We also note that the expressions for the auto- and cross-power spectra involving the fractional galaxy 
number density perturbation $\d_{\rm n}$ are identical in form to those for $\d_{\rm s}$, but with different bias coefficients. 
We will thus no longer distinguish between them.

\subsection{One-loop power spectrum components}

The use of the perturbative form of the shear tensor field given by Eq.~\eqref{eq:SijPT} 
allows us to obtain the one-loop PT expression for the power spectrum as
\eeq{
P^{\rm one-loop}_{ijkl} (\vec k) = P^L_{ijkl} (\vec k) + P^{\rm (22)}_{ijkl} (\vec k)  + P^{\rm (13)}_{ijkl} (\vec k) + P^{\rm (31)}_{ijkl} (\vec k)
+ {\rm c.t.} + {\rm h.d.} + {\rm stochastic},
}
where we noted that appropriate loop counterterms (c.t.) and potential higher derivative (h.d.) bias terms (ones that are not already included in $P^L_{ijkl}$ below) have to be added, 
as well as additional stochastic terms discussed in section~\ref{sec:bias}. 
Individual perturbative contributions are defined as
\eq{
(2\pi)^3 \df^{\rm D}_{\vec k + \vec k'} P^L_{ijkl} (\vec k) &= \la S^{(1)}_{ij} (\vec k) S^{(1)}_{kl} (\vec k')  \ra\,, \\ 
(2\pi)^3 \df^{\rm D}_{\vec k + \vec k'} P^{\rm (22)}_{ijkl} (\vec k) &= \la S^{(2)}_{ij} (\vec k) S^{(2)}_{kl} (\vec k')  \ra\,, \non\\
(2\pi)^3 \df^{\rm D}_{\vec k + \vec k'} P^{\rm (13)}_{ijkl} (\vec k) &=  \la S^{(1)}_{ij} (\vec k) S^{(3)}_{kl} (\vec k')  \ra\,.  \non
}
where the (31) contribution is obtained by replacing the $(ij)$ and $(kl)$ index pairs. The linear order result including the leading higher-derivative term then gives
\eeq{
P^{L+{\rm h.d.}}_{ijkl} (\vec k) =\Big( \tfrac{1}{3}c_L^{\rm s} (k) \dK_{ij} + c_L^{\rm g} (k) \vec Y^{(0)}_{ij}\Big) 
\Big( \tfrac{1}{3} c_L^{\rm s} (k) \dK_{kl} + c_L^{\rm g} (k) \vec Y^{(0)}_{kl} \Big) P_ L (k),
\label{eq:linear_PS}
}
where
\eeq{
a, \in \{{\rm n},\, {\rm s},\, {\rm g} \} : ~~ c_L^{\rm a} (k) = c^{\rm a}_{1} + c^{\rm a}_{R_*, 1} R_*^2 k^2 + \mathcal{O}(R_*^4 k^4),
\label{eq:k-dep_b_1}
}
and $c^{\rm s}_{1}$ is the linear bias of fractional size density while $c^{\rm g}_{1}$ is linear bias of the trace-free shape field.  
Note that in Eq.~\eqref{eq:bias_kernel} we used the two index notation for the bias coefficient $c_{p,q}$, 
but for linear level bias this is of course not required since there is only one coefficient, and thus we suppress 
the second index, i.e. $c^{\rm s}_{1} \equiv c^{\rm s}_{1,1}$. 
Second, the remaining terms of order $R_*^2 k^2$ in the expression above are the leading  size and shape derivative bias terms 
\eeq{
 c_1 c_{R_*} R_*^2 k^2 P_L (k),
}
which contribute to results for $P^{(0)}_{00}$, $P^{(0)}_{02}$, $P^{(0)}_{22}$,
while higher helicity terms do not have such contributions at the order
we work in.

The one-loop contributions (22) and (13) are given by
\eq{
P^{\rm (22)}_{ijkl} (\vec k) 
&= 2  ~ K_{ij}^{(2)}(\vec p, \vec k - \vec p) K_{kl}^{(2)}(\vec p, \vec k - \vec p) P_L (p) P_L (\vec k - \vec p) , \non\\
P^{\rm (13)}_{ijkl} (\vec k)
&= 3 ~ \Big( \tfrac{1}{3} c^{\rm s}_{1} \dK_{ij} + c^{\rm g}_{1} \vec Y^{(0)}_{ij} \Big) P_L (k) K_{kl}^{(3)}(\vec k, \vec p, - \vec p) P_L (p) .
}
Applying the decomposition in Eq.~\eqref{eq:spectra_all}
we get the following table: 
\begin{center}
\begin{table}[h!]
\begin{tabular}{ l |  r | l  } 
 \hline
 Order  & Terms & Integrals \\ \hline \hline
{(lin)} & $P^{ab(q)}_{00}= P^{ab(q)}_{02}~~~$ &   \\
                   & $= P^{ab(q)}_{22} =$ & $\dK_{q0}~c^a_1 c^b_1 P_L (k) $ \\  \hline
{(22)}  
       & $P^{ab(0)}_{00}=~$ & $2~ \tr \Big[ \bm K^{(2)}(\vec p, \vec k - \vec p) \Big] ~ \tr \Big[ \bm K^{(2)}(\vec p, \vec k - \vec p) \Big] P_L (p) P_L (\vec k - \vec p)$ \\ 
       & $P^{ab(0)}_{20}=~$ & $2 ~ \vec Y^{(0)*}_2 . \bm K^{(2)}(\vec p, \vec k - \vec p) ~ \tr \Big[ \bm K^{(2)}(\vec p, \vec k - \vec p) \Big] P_ L (p) P_L (\vec k - \vec p)$ \\         
       & $P^{a{\rm g}(0)}_{02}=~$ & $2 ~\tr \Big[ \bm K^{(2)}(\vec p, \vec k - \vec p) \Big]~ \vec Y^{(0)}_2.\bm K^{(2)}(\vec p, \vec k - \vec p) P_L (p) P_L (\vec k - \vec p)$ \\
       & $P^{{\rm gg}(q)}_{22}=~$ & $ 2 ~ \vec Y^{(q)*}_2 . \bm K^{(2)}(\vec p, \vec k - \vec p)~ \vec Y^{(q)}_2 . \bm K^{(2)}(\vec p, \vec k - \vec p) P_L (p) P_L (\vec k - \vec p)$ \\ \hline         
{(13)} 
      & $P^{ab(0)}_{00} = P^{ab(0)}_{20} = ~$ & $3 ~c^{(a)}_1 ~ \tr \Big[ \bm K^{(3)}(\vec k, \vec p, - \vec p) \Big]P_L (p)P_L (k)$ \\  
      & $P^{ab(q)}_{02} = P^{ab(q)}_{22}= ~$ & $3 ~\dK_{q0} ~ c^{(a)}_1 ~ \vec Y^{(0)*}_2 .  \bm K^{(3)}(\vec k, \vec p, - \vec p) P_L (p)P_L (k)$ \\ \hline    
\end{tabular}
\captionof{table}{One-loop contribution decomposition.}
\label{tb:ps_loops}
\end{table}
\end{center}
We see that only the (22)
contribution gives rise to the $m=\pm 1$ and $m=\pm 2$ helicity modes while linear theory as well as (13) contribute only to the $q=0$ helicity modes.
We can also notice that, in case of $q=1$ and $2$, projections of the type $\vec Y^{(q)*}_{ij} \vec Y^{(-q)*}_{kl}P^{\rm gg}_{ijkl} (\vec k)$
directly probe (22) contributions. That is, these projections are the leading order terms on largest scales and thus directly probe the second kernels $\bm K^{(2)}$.

\subsection{Bias degeneracies and renormalisation of the one-loop power spectrum}
\label{sec:bias_ren}

Counting the operators in table~\ref{tb:Bias_kernels}, we see that there are eleven bias operators (besides stochastic and higher-derivative ones). 
However, given that we are interested here in only one-loop level two-point and tree-level three-point statistics we expect additional degeneracies to appear. 
This will reduce the number of independent operator correlators and consequently also the number of independent bias coefficients relevant for these statistics.
In this section we summarize these degeneracies and give results that contain the maximal non-degenerate set of operators for the one-loop power spectrum.  
For the full study of degeneracy and renormalisation for the one-loop power spectrum we refer the interested reader to appendix~\ref{app:renorm}.

\begin{figure}[t!]
\resizebox{\textwidth}{!}{
\centering
\includegraphics[width=\linewidth]{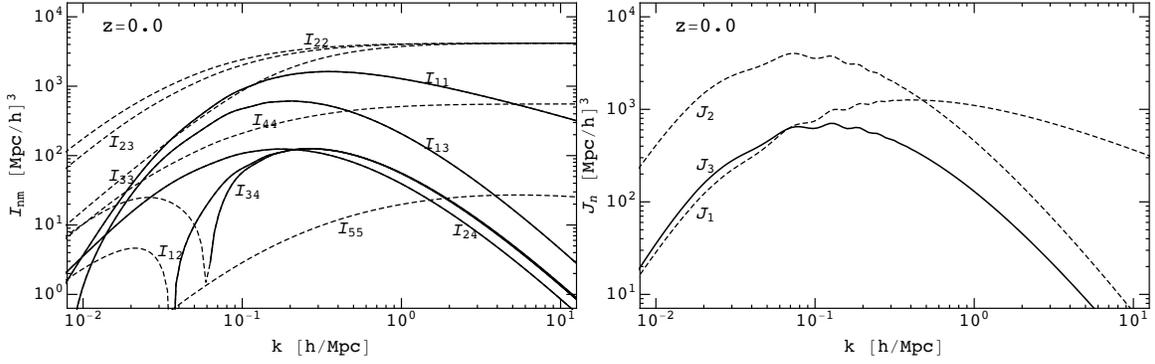}
}
\caption{$I_{nm}(k)$ and $J_n(k)$ contributions that are present in the power spectra of size and shape fields at one-loop order, at $z=0$. 
Dashed lines represent negative contributions while the solid ones are positive. Left panel shows $I_{nm}$
mode coupling terms, obtained from correlating two second order fields. Right panel shows $J_n$, propagator like,
terms that are obtained from correlating the linear and third order fields.}
\label{fig:asymptotic_IandJ}
\end{figure}

After decomposing the second-order operators in table~\ref{tb:Bias_kernels} into helicity components,
we obtain seven independent scalar operators at second order, which we combine into a set of kernels $\bm{\mathcal{K}}^{(2)}(\vec k_1,\vec k_2)$:
\eq{
\bm{ \mathcal K}^{(2)}(\vec{k}_1,\vec{k}_2) \equiv \Big\{  &
\tr \big[ \Pi^{[1](2)} \big], \tr \left[\big( \Pi^{[1]} \big)^2 \right], \left( \tr \left[ \Pi^{[1]} \right]\right)^2, \vec Y^{(0)*}_2(\vec k_{12}) . \big( \Pi^{[1]} \big)^2, \\
&\vec Y^{(1)*}_2(\vec k_{12}) . \big( \Pi^{[1]} \big)^2, \vec Y^{(2)*}_2(\vec k_{12}) . \big( \Pi^{[1]} \big)^2, \vec Y^{(2)*}_2(\vec k_{12}) . \left[ \Pi^{[1]} \tr \big[ \Pi^{[1]} \big] \right] \Big\}(\vec{k}_1,\vec{k}_2). \non
}
Using this operator basis we can evaluate all the contributions to the $(22)$ loop integrals given in table \ref{tb:ps_loops}. 
These contributions are given as the cross-correlations of all the operators in the basis and can be represented by the following integrals
\eeq{
    I_{nm}(k) = \big[\bm{\mathcal K}^{(2)} \big]_n \lb\vec p, \vec k - \vec p\rb \big[ \bm{\mathcal K}^{(2)} \big]_m \lb\vec p, \vec k - \vec p\rb 
P_L (\vec p)P_L (\vec k - \vec p).
\label{eq:I_master_22}
}
Components of the $I_{ij}(k)$ are given in table~\ref{tb:ps_scalar_loops_22}. 
Symmetry of the integral components $I_{ij}(k)=I_{ji}(k)$ is a consequence of the symmetric integrand structure. 
Moreover, these component contributions exhibit further linear dependencies.
Trace and $q=0$ contributions thus give one more dependence expressed by Eq.~\eqref{eq:I_dep_scalar},
while higher helicity correlators give three similar dependencies, given by Eq.~\eqref{eq:I_dep_higher_hel}, after the azimuthal integration is performed.
This finally gives us ten independent integral correlators that need to be evaluated in order to obtain all the $(22)$ contributions in table~\ref{tb:ps_scalar_loops_22}. 
These are
\eeq{
 I_{11}, ~I_{12}, ~I_{13}, ~I_{22}, ~I_{23}, ~I_{24}, ~I_{33}, ~I_{34}, ~I_{44}, ~I_{55} ,
}
explicitly listed in Eq.~\eqref{eq:I_xy_contributions}.
On the left panel of figure~\ref{fig:asymptotic_IandJ} we show the functional form of these terms on the scales of interest 
at $z=0$. We note that all the $I_{nm}$ terms have $\sim k^2$ asymptotic functional dependence on large scales (after the constant part has been subtracted),
except the matter $I_{11}$ term which is $\sim k^4$ (due to the mass conservation equation valid in the dark matter case). 
We subtract the UV-sensitive constant contribution on large scales since this is absorbed 
by the renormalised stochastic power spectrum $P_{\eps}$ in Eq.~\eqref{eq:stoch_power}. We return to this below.
The full biased tracer mode coupling $(22)$ correlators are then determined by two trace, and three trace-free second order bias coefficients
that can be defined as
\eq{
\label{eq:bias_map}
a \in \{ {\rm n},\, {\rm s} \}: ~~ \Big\{ b^{a}_{2,1}, b^{a}_{2,2} \Big\}&, ~~ {\rm where:} ~~  b^{a}_{2,1} = c^a_{2,2} + \tfrac{2}{7} c^a_{2,1}, 
~~~~~~~~  b^{a}_{2,2} =  c^a_{2,3} + \tfrac{5}{7} c^a_{2,1},   \\
{\rm g}: ~ \Big\{ b^{\rm g}_{2,1}, b^{\rm g}_{2,2}, b^{\rm g}_{2,3} \Big\}&, ~~ {\rm where:} ~~b^{\rm g}_{2,1} = c^{\rm g}_{2,1} + c^{\rm g}_{2,2} +c^{\rm g}_{2,3}, 
~~b^{\rm g}_{2,2} = c^{\rm g}_{2,3} + \tfrac{20}{7} c^{\rm g}_{2,1},~~ b^{\rm g}_{2,3} = c^{\rm g}_{2,3}.  \non
}

Let us proceed with investigation of the ${(13)}$ terms in table~\ref{tb:ps_loops}.  
Here, only helicity-zero operators ($q=0$) contribute, appearing in kernel forms
$\tr \big[\bm K^{(3)} \big]$ and $\vec Y^{(0)*}_2 . \bm K^{(3)}$.
As we saw earlier, the reason why only $q=0$ terms contribute to ${(13)}$ term is the coupling of 
linear with third order perturbations. Given that at linear order only scalar ($q=0$) contributions exist,
statistical isotropy ensures only $q=0$ components survive also in all ${(13)}$ terms.
Moreover, the specific form of the kernel input vectors $\vec k$, $\vec p$ and $- \vec p$ gives additional constraints on the 
contributing operators. 
For the trace term, there are five such relations listed in Eq.~\eqref{eq:tr_third_operators_II}, 
leaving six independent operators in the $\tr \big[\bm K^{(3)} \big]$  term. 
The remaining contraction, using the $\vec Y^{(0)*}_2$ basis function, yields similar constraints listed in Eq.~\eqref{eq:Y0_third_operators_II}.
This leaves altogether eight remaining terms in the $\vec Y^{(0)*}_2 . \bm K^{(3)}$ kernel part. 
Using these relations we can re-express the residual contributions in $\bm K^{(3)}$ kernels. 
One particular choice of use of these degeneracy relations is given in table~\ref{tb:ps_scalar_loops_13}, where the surviving independent 
operators are listed with the accompanying bias degeneracies.
Moreover, these operators also exhibit UV-sensitive, potentially divergent, behaviour, which thus needs to be regularised and renormalised. 
In order to isolate these UV-sensitive contributions, we can consider the leading contributions in the low-$k$ Taylor expansion. 
We thus look at the following expression
\eeq{
\lb 3 ~ \int_{\vp} ~ \bm K^{(3)}(\vec k, \vec p, - \vec p) P_L (p)  \rb_{\rm UV}, \non
}
and consider only the set of independent operators given in table~\ref{tb:ps_scalar_loops_13}.
In order to perform this procedure we proclaim the linear bias $c^{\rm s}_1$ parameter to be a `bare' parameter, and to be in fact a 
sum of a finite part and a UV counter term part:
\eeq{
c^{\rm s}_1= \big(c^{\rm s}_1\big)_{\rm fin.} + \big(c^{\rm s}_1\big)_{\rm UV}.
}
The UV counter term above is then chosen to precisely cancel the UV-sensitive kernel parts (explicitly given in Eq.~\eqref{eq:UV_tr}).
This renormalisation procedure generates new degeneracies and after taking these into account we end up with just two independent trace operators
\eeq{
\left \{ \tr \Big[ \Pi^{[1]} \Big],~ \tr \Big[\big( \Pi^{[1]} \big)^2\Big]  \right\}_{\rm fin.}.
\label{eq:tr13_indepe_fnc}
}
This implies one independent third order bias parameter, since $\Pi^{[1]}$ operator multiplies the renormalised linear bias, $\big(c^{\rm s}_1\big)_{\rm fin.}$. 
This result is, as expected, in agreement with earlier obtained one-loop power spectrum results for the density of biased tracers 
(see e.g. \cite{angulo/etal:2015, Fujita2016}).

In analogy to the $I_{nm}$ mode coupling integrals that contribute to ${(22)}$, we can introduce the ${(13)}$ integrals:\footnote{Notice that we never use ordinary Bessel functions in this paper, so no confusion can arise.} 
\eq{
J_{n}(k) = 3
\big[ \bm{\mathcal K}^{(3)} \big]_n \lb \vec k, \vec p, -\vec p\rb P_L (p),
\label{eq:I_master_13}
}
where the relevant integral components are given in Eq~\eqref{eq:J1J2_integral}.
As we shall see,  $J_{n}$ has altogether three independent components, 
even though only two of these contribute to the trace $\tr \big[\bm K^{(3)}\big]$ part.
The third component will contribute only to the trace-free part, as we show below. 

We look next at the trace-free scalar projection $\vec Y^{(0)*}_2 . \bm K^{(3)}$.
The renormalisation procedure is analogous to that for the trace part and detailed explanation is given in and after Eq.~\eqref{eq:renorm_13_Y0K3}. 
It is of interest to also explore the possible dependencies among terms in trace and trace-free basis for the $\bm K^{(3)}$ kernel,
which leads to new relations given in Eq.~\eqref{eq:tr_Y0_degen}.
Finally, for the $\vec Y^{(0)*}_2 . \bm K^{(3)}$ we obtain three independent contributions
\eeq{
\bigg\{  ~  \tr  \Big[ \Pi^{[1]} \Big]_{\rm fin.},  ~~~  ~  \tr  \Big[ \big( \Pi^{[1]} \big)^2 \Big]_{\rm fin.},
 ~~N_0\vec Y^{(0)*}_2 . \Big[ \big( \Pi^{[1]} \big)^2 \Big]_{\rm fin.} \bigg\}.
}
Thus, besides the two independent terms appearing also in trace contributions to the ${(13)}$, trace-free terms, 
we have an additional independent term and thus one additional bias parameter. 
Hence, in addition to the two existing components of the $J_n$ integrals Eq.~\eqref{eq:I_master_13} we 
get a new component given in Eq.~\eqref{eq:J3_integral}. The factor of $N_0$ appears 
because of our choice of normalisation of $\vec Y^{(0)}_2$ tensor basis function. 

Before summarising, notice that the higher-order stochastic terms in
Eq. \eqref{eq:stochBasis}, that we have been neglecting so far, do not lead to additional
contributions. In particular, they are all absorbed in the renormalised values
for $P^{\rm g}_{\eps}$ in Eq.~\eqref{eq:stoch_power}, or similar white noise part in the $P^{\rm s}_{\eps}$ 
case of size and number counts. It is also interesting to note that the cross-correlation power spectrum 
$\la \df(\vk) g_{ij} (\vk') \ra'$  does not have any shot noise contribution at lowest order in $k$, as expected, 
since correlators like $\la \eps (\vk,\tau) \eps_{ij}(\vk',\tau) \ra'$ are zero by symmetry when $k\to 0$. At order $k^2$, there exists a stochastic contribution, however. 

After taking into account all of these degeneracies, functional redundancies and after renormalisation is performed
we obtain the irreducible functional form for trace and trace-free part of ${(13)}$ of the one-loop power spectrum.
Combining these results with results for the ${(22)}$ term we can obtain the full one-loop power spectrum for the biased tracers 
of number densities, sizes and shapes. We have contributions of the following independent bias terms
\eq{
\label{eq:biases}
a, \in \{ {\rm n},\, {\rm s} \} : ~~& \overbrace{ \Big\{ b^{a}_{1} \Big\}}^{P_{11}} \bigcup \overbrace{ \Big\{ b^{a}_{2,1}, b^{a}_{2,2}\Big\}}^{P_{22}} 
\bigcup \overbrace{ \Big\{ b^{a}_{3,1} \Big\}}^{P_{13}} \bigcup \Big\{ b^{a}_{R_*} \Big\} \bigcup \Big\{ {\rm stoch.} \Big\} ,  \\
{\rm g}: ~~& \underbrace{ \Big\{ b^{\rm g}_{1} \Big\}}_{P_{11}} \bigcup \underbrace{ \Big\{ b^{\rm g}_{2,1}, b^{\rm g}_{2,2}, b^{\rm g}_{2,3}\Big\}}_{P_{22}} 
\bigcup \underbrace{ \Big\{ b^{\rm g}_{3,1}, b^{\rm g}_{3,2} \Big\}}_{P_{13}} \bigcup \Big\{ b^{\rm g}_{R_*} \Big\} \bigcup \Big\{ {\rm stoch.} \Big\},  \non
}
with associated one-loop contributions explicitly given in Eq.~\eqref{eq:I_xy_contributions},~\eqref{eq:J1J2_integral} and \eqref{eq:J3_integral}.  
For the one-loop power spectrum of fractional galaxy number density $\df_n$ and fractional galaxy size density perturbation $\df_{\rm s}$ 
we see that we require four bias coefficients (one linear and three non-linear) in addition to the derivative bias terms and stochastic bias terms. 
For the one-loop power spectrum of the trace-free part of galaxy shapes we have two additional (non-linear) bias parameters, that we label $b^{\rm g}_{2,3}$ and $ b^{\rm g}_{3,2}$.
In all cases, the leading derivative contributions are
\eeq{
a,b, \in \{{\rm n},\, {\rm s},\, {\rm g} \} : ~~  b^a_1 b^b_{R_*} R_*^2 k^2 P_L (k),
}
and, as discussed, at one-loop these contribute only to the power spectra with helicity $q=0$. 
Stochastic contributions to the spectra, symbolically added to the parameter list above, consist of white noise contributions 
to the $P^{0}_{00}$ and $P^{q}_{22}$ power spectra. It is important to note however that these white noise contributions 
to the $P^{q}_{22}$ spectra are given by the same parameter for all the helicities $q=0,1$ and $2$.
 
\subsection{Resummation of the IR modes and the BAO}

Standard Eulerian perturbation theory relies on the expansion of density and velocity fields, 
and subsequently correlators of these fields, into perturbations that can be attributed to tidal forces and displacements due to long wavelength modes.
The underlying assumptions is thus that both of these are small. However, on mildly nonlinear scales, the effects of long 
displacement modes can be of order one and thus the simple perturbative expansion is no longer feasible.  
In standard Eulerian PT such contributions mostly cancel out in equal-time correlators, which is a consequence of equivalence principle \cite{peloso/pietroni:2013, kehagias/riotto:2013, creminelli/etal:2013}. 
Nonetheless the effects of these long displacement modes are more prominently visible in the Baryon Acoustic Oscillation (BAO) feature, 
and thus it is of interest to try to handle these displacements in a non-perturbative way. 
It turns out that is possible to perform the resummation of such long (IR) displacement modes, and such procedure is known under the name of IR-resummation.  
For the statistics of number density field the natural setting to perform such IR-resummation is the Lagrangian perturbation theory \cite{senatore/zaldarriaga:2015, Vlah/White/Aviles:2015, vlah/etal:2016, ding/etal:2018}, 
however the equivalent task can be performed also in the Eulerian settings \cite{baldauf/mirbabayi/etal:2015, blas/etal:2016, peloso/pietroni:2017, ivanov/sibiryakov:2018}.

In the simpler rendering of the IR-resummation methods, we first split the linear power spectrum into smooth (nw) and oscillatory parts (w) via
\eeq{
P_L(k) = P^{\rm nw}_L(k) + P^{\rm w}_L(k),
}
where the long displacement resummation now affects only the $P^{\rm w}$ part  \cite{baldauf/mirbabayi/etal:2015, vlah/etal:2016}.
Performing such IR-resummation for the one-loop nonlinear power spectrum, either for dark matter or halo overdensity, 
we get
\eq{
 P^{\rm IR}(k) = P^{\rm nw}_L(k) &+ e^{-\frac{1}{2}\Sigma^2 k^2} \lb 1 + \frac{1}{2}\Sigma^2 k^2 \rb P^{\rm w}_L(k) \\
 &+ P_{\rm 1-loop} \left[  P_L(k) \to  P^{\rm nw}_L(k) + e^{-\frac{1}{2}\Sigma^2 k^2} P^{\rm w}_L(k) \right](k) + \{\rm h.d.\}^{\rm IR} + \ldots, \non
}
where $\Sigma$ is the estimated dispersion of the long mode displacement contributions:
\eeq{
\Sigma^2 = \int_0^\Lambda \frac{dk}{6\pi} ~ \Big[ 1 - j_0(k r_{\rm bao}) +2 j_2 ( k r_{\rm bao}) \Big] P_L(k),
}
and $\Lambda$ is the characteristic scale of the IR mode splitting, although in practice this scale can 
be pushed to arbitrarily small scales, given that contributions to the integral from high $k$ end of $P_L$ is small.
This procedure is equivalent in case of matter and galaxy power spectra, since at leading order of soft modes doing the displacing, 
the displacement field is the same for both matter and biased tracers (due to the equivalence principle). 

In case of the shape spectra and cross-spectra featuring in e.g. Eq.~\eqref{eq:spectra_all} we have 
several spectra $P^{(0)}_{02}$, $P^{(q)}_{22}$ that each follow the same PT expansion as presented above. 
Thus the IR-resummation can be organised in the analogous manner, and we can write a general expression for each of the $P^{(0)}_{00}$, $P^{(0)}_{02}$, $P^{(q)}_{22}$ power spectra 
\eq{
\big[P^{\rm IR}\big]_{\ell \ell'}^{(q)}(k) = \big[ P^{\rm nw}_L & \big]_{\ell \ell'}^{(q)}  (k) + e^{-\frac{1}{2}\Sigma^2 k^2} \lb 1 + \frac{1}{2}\Sigma^2 k^2 \rb \big[ P^{\rm w}_L(k) \big]_{\ell \ell'}^{(q)}  \\
 &+ \bigg[ P_{\rm 1-loop} \left[  P_L(k) \to  P^{\rm nw}_L(k) + e^{-\frac{1}{2}\Sigma^2 k^2} P^{\rm w}_L(k) \right] \bigg]_{\ell \ell'}^{(q)} (k) + \{\rm c.t.\}^{\rm IR} + \ldots. \non
}
The displacement dispersion $\Sigma^2$ can again be estimated in the same way as earlier. 
The IR-resummation procedure will guarantee that the shape of the BAO oscillations is 
described correctly in all of the above spectra, and corresponding correlation functions.

Alternatively, one could perform the full IR-resummation as described in \cite{senatore/zaldarriaga:2015, Vlah/White/Aviles:2015},
which is suitable to apply on Eulerian PT, without resorting to the splitting of linear power spectrum into the wiggle and no-wiggle parts. 
On the other hand, if we start from the bias expansion in Lagrangian coordinates IR-resummation is naturally done by keeping the long 
displacements exponentiated in the power spectrum and correlator estimation \cite{vlah/etal:2016}.

\subsection{Three-point functions}
\label{sec:bis_bias+sto}

We turn next to the three-point function of shapes as well as number and size correlations. 
Even though we will consider only the tree-level result for the bispectrum, the decomposition in spherical tensors that we perform here is valid nonlinearly for any order in PT, 
as was the case for the power spectrum. We saw how the statistical homogeneity, isotropy, and parity invariance contributed in constraining the form of the power 
spectrum of rank two tensor fields. We would like to repeat this exercise here for the case of the bispectrum to obtain the minimal basis template constrained by these symmetries. 
For the bispectrum we correlate three tensor fields, which each contain trace and trace-free parts and are expanded as in Eq.~\eqref{eq:SijPT}. This gives 
\eq{
\la S^{\alpha}_{ij} (\vec k_1)  S^{\beta}_{kl} (\vec k_2) S^{\gamma}_{rs} (\vec k_3) \ra = (2\pi)^3 \df^{\rm D}_{\vec k_1 + \vec k_2 + \vec k_3} B^{\alpha \beta \gamma}_{ijklrs} (\vec k_1, \vec k_2, \vec k_3).
}
where ($\alpha$,$\beta$,$\gamma$) are again the labels of different bias samples (e.g. different mass halos, LRG's, etc), that we drop below.
For more extensive discussion of isotropy and parity invariance we refer the reader to appendix~\ref{app:bispectrum_I}.
As we did in the case of the power spectrum, it is convenient to introduce the bispectrum of spherical tensor components of the $S_{ij}$ tensors, given by
\eq{
\la S_{\ell_1}^{(m_1)}(\vec k_1) ~S_{\ell_2}^{(m_2)}(\vec k_2)~S_{\ell_3}^{(m_3)}(\vec k_3) \ra = (2\pi)^3 \df^{\rm D}_{\vec k_1 + \vec k_2 + \vec k_3} B^{(m_1 m_2 m_3)}_{\ell_1 \ell_2 \ell_3} (\vec k_1, \vec k_2, \vec k_3).
\label{eq:Bellellellm}
}
This provides us with the decomposition of the full bispectrum in terms of the $\vec Y^{(m)}$ basis tensors:
\eq{
B_{ijklrs} (\vec k_1, \vec k_2, \vec k_3)
&= \frac{1}{27} \dK_{ij} \dK_{kl}\dK_{rs}  B^{(0,0,0)}_{000} (\vec k_1, \vec k_2, \vec k_3) \label{eq:bis_decomp_0}  \\
&~~~ +   \frac{1}{9}  \dK_{ij}  \dK_{kl} \sum_{m_3=-2}^2 \vec Y^{(m_3)}_{rs}\big( \vhat  k_3 \big)  B^{(0,0,m_3)}_{002} (\vec k_1, \vec k_2, \vec k_3)
 +  {\rm 2~cycle }\non\\
&~~~ +   \frac{1}{3}  \dK_{ij} \sum_{\substack{m_i=-2 \\ i=(2,3)}}^2 \vec Y^{(m_2)}_{kl}\big( \vhat  k_2 \big) \vec Y^{(m_3)}_{rs}\big( \vhat  k_3 \big) 
B^{(0,m_2,m_3)}_{0 2 2} (\vec k_1, \vec k_2, \vec k_3)
 +  {\rm 2~cycle } \non\\
&~~~ + \sum_{\substack{m_i=-2 \\ i=(1,2,3)}}^2 \vec Y^{(m_1)}_{ij}\big( \vhat  k_1 \big) \vec Y^{(m_2)}_{kl}\big( \vhat  k_2 \big) \vec Y^{(m_3)}_{rs} \big( \vhat  k_3 \big) 
B^{(m_1, m_2, m_3)}_{222} (\vec k_1, \vec k_2, \vec k_3). \non
}
It is not as trivial to employ statistical rotation and parity invariance in 
bispectrum as is for the power spectrum, and we refer the interested reader to appendix~\ref{app:bispectrum_I}
for details. Here we just state that the condition following from these symmetry requirements reads
\eeq{
B^{(m_1,m_2,m_3)}_{\ell_1 \ell_2 \ell_3} = (-1)^{m_1+m_2+m_3} B^{(-m_1,-m_2,-m_3)}_{\ell_1 \ell_2 \ell_3}.
}
Using this we can significantly reduce the number of the independent bispectra of spherical tensor components in the expansion above. 
For the scalar-scalar-tensor $B^{(0,0,m_3)}_{002}$ part we thus have only three independent contributions, while 
the scalar-tensor-tensor $B^{(0,m_2,m_3)}_{022}$ has 13 independent contributions. The tensor-tensor-tensor part has, as expected, 
the richest structure yielding 63 independent contributions. For the explicit decomposition into these independent components 
we refer to Eq.~\eqref{eq:bis_sst}, \eqref{eq:bis_stt} and \eqref{eq:bis_ttt}.
Similar as in the case of power spectrum, all dynamical effects and biasing expansion are contained in these $B^{(m_1 m_2 m_3)}_{\ell_1 \ell_2 \ell_3} $ bispectra of spherical components, while the index structure is captured by the basis tensors $\vec Y^{(m)}$.
We note again that the decomposition above is valid for the fully nonlinear bispectrum and does not rely on PT expansion.
At the end of the next section we will return to the bispectrum and give the explicit tree-level computation in the form of our bias expansion in Eq.~\eqref{eq:EulBasis_TF}.

The observables that we can construct out of these tensors are auto- and cross-correlators of galaxy densities, sizes and shapes that can be obtained 
by taking the trace $\df_{\rm s} = { \rm tr} \big[ \vec S \big] $ and trace-free $g_{ij} =  { \rm TF} \big[ \vec S \big]_{ij}$ components. 
Combining expressions above we have
\eq{
B^{abc}(\vec k_1, \vec k_2, \vec k_3)  & = B^{abc,(0)}_{000} (\vec k_1, \vec k_2, \vec k_3) ,\quad\mbox{where}\quad  a,\,b,\,c \in \{ {\rm n},\, {\rm s} \}  \\
B_{ij}^{ab{\rm g}}(\vec k_1, \vec k_2, \vec k_3)  & = \sum_{m=-2}^2 \vec Y^{(0)}_{ij}\big( \vhat  k_3 \big) B^{ab{\rm g},(m)}_{002} (\vec k_1, \vec k_2, \vec k_3), \non\\
B_{ijkl}^{a{\rm g}{\rm g}}(\vec k_1, \vec k_2, \vec k_3)  & = 
\sum_{\substack{m_i=-2 \\ i=(2,3)}}^2 \vec Y^{(m_2)}_{kl}\big( \vhat  k_2 \big) \vec Y^{(m_3)}_{rs}\big( \vhat  k_3 \big) B^{a{\rm g}{\rm g},(m_2,m_3)}_{0 2 2} (\vec k_1, \vec k_2, \vec k_3), \non\\
B^{{\rm g}{\rm g}{\rm g}}_{ijklrs} (\vec k_1, \vec k_2, \vec k_3) & = 
 \sum_{\substack{m_i=-2 \\ i=(1,2,3)}}^2 \vec Y^{(m_1)}_{ij}\big( \vhat  k_1 \big) \vec Y^{(m_2)}_{kl}\big( \vhat  k_2 \big) \vec Y^{(m_3)}_{rs} \big( \vhat  k_3 \big) 
B^{{\rm g}{\rm g}{\rm g}, (m_1, m_2, m_3)}_{222} (\vec k_1, \vec k_2, \vec k_3), \non
\label{eq:spectra_all}
}
where more explicit decompositions of these bispectra, in terms of minimal number of components, are given in Eq.~\eqref{eq:bis_sst}, \eqref{eq:bis_stt} and \eqref{eq:bis_ttt}.
Note that helicity $\pm 1$ and $\pm 2$ can, in principle, contribute to any of the shape bispectra where at least one shape field is correlated.
At first sight, this might seem unexpected, especially for the size-size-shape contributions, given that in the power spectrum, rotational invariance 
forbade existence of such higher helicity contributions in the cross-correlations with scalar fields. In the bispectrum, however, these symmetry constraints are not as strong,
and higher helicity modes can contribute even in cross-correlations with a single shape field. Naturally, further simplifications on large scales can be obtained once PT is employed,
but those emerge from dynamical and physical reasons rather than symmetries. As expected, the expressions for correlators of the fractional galaxy number 
density perturbation $\d_{\rm n}$ bispectra look equivalent to the $\d_{\rm s}$ bispectra, except for the respective bias coefficients. In the next section, we 
summarise the PT results at tree-level in terms of the above decomposition, while all the calculation details are delegated to appendix~\ref{app:bispectrum_II}.

Before that, we discuss the leading stochastic contributions to the bispectrum, which are less trivial than in the power spectrum case.
Starting from the simplest size-size-size correlator we typically have two type of contributions:
$\la \eps(\vk_1) \eps(\vk_2) \eps(\vk_3) \ra' $ and also $\la \d_{\rm m}(\vk_1) \eps(\vk_2) [\eps  \d_{\rm m}](\vk_3) \ra' \sim {\rm const} ~ P_L(k_2)$. 
Moving on to the size-size-shape correlator we get only one contribution where stochastic operators can contribute at leading order,
\eeq{
 \la  \d_{\rm s} (\vk_1)  \d_{\rm s} (\vk_2) g_{ij}(\vk_3) \ra'  \supset \la \eps(\vk_1)  \d_{\rm m} (\vk_2) \left[ \eps_{\Pi^{[1]}} \TF[\Pi^{[1]}]_{ij} \right] (\vk_3)  \ra'  \sim {\rm const.} ~ P_{L}(k_2)  \vec Y^{(0)}_{ij}( \vhat  k_2 )\,,
 \label{eq:bis_sto_ssg}
}
and, of course, the similar one if we exchange $\vec k_1$ and $\vec k_2$.
This contribution is similar to the second contribution in the size-size-size case above. The size-shape-shape bispectrum has the richest structure in terms of  the stochastic fields. First we have contribution from auto-correlation of stochastic components
\eeq{
\la  \d_{\rm s} (\vk_1) g_{ij}(\vk_2)g_{kl}(\vk_3) \ra'  \supset \la \eps(\vk_1) \eps_{ij}(\vk_2) \eps_{kl}(\vk_3) \ra'  \sim {\rm const.} \left( \dK_{ik}\dK_{jl} + \dK_{il}\dK_{jk} -  \frac{2}{3} \dK_{ij} \dK_{kl}\right) \,.
\label{eq:bis_sto_sgg_1}
}
In addition, there are contributions from the second order stochastic operators 
\eq{
\label{eq:bis_sto_sgg_2}
\la \d_{\rm s} (\vk_1) g_{ij}(\vk_2)g_{kl}(\vk_3) \ra'  &\supset \la \d_{\rm m} (\vk_1) \eps_{ij}(\vk_2) \left[ \eps_{kl}^\d \d_{\rm m} \right] (\vk_3) \ra'\, \\
&\sim {\rm const.} \left( \dK_{ik}\dK_{jl} + \dK_{il}\dK_{jk} -  \frac{2}{3} \dK_{ij} \dK_{kl}\right) P_{L}(k_1)\, , \non\\
\la \d_{\rm s} (\vk_1) g_{ij}(\vk_2)g_{kl}(\vk_3) \ra'  &\supset \la \eps(\vk_1) \TF[\Pi^{[1]}]_{ij}(\vk_2) \left[ \eps_{\Pi^{[1]}} \TF[\Pi^{[1]}]_{kl} \right] (\vk_3) \ra' \non\\
&\sim {\rm const.} ~ P_{L}(k_2)  \vec Y^{(0)}_{ij}( \vhat  k_2 ) \vec Y^{(0)}_{kl}( \vhat k_3 )\,,  \non
}
and the latter contribution can be symmetrised in $\vec k_2$ and $\vec k_3$ variables.
This is an explicit example where the $\eps_O(\vx) O_{ij}(\vx)$ type of operators contribute in a non-degenerate way to the higher $n$-point correlation functions 
from the $\eps_{ij,O}(\vx)O(\vx)$ type of operators, which thus justifies our stochastic operators basis in Eq.~\eqref{eq:stochBasis}.

The last correlator to consider is the shape-shape-shape bispectrum. Here we have only pure tensorial stochastic contributions of $\eps_{ij}$ 
and $ \eps^\d_{ij} \d_{\rm m} $ operators. These contributions are
\eq{
\label{eq:bis_sto_ggg}
\la g_{ij} (\vk_1) g_{kl}(\vk_2)g_{mn}(\vk_3) \ra'  
&\supset \la \eps_{ij}(\vk_1) \eps_{kl}(\vk_2) \eps_{mn}(\vk_3) \ra'\, \\
&\sim {\rm const.} ~ \Delta^{\rm K}_{ij,kl,mn} B_\eps \, , \non\\
\la g_{ij} (\vk_1) g_{kl}(\vk_2)g_{mn}(\vk_3) \ra' 
&\supset \la \TF[\Pi^{[1]}]_{ij}(\vk_1) \left[ \eps_{kl} \d_{\rm m} \right] (\vk_2) \eps_{mn}(\vk_3) \ra'\, \non\\
&\sim {\rm const.} \left( \dK_{km}\dK_{ln} + \dK_{kn}\dK_{lm} -  \frac{2}{3} \dK_{kl} \dK_{mn}\right) P_{L}(k_1)  \vec Y^{(0)}_{ij}( \vhat  k_1 ) \,, \non
}
where the latter contribution is to be symmetrised in all $\vec k_i$ variables.
Here, $B_\eps$ is a white noise tensorial bispectrum amplitude, and we used the $\Delta^{\rm K}_{ij,kl,mn}$ notation for the 
total pair-symmetric, pair-traceless tensor
\eq{
\Delta^{\rm K}_{ij,kl,mn} &= 
\frac{4}{3} \dK_{ij} \dK_{kl} \dK_{mn} 
+ \frac{3}{4} \Big( \dK_{in} \dK_{jk} \dK_{lm} + \dK_{in} \dK_{jl} \dK_{km} + \dK_{im} \dK_{jk} \dK_{ln} + \dK_{im} \dK_{jl} \dK_{kn} \\
&\hspace{4.85cm}+ \dK_{ik} \dK_{jm} \dK_{ln} + \dK_{ik} \dK_{jn} \dK_{lm} + \dK_{il} \dK_{jm} \dK_{kn} + \dK_{il} \dK_{jn} \dK_{km} \Big) \non\\
&\hspace{0.5cm}- \dK_{ij} \dK_{kn} \dK_{lm} - \dK_{ij} \dK_{km} \dK_{ln}
- \dK_{kl} \dK_{in} \dK_{jm} - \dK_{kl} \dK_{im} \dK_{jn}
- \dK_{nm} \dK_{ik} \dK_{jl} - \dK_{nm} \dK_{il} \dK_{jk}\, . \non
}
The contributions above constitute all the stochastic terms that are present at tree level. 
We note that in the expressions above we did not explicitly include all the necessary index and argument permutation, nonetheless, it should be 
clear how these should be added to the total correlators.

\section{Results}
\label{sec:res}

In this section we give explicit expressions for 3D shape power spectra up to one-loop order in PT, and tree-level bispectra. 
We combine the expressions for the spectra given in table~\ref{tb:ps_loops} with independent renormalised biases in Eq.~\eqref{eq:biases}
and explicit integral forms given in  Eq.~\eqref{eq:I_xy_contributions},~\eqref{eq:J1J2_integral} and \eqref{eq:J3_integral}.
Results for the bispectra are presented summarising the results from appendix~\eqref{app:bispectrum_II}.
In order for the results to be as general as possible we also consider the cross-correlations of different types of tracers which we label with $\alpha$, $\beta$ and $\gamma$
\footnote{Note again that these labels are different from $a$ and $b$ labels we used to label either the number density $\df_{\rm n}$ or size density $\df_{\rm s}$ 
and $\df_a$. In this section we have no need in distinguishing between the number density and size since the expressions for size we give are also equally valid for number density.}. 
Thus our formalism distinguishes tracers depending on their type (scalar, tensor) and their internal properties 
(mass, luminosity, metallicity, etc.), which are in turn reflected in the values of the bias coefficients. 

\subsection{One-loop power spectrum}
\label{sec:one-loop res}

We first look at the ${(22)}$ contributions.
Combining all the results given in table~\ref{tb:ps_loops}, with explicit, non-degenerate form of second order kernels 
given in Eq. \eqref{eq:K_tr}, and redefinition of bias terms in Eq. \eqref{eq:bias_b_second} we have 
\eq{
\Big[ P^{\alpha\beta(0)}_{00} \Big]_{(22)}&= 2~ \tr \Big[ \bm{K}_{\alpha}^{(2)}(\vec p, \vec k - \vec p) \Big] ~ \tr \Big[ \bm{K}_{\beta}^{(2)}(\vec p, \vec k - \vec p) \Big] P_L ( p) P_L (|\vec k - \vec p|) \\ 
&= 2 b^{{\rm s}(\alpha)}_{1} b^{{\rm s}(\beta)}_{1} I_{11}(k)
+ 2 b^{{\rm s}\{(\alpha)}_{1} b^{{\rm s}(\beta)\}}_{2,1} I_{12}(k)
+ 2 b^{{\rm s}\{(\alpha)}_{1} b^{{\rm s}(\beta)\}}_{2,2} I_{13}(k) \non\\
&\hspace{3.1cm}+ 2 b^{{\rm s}(\alpha)}_{2,1} b^{{\rm s}(\beta)}_{2,1} I_{22}(k)
+ 2 b^{{\rm s}\{(\alpha)}_{2,1} b^{{\rm s}(\beta)\}}_{2,2} I_{23}(k)
+ 2 b^{{\rm s}(\alpha)}_{2,2} b^{{\rm s}(\beta)}_{2,2} I_{33}(k). \non
}
This result corresponds to the standard density-density bias expansion and can be found in many other references (e.g. \cite{biasreview, assassi/etal, angulo/etal:2015, Fujita2016}). 
Next we consider the ${(22)}$ part of the cross-correlation of density and shape:
\eq{
\Big[P^{\alpha\beta(0)}_{02} = P^{\beta\alpha(0)}_{20} \Big]_{(22)} 
&= 2 ~\tr \Big[ \bm K_{\alpha}^{(2)}(\vec p, \vec k - \vec p) \Big]~ \vec Y^{(0)*}_2. \bm{K}_{\beta}^{(2)}(\vec p, \vec k - \vec p) P_L ( p) P_L (|\vec k - \vec p|) \\ 
&= 2 \sqrt{\tfrac{2}{3}} b^{{\rm s}(\alpha)}_{1} b^{{\rm g}(\beta)}_{1}  I_{11}(k)
+2 \sqrt{\tfrac{2}{3}}  b^{{\rm s}(\alpha)}_{2,1} b^{{\rm g}(\beta)}_{1} I_{12}(k)
+2 \sqrt{\tfrac{2}{3}}  b^{{\rm s}(\alpha)}_{2,2} b^{{\rm g}(\beta)}_{1} I_{13}(k)\non\\
&\hspace{0.5cm} + 2 b^{{\rm s}(\alpha)}_{1} b^{{\rm g}(\beta)}_{2,1} I_{14}(k)
+ 2 b^{{\rm s}(\alpha)}_{2,1} b^{{\rm g}(\beta)}_{2,1} I_{24}(k)
+ 2 b^{{\rm s}(\alpha)}_{2,2} b^{{\rm g}(\beta)}_{2,1} I_{34}(k)\non\\
&\hspace{0.5cm} + b^{{\rm s}(\alpha)}_{1} b^{{\rm g}(\beta)}_{2,2} \sqrt{\tfrac{1}{6}} \lb I_{13}(k) - I_{12}(k) \rb 
+ b^{{\rm s}(\alpha)}_{2,1} b^{{\rm g}(\beta)}_{2,2} \sqrt{\tfrac{1}{6}} \lb I_{23}(k) - I_{22}(k) \rb \non\\
&\hspace{0.5cm}+ b^{{\rm s}(\alpha)}_{2,2} b^{{\rm g}(\beta)}_{2,2} \sqrt{\tfrac{1}{6}} \lb I_{33}(k) - I_{23}(k) \rb,  \non
}
while the scalar part of the shape-shape auto-correlation is given by:
\eq{
\Big[ P^{\alpha\beta(0)}_{22} \Big]_{(22)} 
&= 2~ \vec Y^{(0)*}_2.  \Big[ \bm{K}_{\alpha}^{(2)}(\vec p, \vec k - \vec p) \Big] ~ \vec Y^{(0)*}_2. \Big[ \bm{K}_{\beta}^{(2)}(\vec p, \vec k - \vec p) \Big] P_L ( p) P_L (|\vec k - \vec p|) \\ 
&= \tfrac{4}{3} b^{{\rm g}(\alpha)}_{1} b^{{\rm g}(\beta)}_{1} I_{11}(k)
+ 2 b^{{\rm g}\{(\alpha)}_{1} b^{{\rm g}(\beta)\}}_{2,1} \sqrt{\tfrac{2}{3}}  I_{14}(k)
+  b^{{\rm g}\{(\alpha)}_{1} b^{{\rm g}(\beta)\}}_{2,2}  \tfrac{1}{3}  \lb I_{13}(k) - I_{12}(k) \rb \non\\
&\hspace{0.5cm} + b^{{\rm g}(\alpha)}_{2,2} b^{{\rm g}(\beta)}_{2,2} \tfrac{1}{12} \lb I_{22}(k) - 2 I_{23}(k) + I_{33}(k) \rb
+  b^{{\rm g}\{(\alpha)}_{2,1} b^{{\rm g}(\beta)\}}_{2,2}  \sqrt{\tfrac{1}{6}} \lb I_{34}(k) - I_{24}(k) \rb \non\\
&\hspace{0.5cm} + 2 b^{{\rm g}(\alpha)}_{2,1} b^{{\rm g}(\beta)}_{2,1} I_{44}(k). \non
}
We notice that all these spectra depend only on three bias parameters $b_1,~b_{2,1}$ and $b_{2,2}$, besides the stochastic contributions.
This implies that the scalar power spectra are not sensitive to the third bias operator $b_{2,3}$, which is of course not surprising 
given our choice of basis given in Eq.~\eqref{eq:bias_b_second}. This bias term shows up only in the $q=2$ helicity term.

The $q=1$ and $q=2$ helicities contribute only to the shape-shape auto-correlations. 
For the $q=1$ contributions we have a single contribution
\eq{
\Big[ P^{\alpha\beta(1)}_{22} \Big]_{(22)} &= 2~  \vec Y^{(1)*}_2 . \bm{ K}_{\alpha}^{(2)}(\vec p, \vec k - \vec p)~ \vec Y^{(-1)*}_2 . \bm{ K}_{\beta}^{(2)}(\vec p, \vec k - \vec p) P_L ( p) P_L (|\vec k - \vec p|) \\ 
&= 2 b^{{\rm g}(\alpha)}_{2,1} b^{{\rm g}(\beta)}_{2,1} I_{55}(k). \non
}
At one-loop this term consists thus of the autocorrelation of the operator $\vec Y^{(1)}_2 . ( \Pi^{[1]})^2$. 
It is interesting to note that this $q=1$ projection can then be used to isolate the bias $b^{\rm g}_{2,1}$, i.e. without resorting to higher $n$-point functions, 
and can thus serve as a consistency check of the biasing framework. 
However, as we see from the figure~\ref{fig:asymptotic_IandJ}, on large and mildly nonlinear scales $k<1$Mpc/$h$ this term is suppressed by a few orders of magnitude relative to the other mode coupling $I_{nm}$ terms.
Moreover, we should keep in mind that these are 3D spectra which, when considering the real data survey, eventually need to be projected on the sky. 
These projections further complicate the potential isolation of this term, even though this $q=1$ contribution is the sole one-loop contribution to the
$B$-modes of projected galaxy shapes \cite{vlah/chisari/schmidt2019}, and thus
the sole contaminant from nonlinear intrinsic alignments to searches for
tensor modes using galaxy shapes \cite{GWshear,schmidt/pajer/zaldarriaga,Chisari14}. 
For $q=2$, there are several contributions:
\eq{
\Big[ P^{\alpha\beta(2)}_{22} \Big]_{(22)} &= 2~  \vec Y^{(2)*}_2 . \bm{K}_{\alpha}^{(2)}(\vec p, \vec k - \vec p)~ \vec Y^{(-2)*}_2 .\bm{K}_{\beta}^{(2)}(\vec p, \vec k - \vec p) P_L ( p) P_L (|\vec k - \vec p|) \\ 
&= 2 b^{{\rm g}(\alpha)}_{2,1} b^{{\rm g}(\beta)}_{2,1} I_{66}(k) + 2 b^{{\rm g}\{(\alpha)}_{2,2} b^{{\rm g}(\beta)\}}_{2,3}  \lb I_{67}(k) - I_{66}(k) \rb \non\\
&\hspace{0.5cm}  + 2 b^{{\rm g}(\alpha)}_{2,3} b^{{\rm g}(\beta)}_{2,3} \lb I_{66}(k) - 2 I_{67}(k) + I_{77}(k) \rb. \non
}
As was mentioned before, this is the only one-loop power spectrum channel that depends on the new $b^{\rm g}_{2,3}$ bias parameter. 
All the scale-dependent mode coupling terms $I_{66}$, $I_{67}$ and $I_{77}$ that appear here are given as linear combinations of other $I_{nm}$ terms presented in Eq.~\eqref{eq:I_dep_higher_hel}.
It is however interesting to discuss the scale-dependence of these terms (figure~\ref{fig:asymptotic_I6677}). We see that, at scales where we expect the one-loop results 
to be relevant (at $k\sim0.1$Mpc/$h$ for $z=0$), all of these three functional dependent terms 
in tensor channel are at least one order of magnitude larger than the $I_{55}$ term relevant in the vector channel. 

We proceed to investigate the ${(13)}$ contributions to the one-loop power spectra. 
Here, as discussed earlier, only the $q=0$ part contributes and thus we have the density/size and shape auto- and cross-correlations
\eq{
\Big[ P^{\alpha\beta(0)}_{00} \Big]_{(13)} &= 2 b^{{\rm s}(\alpha)}_{1}    b^{{\rm s}(\beta)}_{1} J_1(k) +  b^{{\rm s}\{(\alpha)}_{1} b^{{\rm s}(\beta)\}}_{3,1} J_2(k), \\
\Big[ P^{\alpha\beta(0)}_{02} \Big]_{(13)} 
&= b^{{\rm s}(\alpha)}_{1}    b^{{\rm g}(\beta)}_{1} \lb 1 + \sqrt{\tfrac{2}{3}} \rb J_1(k) + \lb b^{{\rm s}(\alpha)}_{3,1} b^{{\rm g}(\beta)}_{1} 
+ b^{{\rm s}(\alpha)}_{1} b^{{\rm g}(\beta)}_{3,1} \sqrt{\tfrac{2}{3}}  \rb J_2(k) + b^{{\rm s}(\alpha)}_{1} b^{{\rm g}(\beta)}_{3,2}  J_3(k), \non\\
\Big[ P^{\alpha\beta(0)}_{22} \Big]_{(13)} 
&= 2 b^{{\rm g}(\alpha)}_{1} b^{{\rm g}(\beta)}_{1} \sqrt{\tfrac{2}{3}}J_1(k) + b^{{\rm g}\{(\alpha)}_{1} b^{{\rm g}(\beta)\}}_{3,1}  \sqrt{\tfrac{2}{3}} J_2(k) 
+ b^{{\rm g}\{(\alpha)}_{1} b^{{\rm g}(\beta)\}}_{3,2}  \sqrt{\tfrac{2}{3}} J_3(k). \non
}
Integrals $J_1$, $J_2$ and $J_3$ above are derived in appendix~\ref{app:renorm} and the explicit expressions can be found in Eq.~\eqref{eq:J1J2_integral} and \eqref{eq:J3_integral}.
The factor of $\sqrt{2/3}$ comes from the normalisation of the $N_0$.
We also note that both cross-size-shape and auto-shape-shape power spectra have a new bias dependence $b^{{\rm g}}_{3,2}$.
From the left panel of figure~\ref{fig:asymptotic_IandJ} we see that the contribution of $J_3$ term is comparable to $J_1$ term, but is 
suppressed relative to the $J_2$ on scales of interest ($k<0.15$Mpc/$h$ for $z=0$).
 
These ${(22)}$ and ${(13)}$ terms, in addition to stochastic and derivative contributions, can now be combined to give galaxy number, size and shape two-point statistics, 
up to one-loop. 
\vspace{0.2cm}\\
\textbf{Size-size auto- and cross-correlations} are given by
\eeq{
P^{\rm ss}_{\alpha\beta}(k) = P^{\alpha\beta(0)}_{00} (k) = \Big[P^{\alpha\beta(0)}_{00} \Big]_L (k) +  \Big[ P^{\alpha\beta(0)}_{00} \Big]_{22} (k) 
+ \Big[ P^{\alpha\beta(0)}_{00} \Big]_{13} (k) + \Big[ P^{\alpha\beta(0)}_{00} \Big]_{\eps} (k) + \ldots,
}
where the linear term includes also the leading derivative contributions 
\eeq{
 \Big[P^{\alpha\beta(0)}_{00} \Big]_L (k)  = \lb b^{{\rm s}(\alpha)}_{1}  b^{{\rm s}(\beta)}_{1} + b^{{\rm s}(\{\alpha)}_{1}  b^{{\rm s}(\beta\})}_{R_*} R_*^2 k^2 \rb P_L (k) + \mathcal O\lb R_*^4k^4 \rb,
}
as explicitly given in Eq.~\eqref{eq:linear_PS} and \eqref{eq:k-dep_b_1}, and we include the sum over the symmetrised indices $\alpha$ and $\beta$.
Note that $b^{{\rm s}}_{R_*}$ parameter is also degenerate with the leading counterterms to the (13) loops and thus can be considered as a parameter that also 
captures these renormalisation effects. No new counterterms thus need to be included at this PT order. 
The leading stochastic contributions is given just by the shot noise term
\eeq{
\Big[ P^{\alpha\beta(0)}_{00} \Big]_{\eps} (k) = P^{\alpha\beta}_{{\rm s}, \eps} + \mathcal O(R_*^2 k^2), 
}
where $P^{\alpha\beta}_{{\rm s}, \eps} = {\rm const.}$, as discussed in Sec. \ref{app:stoch}.
\vspace{0.2cm}\\
\textbf{Size-shape cross-correlation} have a simple tensorial structure
\eeq{
\Big[ P^{\rm sg}_{\alpha\beta} (\vec k) \Big]_{ij} = P^{\alpha\beta(0)}_{02}(k) \vec Y^{(0)}_{ij}(\vhat  k),
}
where we have
\eeq{
P^{\alpha\beta(0)}_{02}(k) = \Big[P^{\alpha\beta(0)}_{02} \Big]_L (k) +  \Big[ P^{\alpha\beta(0)}_{02} \Big]_{22} (k) 
+ \Big[ P^{\alpha\beta(0)}_{02} \Big]_{13} (k) + \ldots.
}
The linear term again includes leading derivative contributions 
\eeq{
 \Big[P^{\alpha\beta(0)}_{02} \Big]_L (k)  = \lb b^{{\rm s}(\alpha)}_{1}  b^{{\rm g}(\beta)}_{1} 
 + b^{{\rm s}(\{\alpha)}_{1}  b^{{\rm g}(\beta\})}_{R_*} R_*^2 k^2 \rb P_L (k) + \mathcal O\lb R_*^4k^4 P_L(k) \rb,
}
while the shot noise contributions vanish due to symmetries at leading order in derivatives (see the discussion related to Eq. \eqref{eq:biases}),
no matter which tracers are considered. 
Notice that the term proportional to $b^{{\rm s}(\alpha)}_{1}  b^{{\rm g}(\beta)}_{1}$ corresponds to the case of the LA model \cite{Catelan01}
evaluated in linear theory.
The stochastic contributions in the power spectrum start with the leading 
derivative expansion term  $\sim R_*^2 k^2$, which come from the off-diagonal derivative terms 
$\propto R_*^2 (\partial_i \partial_j - \frac13 \delta_{ij} \nabla^2) \eps(\vec x)$ in the expansion of stochastic fields.\footnote{The 
leading derivative contributions to the stochastic trace-free tensor field come from operators
\eeq{
 R_*^2 \nabla^2 \eps_{ij}(\vec x),~~~ R_*^2(\partial_i \partial_j - \frac13 \delta_{ij} \nabla^2) \eps(\vec x). 
}
In cross-correlations of scalar and tensor fields $\la \df_{\rm s} g_{ij} \ra$ the first term above does not contribute, 
for essentially same reasons as in the case of white noise contribution. The second term however gives a 
non-vanishing contribution
\eeq{
\la \df_{\rm s} g_{ij} \ra' \supset \la \eps \eps_{ij} \ra' \sim {\rm const.}\, R_*^2 (k_i k_j - \delta_{ij} k^2/3),
}
which then contributes to the power spectrum $P^{(0)}_{02}(k) \supset {\rm const.}~\vec Y^{(0)*}_{ij}(\vhat k) R_*^2 (k_i k_j - \delta_{ij} k^2/3) = {\rm const.} ~R_*^2 k^2$.
Note that, one is in fact forced to introduce such terms when considering the UV sensitivity of the mode coupling (22) loop integrals, which 
means that ``const.'' parameter above can also be interpreted as a renormalised stochastic counterterm.
}
\vspace{0.2cm}\\
\textbf{Shape-shape auto- and cross-correlations} are given by the expression
\eeq{
\Big[ P^{\rm gg}_{\alpha\beta} (\vec k) \Big]_{ijlm} = \vec Y^{(0)}_{ij} \vec Y^{(0)}_{lm}  P^{\alpha\beta(0)}_{22}(k) + \sum_{q=1}^2  \vec Y^{\{(q)}_{ij} \vec Y^{(-q)\}}_{lm} P^{\alpha\beta(q)}_{22}(k), 
}
where we have
\eq{
P^{\alpha\beta(0)}_{22}(k) &=  \Big[P^{\alpha\beta(0)}_{22} \Big]_L (k) +  \Big[ P^{\alpha\beta(0)}_{22} \Big]_{22} (k) + \Big[ P^{\alpha\beta(0)}_{22} \Big]_{13} (k) + \Big[ P^{\alpha\beta(0)}_{22} \Big]_{\eps} (k)  + \ldots , \\
P^{\alpha\beta(1)}_{22}(k) &=  \Big[P^{\alpha\beta(1)}_{22} \Big]_{22} (k)  + \Big[ P^{\alpha\beta(1)}_{22} \Big]_{\eps} (k)  + \ldots , \non\\
P^{\alpha\beta(2)}_{22}(k) &=  \Big[P^{\alpha\beta(2)}_{22} \Big]_{22} (k)  + \Big[ P^{\alpha\beta(2)}_{22} \Big]_{\eps} (k)  + \ldots . \non
}
The linear term contributes only to the first equation above and contains also the leading derivative contributions 
\eeq{
 \Big[P^{\alpha\beta(0)}_{22} \Big]_L (k)  = \lb b^{{\rm g}(\alpha)}_{1}  b^{{\rm g}(\beta)}_{1} + b^{{\rm g}(\{\alpha)}_{1}  b^{{\rm g}(\beta\})}_{R_*} R_*^2 k^2 \rb P_L (k) + \mathcal O\lb R_*^4k^4 \rb.
}
Here again the term proportional to $ b^{{\rm g}(\alpha)}_{1}  b^{{\rm g}(\beta)}_{1}$ is the one considered in the LA model and evaluated in linear theory.
Leading stochasticity is given by Eq.~\eqref{eq:stoch_power}, while no other operators in Eq.~\eqref{eq:stochBasis}
yield new contributions (see also the discussion related to Eq. \eqref{eq:biases}). 
Interestingly, all the helicity contributions, $q=0,1,2$, have the same functional form
\eeq{
 \Big[ P^{\alpha\beta(q)}_{22} \Big]_{\eps} (k) = 2 P^{\alpha\beta}_{{\rm g}, \eps} + \mathcal O(R_*^2 k^2),
}
where $P^{\alpha\beta}_{{\rm g}, \eps} \sim {\rm const.} $ is the simple shot noise contribution. 
This fact can potentially also be used as a cross-check of possible systematic errors in the data. 

\begin{figure}
\resizebox{\textwidth}{!}{
\centering
\includegraphics[width=\linewidth]{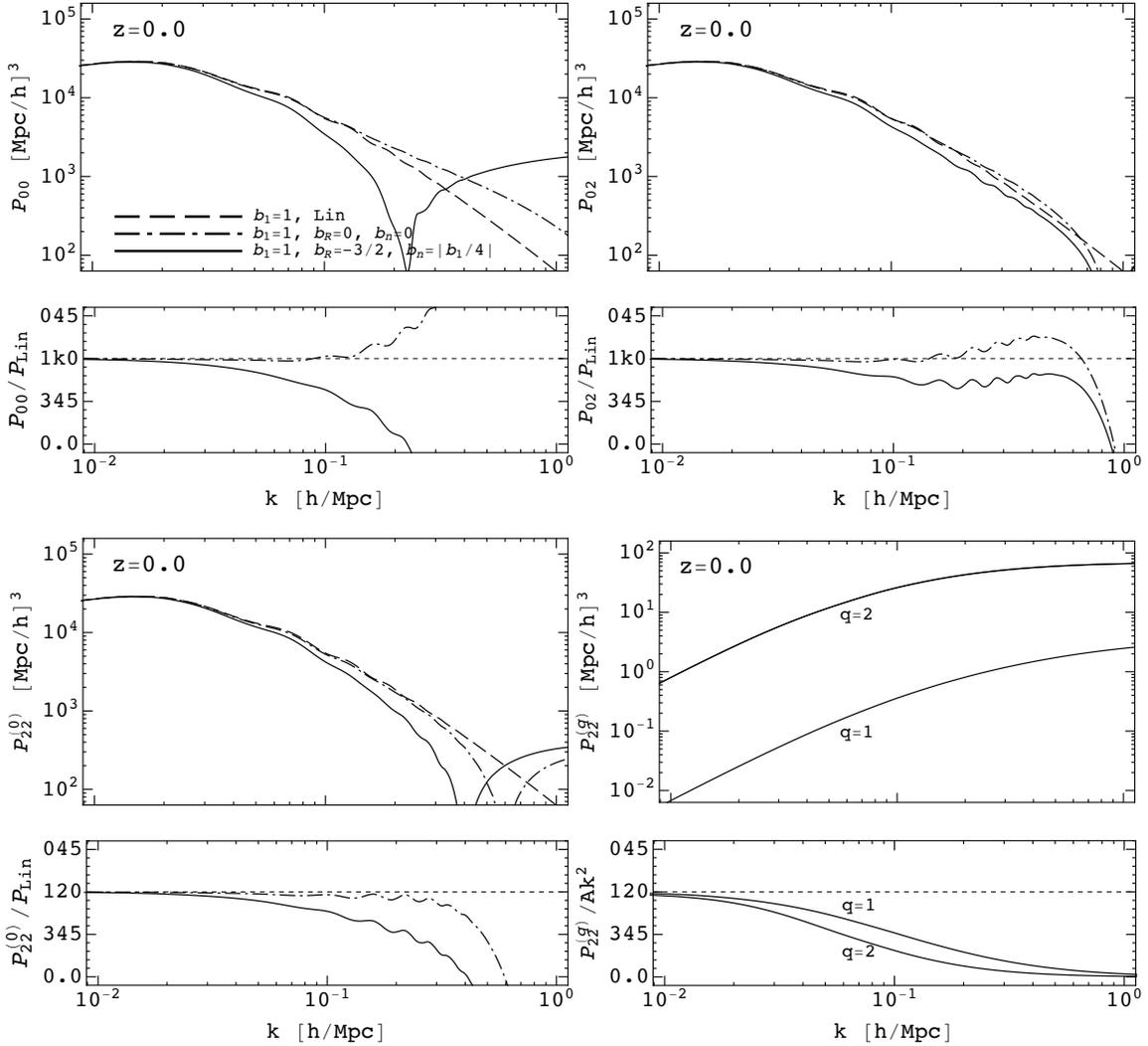}
}
\caption{The size and shape auto- and cross- power spectra. In each of the four panels, the upper plot shows the power spectra themselves, while the lower plot shows the ratio to the linear theory or leading low-$k$ dependence (in case of $q=1$ and $2$). 
In all configurations we use characteristic bias values $b_1=1$, $b_{R_*}= - 3/2$ (and $R_* = 1 {\rm Mpc}/h$), 
as well as $b_{n} = b_1/4$ for all of the other one-loop bias contributions. In all cases, we neglect the stochastic bias contributions.
In the \textit{upper-left} panel, we show the size-size contribution $P^{(0)}_{00}$. Dashed lines are linear theory predictions 
containing only $b_1$ bias parameter, solid lines are nonlinear predictions but again using only $b_1$ bias parameter, 
while the dot-dashed lines show the full one-loop bias result. In the \textit{upper-right} panel, we show size-shape power spectrum $P^{(0)}_{02}$,
containing only helicity zero component. Dashed, solid, and dot-dashed lines again correspond to contributions as in the previous panel. 
\textit{Lower-left} panel shows the helicity zero component of the shape-shape power spectrum $P^{(0)}_{22}$, 
where again, three different lines are the same as in the previous panels.
Lastly, the \textit{lower-right} panel shows helicity one and two components of the shape-shape power spectrum $P^{(1)}_{22}$ and $P^{(2)}_{22}$.
Here the leading order contributions start at one-loop with the characteristic mode coupling $\sim k^2$ dependence at large scales.
Note that the amplitude of the helicity one contribution, which is the only one contributing to the $C_{BB}$ angular spectrum, is approximately 
two orders of magnitude suppressed relative to the helicity two case (which contributes to the $C_{EE}$). 
}
\label{fig:powers_spectra}
\end{figure}

In figure~\ref{fig:powers_spectra} we show auto
power spectra for $P^{(0)}_{00}$, $P^{(0)}_{02}$, $P^{(0)}_{22}$, $P^{(1)}_{22}$ and $P^{(2)}_{22}$. We present the total one-loop power spectra for an arbitrary choice of bias coefficients. In figures, we fix $b_1=1$ and all the 
plots are for $z=0$. (Validation against simulations and observations are left for future work.) In the larger top panels we show several lines; the dashed line corresponds to the linear theory (the LA model \cite{Catelan01}), the dot-dashed corresponds 
to one-loop results where only $b^{\rm s/g}_1$ contributions is kept and the rest of bias parameters are set to zero, and the solid line is the 
full one-loop result with bias choice $b^{\rm s/g}_n = b_1/4$ and $b^{\rm s/g}_{R_*}=-3/2$  fixed (with the scale $R_*$ set to 1 ${\rm Mpc}/h$), and no stochastic contributions.
In the lower panels of figure~\ref{fig:powers_spectra}, we show the ratios to linear theory (except for the $P^{(1)}_{22}$ and $P^{(2)}_{22}$ case where there is no 
linear contribution). 
These plots indicate the relative size of these one-loop contributions when order-unity values for the bias parameters are assumed. 

\begin{figure}[t!]
\centering
\includegraphics[width=0.7\linewidth]{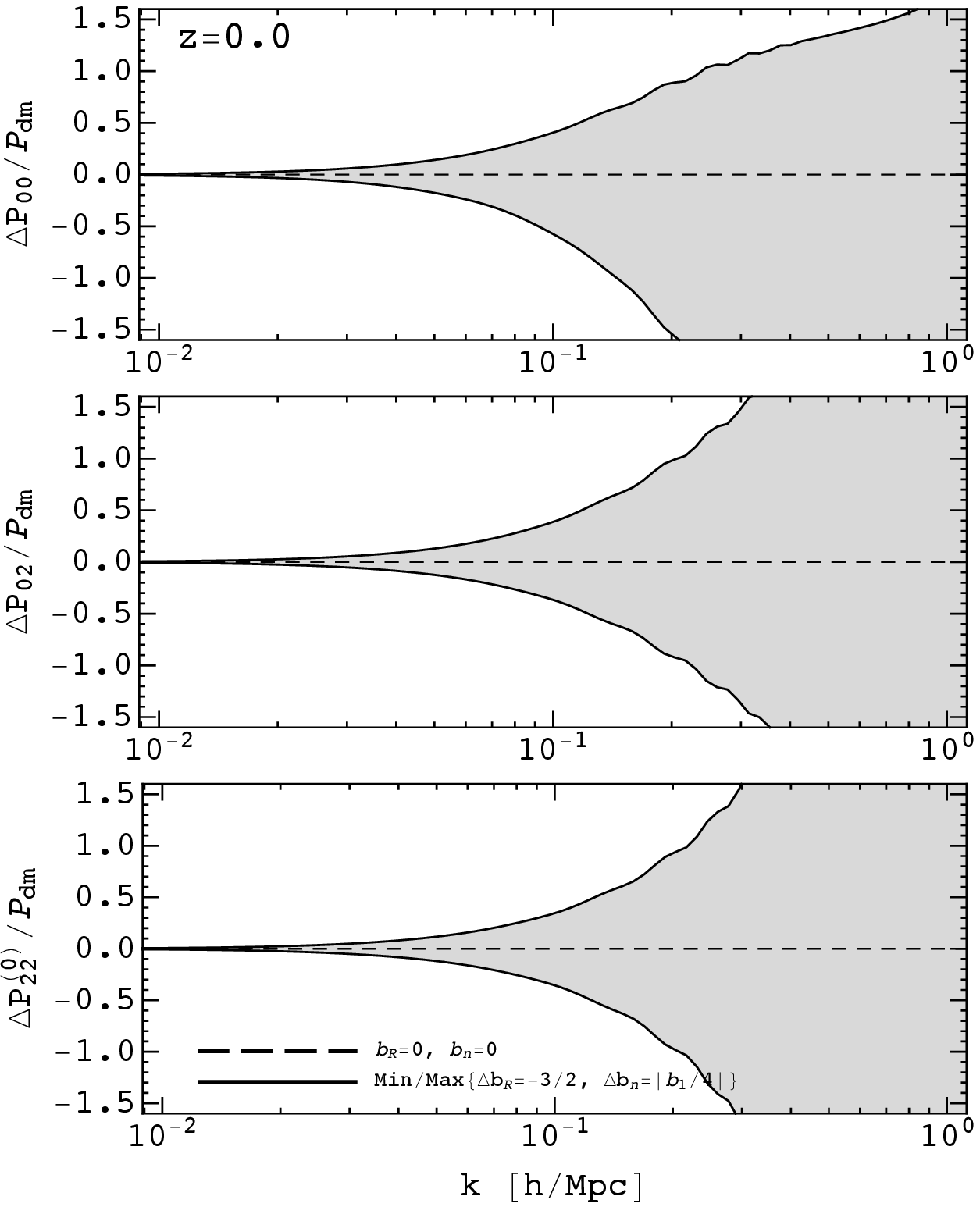}
\caption{Relative contributions of the one-loop bias contributions compared to the dark matter one-loop 
power spectra for $P^{(0)}_{00}$, $P^{(0)}_{02}$, $P^{(0)}_{22}$. We again used $b_1=1$, $b_{R_*}= - 3/2$ 
and for the rest of the biases we use the values $|b_n| = b_1/4$. The sign of bias values $b_n$ are adjusted 
so that the grey area represents the band of potential bias one-loop contributions where $|b_n|<b_1/4$.  
In all cases we neglect the stochastic bias contributions. Effectively this provides an estimate of the validity regime of linear alignment model, provided our bias coefficient choice is realistic.}
\label{fig:asymptotic}
\end{figure}

An alternative way to estimate the relevance of the one-loop contribution is to 
treat the loop corrections as a theoretical error estimates. These we can again do by assuming the order of magnitude values for all 
of the bias parameters. We assume that the higher bias coefficients are approximately of the same order as the linear bias. 
Even though this might not be the true value for each of the bias coefficients, the expectation is 
that for most of coefficients this will be the case. This allows us to estimate the band of relevance where we expect the 
contributions for the one-loop power spectrum to be of importance. Using again the values 
$b^{\rm s/g}_1=1$ and $b^{\rm s/g}_{R_*}=-3/2$ (with $R_* = 1 {\rm Mpc}/h$) we estimate the values of the $b^{\rm s/g}_n$, 
with a limit of $|b^{\rm s/g}_n| < b_1/4$ imposed, such that deviation from the linear theory on the scales of interest ( $k<0.15$Mpc/$h$ at $z=0$)
is maximal. In figure~\ref{fig:asymptotic} we show these estimated error band regions for the three scalar spectrum 
contributions $P^{(0)}_{00}$, $P^{(0)}_{02}$, $P^{(0)}_{22}$, since only scalar spectra have a linear order contribution.
We see that in all three cases one-loop corrections behave similarly and thus we can expect approximately the same accuracy 
in comparison to the e.g. N-body simulations or data from galaxy surveys.
At higher redshift, the accuracy is expected to improve at a given scale, given that there has been less time for structure to grow.

We should also note that the scale $R_*$ controlling the expansion in derivatives (powers of $k^2$) could be substantially different 
from the scale controlling loop corrections. This scale depends on the tracer \cite{vanUitert} and, in particular, also captures non-gravitational effects such as pressure forces, radiative transfer, the choice of shape measurement method or the wavelength of observation \cite{Singh16,Georgiou18,Georgiou19}.

\subsection{Tree-level bispectrum}

Here we give expressions for the auto- and cross-bispectra at tree-level in PT. 
For details we refer the reader to equation~\eqref{app:bispectrum_II}, while 
here we summarize the main results and give explicit expressions for all the possible contributions. 
We consider all possible three-point correlations of galaxy sizes and shapes\footnote{Noting again that correlators 
considering galaxy density $\df_{\rm n}$ are equivalent to the ones with galaxy size $\df_{\rm s}$, with same bias structure but 
different bias values.}:
\vspace{0.2cm}\\
\textbf{Size-size-size correlations} are given by scalar expression\\
\eeq{
B^{\rm sss} (\vec k_1,\vec k_2,\vec k_3) =  B^{(0)}_{000} (\vec k_1, \vec k_2, \vec k_3).
}
This is the usual scalar (for example, number-count) bispectrum at tree-level. These results have been extensively investigated in many earlier works 
(see e.g. \cite{chan/scoccimarro/sheth:2012, baldauf/etal:2012, angulo/etal:2015, Fujita2016, biasreview, pkgs, Eggemeier2019} and references therein, 
for recent discussions). The deterministic bias contributions, can be organised and written as three contributions:
\eeq{
B^{\alpha\beta\gamma,(0)}_{000} (\vec k_1, \vec k_2, \vec k_3) 
= \mathcal F^{\alpha\beta\gamma,(0)}_{0} (\vec k_1, \vec k_2) + \mathcal F^{\beta\gamma\alpha,(0)}_{0} (\vec k_2, \vec k_3) + \mathcal F^{\gamma\alpha\beta,(0)}_{0} (\vec k_3, \vec k_1),
}
where the kernel for size like tracers is given by
\eq{
\label{eq:bis_kernel_F0}
\mathcal F^{\alpha\beta\gamma,(0)}_{0} (\vec k_1, \vec k_2) &= 2 b^{\rm s(\alpha)}_1 b^{\rm s(\beta)}_1 \tr \big[ \bm K_{\gamma}^{(2)}(\vec k_1,\vec k_2) \big] P_L (\vec k_1) P_L (\vec k_2) \\
&=2 b^{\rm s(\alpha)}_1 b^{\rm s(\beta)}_1 \lb b^{\rm s,(\gamma)}_1 F^{(2)} (\vec k_1, \vec k_2) +   b^{\rm s,(\gamma)}_{2,1} \lb \vhat k_1 . \vhat k_2 \rb^2 +  b^{\rm s,(\gamma)}_{2,2} \rb P_L (k_1) P_L (k_2). \non
}
This kernel, besides the dependence on the $b^{\rm s}_1$ bias parameter, depends also on two other second order size bias parameters 
$b^{\rm s}_{2,1}$ and $b^{\rm s}_{2,2}$. The explicit dependence of the kernel at tree-level PT is given in Eq~\eqref{eq:trace_K2} 
and corresponding operators are defined in Table~\ref{tb:Bias_kernels}. 
In addition to these deterministic quadratic order biases, tree-level results also have contribution from stochastic 
field components discussed in Sec.~\ref{sec:bis_bias+sto}. 
The two contributions can be written as
\eq{
B^{\alpha\beta\gamma,(0)}_{000}(\vec k_1, \vec k_2, \vec k_3) \supset ~  {\rm const.}^{\alpha\beta\gamma}_1 + {\rm const.}^{\alpha\beta\gamma}_2 ~ P_{L}(k_1) & +  {\rm 2~cycle}~.
}
\vspace{0.2cm}\\
\textbf{Size-size-shape correlation} is the simplest bispectrum sensitive to the shape fields. 
It consists of correlating one shape and two size fields, and according to Eq.~\eqref{eq:spectra_all} and \eqref{eq:bis_sst}
can be expanded in three independent bispectra of spherical tensor components 
\eq{
B^{\rm ssg}_{ij} (\vec k_1,\vec k_2,\vec k_3) 
&= \sum_{m=-2}^2 \vec Y^{(m)}_{ij}\big( \vhat  k_3 \big)  B^{(0)}_{002} (\vec k_1, \vec k_2, \vec k_3) \\
&= \vec Y^{(0)}_{ij}\big( \vhat  k_3 \big)  B^{(m)}_{002} (\vec k_1, \vec k_2, \vec k_3)
+ 2 \sum_{m=1}^2  {\rm Re} \left[ \vec Y^{(m)}_{ij}\big( \vhat  k_3 \big) \right]   B^{(m)}_{002} (\vec k_1, \vec k_2, \vec k_3). \non
}
This is the term where all four shape bias coefficients can appear in addition to the usual scalar (size) bias coefficients.
We can organise these deterministic bias contributions as follows
\eeq{
B^{\alpha\beta\gamma,(m)}_{002} (\vec k_1, \vec k_2, \vec k_3) 
= \mathcal F^{\alpha\beta\gamma,(m)}_{2} (\vec k_1, \vec k_2) + \tilde \df^{\rm K}_{0,m} \mathcal F^{\beta\gamma\alpha,(0)}_{0} (\vec k_2, \vec k_3) + \tilde \df^{\rm K}_{0,m} \mathcal F^{\gamma\alpha\beta,(0)}_{0} (\vec k_3, \vec k_1)
 + {\rm stoch.},
}
where $\tilde \df^{\rm K}_{0,m} = N_0^{-1} \dK_{0,m}$ is the normalised Kronecker delta function.
In addition to the scalar contributions given in Eq.~\eqref{eq:bis_kernel_F0} we have 
tensor (shear) contributions given by the new kernel
\eq{
\mathcal F^{\alpha\beta\gamma,(m)}_{2} (\vec k_1, \vec k_2) &= 2 b^{\rm s(\alpha)}_1 b^{\rm s(\beta)}_1 \vec Y^{(m)}_{2} \big( \vhat  k_3 \big) . \bm K_{\gamma}^{(2)}(\vec k_1,\vec k_2) P_L (\vec k_1) P_L (\vec k_2).
\label{eq:bis_kernel_F2}
}
The explicit form of this kernel is at tree-level PT is given in Eq.~\eqref{eq:Y0_K2} with corresponding operators given in Table~\ref{tb:Bias_kernels}. 
Besides the dependence on the linear $b^{\rm g}_1$ bias parameter, this shape kernel depends on additional three, second order, shape bias parameters 
$b^{\rm g}_{2,1}$, $b^{\rm g}_{2,2}$ and $b^{\rm g}_{2,3}$.  In order to show explicitly these result we can consider this bispectrum in the plane 
parallel approximation, i.e. we consider only the triangle configurations $\vec k_1,$ $\vec k_2,$ and $\vec k_3$ that lie in the plane parallel to the line of sight. 
Furthermore, we can assume that the $\vec k_3$ vector lies along the $\hat x$ axes in the plane, while $\vec e_1$ is along the $\hat y$ plane axes.
This gives us contributions
\eq{
\mathcal F^{\alpha\beta\gamma,(0)}_{2} (\vec k_1, \vec k_2) &= 2 b^{\rm s(\alpha)}_1 b^{\rm s(\beta)}_1 \frac{2  N_0}{3} \Bigg[ c^{\rm g(\gamma)}_1 F^{(2)} (\vec k_1, \vec k_2) \\
&\hspace{1cm} +  c^{\rm g(\gamma)}_{2,1} \Bigg( \frac{5}{7} \lb1- \lb \vhat k_1 . \vhat k_2\rb^2\rb
+ \frac{1}{4} \lb \hat k_{1x}^2 - \hat k_{1y}^2 + \hat k_{2x}^2 - \hat k_{2y}^2 \rb \non\\
&\hspace{9cm} +  \frac{1}{2} \lb \vhat k_1. \vhat k_2\rb \hat k_{1x} \hat k_{2x} \Bigg) \non\\
&\hspace{1cm} + c^{\rm g(\gamma)}_{2,2} \lb \frac{1}{4} \lb \hat k_{1x}^2 - \hat k_{1y}^2 + \hat k_{2x}^2 - \hat k_{2y}^2 \rb + \frac{1}{2} \lb \vhat k_1. \vhat k_2\rb \hat k_{1x} \hat k_{2x} \rb \non\\
&\hspace{1cm} + c^{\rm g(\gamma)}_{2,3} \frac{1}{4} \lb 2\hat k_{1x}^2 - \hat k_{1y}^2 + 2\hat k_{2x}^2 - \hat k_{2y}^2 \rb \Bigg] P_L (k_1) P_L (k_2), \non\\
\mathcal F^{\alpha\beta\gamma,(1)}_{2} (\vec k_1, \vec k_2) &= 
- 2 b^{\rm s(\alpha)}_1 b^{\rm s(\beta)}_1 \frac{N_1}{\sqrt{2}} \lb c^{\rm g(\gamma)}_{2,1} + c^{\rm g(\gamma)}_{2,2}+ c^{\rm g(\gamma)}_{2,3} \rb \lb \hat k_{1x} \hat k_{1y} +  \hat k_{2x} \hat k_{2y} \rb P_L (k_1) P_L (k_2), \non\\
\mathcal F^{\alpha\beta\gamma,(2)}_{2} (\vec k_1, \vec k_2) &= 
2 b^{\rm s(\alpha)}_1 b^{\rm s(\beta)}_1 N_2 \Bigg[ \lb c^{\rm g(\gamma)}_{2,1} + c^{\rm g(\gamma)}_{2,2} \rb \frac{1}{2} \lb \vhat k_1. \vhat k_2\rb \hat k_{1y} \hat k_{2y} \non\\
&\hspace{6.5cm}+ c^{\rm g(\gamma)}_{2,3} \frac{1}{4} \lb \hat k_{1y}^2 + \hat k_{2y}^2 \rb \Bigg] P_L (k_1) P_L (k_2), \non
}
where the relations between the $c^{\rm g}$ and $b^{\rm g}$ bias coefficients is given in Eq.~\eqref{eq:bias_map}.
However, besides these deterministic bias contributions given explicitly in appendix~\ref{app:bispectrum_II} 
we need to consider also stochastic contribution given in Eq.~\eqref{eq:bis_sto_ssg}.
This contribution can be projected to the bispectrum of spherical tensor components $B^{(0)}_{002}$, 
and it is proportional to the scalar product of two spherical tensors
\eq{
B^{\alpha\beta\gamma,(0)}_{002}(\vec k_1, \vec k_2, \vec k_3) \supset {\rm const.}^{\alpha\beta\gamma} ~ P_{L}(k_2) \lb \vec Y^{(0)} \big( \vhat  k_2 \big).\vec Y^{(0)} \big( \vhat  k_3 \big) \rb
\propto \left[ \lb \vhat  k_2. \vhat  k_3\rb^2  - \frac{1}{3}\right] P_{L}(k_2)\, ,
}
and the similar contribution also comes from exchanging $\vec k_1$ and $\vec k_2$.

It is convenient to perform the change of basis and to define the electric $E$ and magnetic $B$ 
components of the $B^{\rm ssg}_{ij}$ bispectrum by introducing the components of the shear field 
$E= (g_{xx} - g_{yy})/2$, $B= g_{xy}=g_{yx}$ in the plane parallel approximation. 
We will discuss these projection effects in various bases, including the full sky treatment, in more detail 
in the accompanying paper \cite{vlah/chisari/schmidt2019}. Here however, motivated by making the connection to the recent results in \cite{Schmitz18}, 
we stick to this basis. In this new basis, using the coordinate frame set up from above, we obtain 
two independent $E$ and $B$ contributions
\eq{
B^{\alpha\beta\gamma,{\rm ssg}}_{E} (\vec k_1, \vec k_2, \vec k_3) & = \frac{1}{2} N_0 B^{\alpha\beta\gamma,(0)}_{002} (\vec k_1, \vec k_2, \vec k_3) -  \frac{1}{2} N_2 B^{\alpha\beta\gamma,(2)}_{002} (\vec k_1, \vec k_2, \vec k_3) \\
&=  b^{\rm s(\beta)}_1 b^{\rm g(\gamma)}_1 \lb b^{\rm s(\alpha)}_1 F^{(2)} (\vec k_2, \vec k_3) +   b^{\rm s(\alpha)}_{2,1} \lb \vhat k_2 . \vhat k_3 \rb^2 +  b^{\rm s(\alpha)}_{2,2} \rb P_L (k_2) P_L (k_3)  \non\\
&~~~~ + b^{\rm s(\alpha)}_1 b^{\rm g(\gamma)}_1 \lb b^{\rm s(\beta)}_1 F^{(2)} (\vec k_3, \vec k_1) +   b^{\rm s(\beta)}_{2,1} \lb \vhat k_3 . \vhat k_1 \rb^2 +  b^{\rm s(\beta)}_{2,2} \rb P_L (k_3) P_L (k_1) \non\\
&~~~~ + b^{\rm s(\alpha)}_1b^{\rm s(\beta)}_1 \bigg( b^{\rm g(\gamma)}_1 F^{(2)} (\vec k_1, \vec k_2) +\lb b^{\rm g(\gamma)}_{2,2} - b^{\rm g(\gamma)}_{2,3} \rb \frac{1}{4} \lb 1- \lb \vhat k_1 . \vhat k_2\rb^2 \rb \non\\
&\hspace{4.45cm}+ b^{\rm g(\gamma)}_{2,1} \Big[ f_E(\vec k_1) +  f_E(\vec k_2) \Big] \bigg)  P_L (k_1) P_L (k_2) \non\\
&~~~~ +  {\rm const.}^{\alpha\beta\gamma} \lb \left[ \lb \vhat  k_1. \vhat  k_3\rb^2  - \frac{1}{3}\right] P_{L}(k_1) + \left[ \lb \vhat  k_2. \vhat  k_3\rb^2  - \frac{1}{3}\right] P_{L}(k_2) \rb, \non\\
B^{\alpha\beta\gamma,\rm ssg}_{B} (\vec k_1, \vec k_2, \vec k_3) & =  - \sqrt{2} N_1 B^{(1)}_{002} (\vec k_1, \vec k_2, \vec k_3)  \non\\
&= b^{\rm s(\alpha)}_1b^{\rm s(\beta)}_1 b^{\rm g(\gamma)}_{2,1} \Big[ f_B(\vec k_1) +  f_B(\vec k_2) \Big] P_L (k_1) P_L (k_2). \non
}
where, following \cite{Hirata04}, we used abbreviations $f_E(\vec k) = (\hat k^2_{x} - \hat k^2_{y})/2$ and $f_B(\vec k) = \hat k_{x}\hat k_{y}$, 
and we replaced $c^{\rm g}$ bias coefficients with $b^{\rm g}$ biases using the map given in Eq.~\eqref{eq:bias_map}. 
Note that from this projected bispectrum alone one can constrain the combination of bias parameters $b^{\rm g}_1$, $b^{\rm g}_{2,1}$ and 
$b^{\rm g}_{2,2} - b^{\rm g}_{2,3}$.
As mentioned, this bispectrum configuration has been studied in reference \cite{Schmitz18} (see Eq. (2.34) and (2.35)), using equivalent projections on the plane, 
in order to constrain the second order bias parameters. Our results agree with theirs, up to trivial bias coefficient redefinitions and a contribution in $B^{\rm ssg}_{E}$ 
given by $c_s b_1^3 D^4 (1+z) P(k_1) P(k_2)$ (first term in the third line of their Eq. (2.34)), which does not appear in our results; indeed, each contribution to the 
tree-level bispectrum should come with precisely three bias coefficients.
We also give the stochastic term contributing to the  
$B^{\rm ssg}_{E}$ bispectrum component in the last line.
\vspace{0.2cm}\\
\textbf{Size-shape-shape correlation} is the next bispectrum we consider here. 
Once at least two shape fields are correlated, higher helicities $m=\pm1$ and $\pm2$ start contributing in richer ways.
We can decompose this bispectrum as
\eq{
B^{\rm sgg}_{ijkl} (\vec k_1,\vec k_2,\vec k_3) 
= \sum_{\substack{m_i=-2 \\ i=(2,3)}}^2 \vec Y^{(m_2)}_{kl}\big( \vhat  k_2 \big) \vec Y^{(m_3)}_{rs}\big( \vhat  k_3 \big) B^{(m_2,m_3)}_{0 2 2} (\vec k_1, \vec k_2, \vec k_3),
}
where the explicit expansion can be written in terms of 13 independent contributions $B^{(m_2,m_3)}_{0 2 2}$, as given in Eq.~\eqref{eq:bis_stt}.
However, since at least two linear PT level fields need to be correlated and those carry only scalar contributions, we can further simplify these results at the tree-level PT. 
It follows that only bispectra with only one $m_i$ different from zero can contribute. In other words, contributions like $B^{(\pm1,\pm1)}_{0 2 2}$,
$B^{(\pm1,\pm2)}_{0 2 2}$ and $B^{(\pm2,\pm2)}_{0 2 2}$ appear only at PT orders higher than tree-level. 
These reduces the number of independent terms to altogether five contributions, of $B^{(0,m_3)}_{0 2 2}$ and $B^{(m_2,0)}_{0 2 2}$ form.
In terms of the bias expansion, these two contributions have equivalent functional forms, up to the exchange of $\vec k_2$ and $\vec k_3$ variables.
Thus for the tree-level result we can further simplify the decomposition above and write
\eq{
B^{\rm sgg}_{ijkl} (\vec k_1,\vec k_2,\vec k_3) 
= \vec Y^{(0)}_{ij} \big( \vhat  k_2 \big) \vec Y^{(0)}_{kl}\big( \vhat  k_3 \big) B^{(0,0)}_{0 2 2} 
+ 2 \sum_{m=1}^2 \bigg(& {\rm Re} \left[  \vec Y^{(m)}_{ij} \big( \vhat  k_2 \big) \vec Y^{(0)}_{kl}\big( \vhat  k_3 \big) \right]  B^{(m,0)}_{0 2 2} \\
&+ {\rm Re} \left[  \vec Y^{(0)}_{ij}\big( \vhat  k_2 \big) \vec Y^{(m)}_{kl} \big( \vhat  k_3 \big) \right] B^{(0,m)}_{0 2 2} \bigg). \non
}
At tree PT order, deterministic bias contributions are given in terms of kernels $\mathcal F^{(0)}_{0}$ and $\mathcal F^{(m)}_{2}$ that we already 
introduced above Eqs.~\eqref{eq:bis_kernel_F0} and \eqref{eq:bis_kernel_F2} respectively.  We have
\eq{ 
B^{\alpha\beta\gamma,(m_2,m_3)}_{022} (\vec k_1, \vec k_2, \vec k_3) &= \tilde \df^{\rm K}_{0,m_2} \tilde \df^{\rm K}_{0,m_3}  \mathcal F^{\alpha\beta\gamma,(0)}_{0} (\vec k_2, \vec k_3)
  + \tilde \df^{\rm K}_{0,m_2} \mathcal F^{\beta\gamma\alpha,(m_3)}_{2} (\vec k_1, \vec k_2)  \\
&\hspace{5.0cm}+ \tilde \df^{\rm K}_{0,m_3} \mathcal F^{\gamma\alpha\beta,(m_2)}_{2} (\vec k_1, \vec k_3) + {\rm stoch.}, \non
}
where Kronecker delta functions reflect the constraints of the tree-level PT result discussed above.
These deterministic biases are also more extensively discussed in appendix~\ref{app:bispectrum_II}.

Besides deterministic bias contributions, this bispectrum has the richest stochastic structure given by three distinct 
terms in Eq.~\eqref{eq:bis_sto_sgg_1} and \eqref{eq:bis_sto_sgg_2}. This contribution can be projected to the 
bispectrum of spherical tensor components $B^{(0,0)}_{0 2 2}$, and it is proportional 
to the scalar product of two spherical tensors
\eq{
B^{\alpha\beta\gamma,(m_2,m_3)}_{0 2 2}(\vec k_1, \vec k_2, \vec k_3) 
\supset \tilde \df^{\rm K}_{0,m_2} & \tilde \df^{\rm K}_{0,m_3} \Big( {\rm const.}^{\alpha\beta\gamma}_1 + {\rm const.}^{\alpha\beta\gamma}_2 ~ P_{L}(k_1)  \\
&+ {\rm const.}^{\alpha\beta\gamma}_3 ~ \big[ P_{L}(k_2) + P_{L}(k_3) \big] \Big) \lb \vec Y^{(0)} \big( \vhat  k_2 \big).\vec Y^{(0)} \big( \vhat  k_3 \big) \rb.\non
}
As indicated by the Kronecker delta functions, it contributes only to the $q=0$ component, while higher helicity 
components are free of constant, Poisson noise-like, contributions. For these, the stochasticity starts with the derivative terms
like the ones discussed earlier when we looked at the power spectrum.
\vspace{0.2cm}\\
\textbf{Shape-shape-shape correlation} is the shape auto-bispectrum and consists of only shape bias parameters and operators. 
It has the richest correlator structure where in principle all the helicities can contribute. However, for the same reason as for the 
size-shape-shape correlator, only components with one non-zero helicity index $m_i$ survive at tree level. We can decompose the 
auto-tensor bispectrum as 
\eq{
B^{\rm ggg}_{ijklrs} (\vec k_1,\vec k_2,\vec k_3)& = 
\sum_{\substack{m_i=-2 \\ i=(1,2,3)}}^2 \vec Y^{(m_1)}_{ij}\big( \vhat  k_1 \big) \vec Y^{(m_2)}_{kl}\big( \vhat  k_2 \big) \vec Y^{(m_3)}_{rs} \big( \vhat  k_3 \big)
B^{(m_1, m_2, m_3)}_{222} (\vec k_1, \vec k_2, \vec k_3).
}
As argued above, in the expansion given in Eq.~\eqref{eq:bis_ttt} only terms with only one nonzero $m_i$ contribute and we have 
\eq{
B^{\rm ggg}_{ijklrs} (\vec k_1,\vec k_2,\vec k_3) =  \vec Y^{(0)}_{ij}\big( \vhat  k_1 \big) &\vec Y^{(0)}_{kl}\big( \vhat  k_2 \big) \vec Y^{(0)}_{rs}  \big( \vhat  k_3 \big)  B^{(0, 0,0)}_{222} \\
&+ 2  \sum_{m=1}^2 \bigg( {\rm Re} \left[ \vec Y^{(0)}_{ij}\big( \vhat  k_1 \big) \vec Y^{(0)}_{kl} \big( \vhat  k_2 \big) \vec Y^{(m)}_{rs} \big( \vhat  k_3 \big) \right] B^{(0, 0,m)}_{222} + {\rm 2~cycle} \bigg). \non
}
This gives us just seven independent contributions, which is a drastic reduction from the full case of 63 independendt contributions. 
Deterministic bias in these contributions consist only of $\mathcal F^{(m)}_{2}$ kernels 
given in Eq.~\eqref{eq:bis_kernel_F2}, and component bispectra can be written as 
\eq{
B^{\alpha\beta\gamma,(m_1,m_2,m_3)}_{222} (\vec k_1, \vec k_2, \vec k_3) 
&= \tilde \df^{\rm K}_{0,m_1} \tilde \df^{\rm K}_{0,m_2}  \mathcal F^{\alpha\beta\gamma,(m_3)}_{2} (\vec k_1, \vec k_2) \\
&~~~ + \tilde \df^{\rm K}_{0,m_1} \tilde \df^{\rm K}_{0,m_3}  \mathcal F^{\beta\gamma\alpha,(m_2)}_{2} (\vec k_2, \vec k_3) 
+ \tilde \df^{\rm K}_{0,m_2} \tilde \df^{\rm K}_{0,m_3}  \mathcal F^{\gamma\alpha\beta,(m_1)}_{2} (\vec k_3, \vec k_1) \non\\
&~~~ + {\rm stoch.} \non
}
As earlier, explicit derivations are given in appendix~\ref{app:bispectrum_II}.
The structure of the kernel and the Kronecker deltas again reflect the constraints of the tree-level PT, 
showing that only one $m_i$ can be of helicity $\pm1$ or $\pm2$ at the time.

In addition to the deterministic contributions size auto-bispectrum has two distinct constant stochastic contributions given in Eq.~\eqref{eq:bis_sto_ggg}.
This contribution can be projected to the bispectrum of spherical tensor components, where they contribute only to the helicity zero term $B^{(0,0,0)}_{2 2 2}$. 
These are proportional to the scalar product of spherical tensors
\eq{
B^{\alpha\beta\gamma,(m_1,m_2,m_3)}_{2 2 2}(\vec k_1, \vec k_2, \vec k_3) 
\supset ~ \tilde \df^{\rm K}_{0,m_1} & \tilde \df^{\rm K}_{0,m_2} \tilde \df^{\rm K}_{0,m_3}  \bigg(
{\rm const.}^{\alpha\beta\gamma}_2 \lb \vec Y^{(0)} \big( \vhat  k_1 \big).\vec Y^{(0)} \big( \vhat  k_2 \big).\vec Y^{(0)} \big( \vhat  k_3 \big) \rb \\
& +  {\rm const.}^{\alpha\beta\gamma}_1 ~ P_{L}(k_1) \lb \vec Y^{(0)} \big( \vhat  k_2 \big).\vec Y^{(0)} \big( \vhat  k_3 \big) \rb +  {\rm 2~cycle } \bigg), \non
}
while the stochasticity of the higher helicity contributions starts with the $k-$dependent terms, similar to what we discussed in 
the case of the one-loop power spectrum.

\section{Projection and redshift-space distortions}
\label{sec:rsd}

Defining a projection operator onto the sky
\be
\P_{ij} = \dK_{ij} - \hat x_i\hat x_j\,,
\ee
where $\hat{\vx}$ denotes the direction of the line of sight, 
the measured shape $\gamma_{ij}$ is given by the trace-free part of the sky-projected shape tensor \cite{schmidt/chisari/dvorkin}:
\be
\gamma_{ij}(\vx,z) = \TF\left[\P_i^{\  k}\P_j^{\  l} g_{kl}(\vx,\tau[z])\right]
+ \gamma_{G, ij}(\vx,\tau[z])\,,
\label{eq:proj}
\ee
where $\gamma_{G,ij}$ denotes the weak lensing shear, which at linear order
simply adds to the intrinsic shape. 
We shall address the effects of projection in the accompanying 
paper \cite{vlah/chisari/schmidt2019}, but emphasise that the spherical-tensor decomposition 
makes their implementation especially simple and conceptually clear.

The expression in \refeq{proj} above is only correct at leading order, since the true relation between intrinsic shape, shear, and observed shear is nonlinear.
Further, redshift-space distortions need to be taken into account 
in case of spectroscopic galaxy samples
(see e.g. \cite{Singh15,okumura++:2017, okumura++:2019} for linear theory treatments). Finally, there are
``GR corrections'' to the projection effects similar to the galaxy density
going beyond the usual ``Kaiser+magnification bias'' formula.  
These subleading projection contributions, which contain the effect of
Doppler boost on the observed galaxy shape and other interesting effects, 
could be straightforwardly derived following e.g. \cite{schmidt/jeong:2012} 
(see also references therein). We defer all of these additional contributions to future work. 

\section{Conclusions}
\label{sec:concl}

In this paper, we have developed a consistent perturbative framework to study the statistics of extended objects. 
In particular, we focused on the statistics of objects that can be described with tensor fields of rank two. 
As a most obvious application, the formalism describes galaxy shape correlations, 
i.e. intrinsic alignments, as well as cross-correlations with scalar fields like galaxy number density or galaxy size fields.
Our key findings can be summarised as follows:
\begin{itemize}
\item As in the case of galaxy clustering, we should disentangle intrinsic
physical effects (from the point of view of an observer comoving with
the galaxy) from projection effects which deal with the mapping from
the galaxy's rest frame to the observer's coordinates. In the context
of galaxy shapes, this means that we describe physical alignment effects
in terms of the 3D galaxy shape $g_{ij}$ in its rest frame, \emph{before} any projection on the sky ($\gamma_{ij}$).
\item We allow for all local physical observables to appear in the
expansion of 3D galaxy shapes.  This includes the density and tidal field
along the fluid trajectory, as well as spatial derivatives thereof \cite{senatore:2015,MSZ,angulo/etal:2015}.  
The latter (higher derivative terms) are suppressed by an additional scale $R_*$
(non-locality scale) which depends on the galaxy sample, but is expected to
be of order the Lagrangian radius of the parent halos.
The density and tidal field along the fluid trajectory can be
expanded in terms of convective time derivatives which can then be reordered
into terms of successively higher order in perturbation theory \cite{MSZ}.
The additional nonlocal bias terms generated by primordial non-Gaussianity
can also be included in this formalism along the lines of \cite{Chisari14,schmidt/chisari/dvorkin,Chisari16b}.
In particular, our expansion allows for a consistent derivation of the three-point function of galaxy shapes including primordial non-Gaussianity, which is left for future work.
\item The case of shapes is exactly analogous to that of the galaxy number density,
the only difference being that instead of a 3-scalar, we are now dealing 
with a trace-free, symmetric 3-tensor, $g_{ij}$. 
The bias expansion of galaxy intrinsic sizes is on the other hand equivalent to that of galaxy number density
(with the same number of bias parameters and operator basis). 
While the latter are not as well developed in the context of current experiments,
the projection on the sky plane mixes 3D trace and trace-free parts, so that bias expansions for both components are necessary in order to consistently describe projected galaxy shapes. Further, the results could prove useful for the application of intrinsic size correlations 
in galaxy evolution studies and to model their contamination to the cosmological magnification of galaxy sizes in the future.
\item Any symmetric three-tensor can be decomposed into spherical tensors, 
which gives us a natural framework to study correlations and utilize symmetries 
like statistical isotropy and parity, independently of the details of dynamics and bias expansion. 
The choice of spherical tensor basis also charts the path towards projecting the correlators 
in 3D onto the sky sphere where they are observed, which will be addressed separately in \cite{vlah/chisari/schmidt2019}.
\item We use the bias expansion and perturbation theory to compute the one-loop power spectrum 
and tree-level bispectrum statistics in the rest frame. The expression we derive is valid for all auto- and 
cross-correlations of galaxy number density, sizes, and shapes. We find that in order to have 
a closed bias renormalisation scheme, shape bias requires independent bias coefficients from galaxy 
number density or galaxy size (appendix~\ref{app:reno}).
That is, even if one were to observe the intrinsic three-dimensional shape tensor of galaxies, one would have to allow for different coefficients in the bias expansion of its trace and trace-free parts. This is expected because these two components transform differently under rotations. They thus do not respond to a specific configuration of large-scale modes in the same way.
\item The bias coefficients contributing to the one-loop power spectrum and tree-level bispectrum statistics are
\eq{
a, \in \{ {\rm n},\, {\rm s} \} : ~~& \overbrace{ \Big\{ b^{a}_{1} \Big\}}^{P_{11}} \bigcup \overbrace{ \Big\{ b^{a}_{2,1}, b^{a}_{2,2}\Big\}}^{P_{22}} 
\bigcup \overbrace{ \Big\{ b^{a}_{3,1} \Big\}}^{P_{13}} \bigcup \Big\{ b^{a}_{R_*} \Big\} \bigcup \Big\{  {\rm stoch.} \Big\} ,  \non\\
{\rm g}: ~~& \underbrace{ \Big\{ b^{\rm g}_{1} \Big\}}_{P_{11}} \bigcup \underbrace{ \Big\{ b^{\rm g}_{2,1}, b^{\rm g}_{2,2}, b^{\rm g}_{2,3}\Big\}}_{P_{22}}
 \bigcup \underbrace{ \Big\{ b^{\rm g}_{3,1}, b^{\rm g}_{3,2} \Big\}}_{P_{13}} \bigcup \Big\{ b^{\rm g}_{R_*} \Big\} \bigcup \Big\{  {\rm stoch.} \Big\}\, .  \non
\label{eq:biases}
}
For the fractional galaxy number density $\df_{\rm n}$ and fractional galaxy size perturbation $\df_{\rm s}$, four 
deterministic bias coefficients (one linear and three non-linear) are required. 
For the power spectrum of the trace-free part of galaxy shapes, 
we have two additional (non-linear) bias parameters that we label $b^{\rm g}_{2,3}$ and $ b^{\rm g}_{3,2}$.
For each tracer we add the leading higher-derivative bias parameter, and the associated operator we treat as a third order term 
in the power spectrum.
Shape-shape power spectra with helicity different from zero have contributions from only mode coupling terms and thus need only second order bias terms. 
Due to this counting, these higher-derivative terms thus do not contribute to the leading, tree-level, bispectra. 
In the case of the tree-level bispectra, only second-order bias parameters contribute, entering in all the auto- and cross-correlations.
The range of scales where this perturbative approach is expected to be applicable is the same as that for the clustering of biased tracers studied extensively in the literature. 

\item We also explore the contributions arising from the stochasticity in the bias relation. In the power spectrum, 
the dominant contributions at large scales come from the shot noise-like spectrum (one parameter for each cross-spectrum),
except for the $P^{0}_{02}$ component which does not have a shot noise-like contribution. 
In shape-shape power spectrum, these arise from auto-correlations of the $\eps_{ij}$ stochastic field, 
while the higher stochastic operator contribution only renormalises the shot noise spectrum. 
In the case of the shape-shape power spectra all three helicity contributions $q=0,1,2$ have the same shot noise spectrum, providing a window for consistency checks. 
In the bispectrum, the landscape is a bit richer, and besides purely stochastic contributions, higher order stochastic operators 
$\eps^\d_{ij} \tr[\Pi^{[1]}],~\eps_{\Pi^{[1]}} \TF[\Pi^{[1]}]_{ij}$ also give non-degenerate contributions already 
at tree-level. We thus need two, one, three, and two stochastic parameters in the size-size-size, size-size-shape, 
size-shape-shape and shape-shape-shape contributions, respectively. 
\end{itemize}

One of the advantages of the general, EFT-based expansion presented here is that it applies regardless of the precise definition of the shape field, 
e.g. whether number-density weighting is included, how the galaxies are selected, how the shape is measured or even if the objects being studied are galaxies at all (e.g. clusters of galaxies also display significant alignments \cite{vanUitert}). All of these effects are absorbed by the free coefficients in the bias expansion.
In addition, if one were to study shape correlations of spectroscopic galaxies, redshift space distortions need to be considered, and our formalism can be 
straightforwardly extended to include these effects. 

Last but not least, we stress that the formalism we have developed for the correlations 
of the description of galaxy intrinsic alignments goes beyond this specific physical application. 
In particular, the treatment and decomposition of tensor fields in terms of the tensor spherical harmonics
are, to our knowledge, pioneered in this paper. This allows one to modularise and study separately the dynamical effects, 
like nonlinear clustering and bias expansion, from the symmetry properties, like isotropy and parity. 
The significance and benefit of this approach become especially apparent once additional physical 
and statistical phenomena need to be considered. As an example, such decomposition simplifies the 
treatment of the projection effects (that we address in \cite{vlah/chisari/schmidt2019}), 
and it could be useful in the full sky treatment of other projection effects, like redshift space distortions.

\acknowledgments

We would like to thank Jonathan Blazek, Alvise Raccanelli and Marko Simonovi\' c for useful discussions. 
ZV is especially grateful to Sergey Sibiryakov for many illuminating conversations and discussions.
NEC has been supported by a Royal Astronomical Society Research Fellowship for the duration of this work. 
FS is supported by Starting Grant (ERC-2015-STG 678652) ``GrInflaGal'' of the European Research Council.

\clearpage
\appendix

\section{Helicity decomposition} 
\label{app:SVT}

The tensor field of our interest $S_{ij} (\vec x)$ is, by construction, a symmetric tensor. This means that, 
at each the ($t = {\rm const}$), spatial hyper-surfaces are homogeneous and isotropic and we 
can perform harmonic analysis. A spatial tensor field can be decomposed 
into irreducible components under rotations and translations, which then 
evolve independently. In flat space this is nothing but the Fourier decomposition, where
by considering an arbitrary Fourier mode $\vec k$ on a given slice we can choose 
two normalised vectors $\vec e_1$ and $\vec e_2$ perpendicular to $\vec k$, 
so that the set $(\vec e_1,\vec e_2, \vec e_k)$ constitutes an orthonormal basis.
For a given vector mode $\vec k$, this basis can be constructed just by choosing another 
random and non-collinear direction to $\vhat k$, that we can call $\vnhat$.
The ortonormal basis is given as 
\eeq{
\vec e_1= \frac{\vec k \times \hat {\vec n}}{|\vec k \times \hat {\vec n} |},~~ \vec e_2 = \hat {\vec k} \times \vec e_1~~ \vec e_k = \hat {\vec k}.
\label{eq:basisA}
}
In this basis each vector can be decomposed as $\vec V = V^1 \vec e_1 + V^2 \vec e_2 + V^k \vec e_k
= V_1 \vec e^1 + V_2 \vec e^2 + V_k \vec e^k$, and the covariant and contravariant cartesian basis
are equivalent, i.e. $\vec e_i = \vec e^i$. It also is useful to introduce the (covariant) helicity basis $\vec e_0 = \vec e_k$,
$\vec e_\pm =  \mp \tfrac{1}{\sqrt 2} \lb \vec e_1 \pm i \vec e_2 \rb$, where the
relations between the covariant and contravariant basis vectors read
\eeq{
\vec e_i = \vec e^{i*}, ~~ \vec e^i = \vec e_i^{*}, ~~ \vec e_i = (-1)^i \vec e^{-i}, ~~ \vec e^i = (-1)^i \vec e_{-i},
}
and form a complex orthonormal basis $\vec e_i . \vec e^j = \vec e_i . \vec e_j^* = \dK_{ij}$.
The contravariant basis is explicitly given by $\vec e^0 = \vec e_k$ and $\vec e^\pm =  \mp \tfrac{1}{\sqrt 2} \lb \vec e_1 \mp i \vec e_2 \rb$.
Any vector can be expanded in terms of basis vectors, i.e., written as $\vec V = V_i \vec e^i = V^i \vec e_i$, and the relations 
between covariant and contravariant spherical components are given by $V_i = (-1)^i V^{-i}$ and $V^i = (-1)^i V_{-i}$.
If $\vec V$ is a real vector i.e., if $\vec V^*= \vec V$, then $V^*_i= V^i$ and $V^{i*}= V_i$. On the other hand, if 
$\vec V$ is complex, then $V^*_i= (\vec V^*)^i$ and $V^{i*} = (\vec V^*)_i$, where vector components are $V_i = \vec V . \vec e_i$
and $V^i = \vec V . \vec e^i$.

Let us first decompose our tensor field in the trace and trace-free part:
\eeq{
T_{ij} (\vec k) =  \tfrac{1}{3} \dK_{ij} \tr \big[ T \big] (\vec k) + \TF\big[T \big]_{ij} (\vec k).
}
We can now decompose any field in the sets of (contravariant) harmonic tensor basis: 
\eq{
\label{eq:Y_basisA}
&\mbox{helicity-0:} \qquad  Y^{(0)} = 1, \\
&\mbox{helicity-1:} \qquad \vec Y^{(0)}_i = \hat k_i, \hspace{1.4cm} \vec Y^{(\pm 1)}_i =  e_i^{\pm}, \non\\
&\mbox{helicity-2:} \qquad \vec Y^{(0)}_{ij} = N_0 \Big( \hat k_i \hat k_j - \tfrac{1}{3} \dK_{ij} \Big), 
\qquad \vec Y^{(\pm 1)}_{ij} = N_1 \Big( \hat k_j e_i^{\pm} + \hat k_i e_j^{\pm} \Big),
\qquad \vec Y^{(\pm 2)}_{ij} = N_2 e_i^{\pm} e_j^{\pm}, \non
}
where $N_{0,1,2}=\left\{ \sqrt{\tfrac{3}{2}}, \sqrt{\frac{1}{2}}, 1\right\}$ so that the basis is orthonormal. 
The relation to the covariant basis is given by
\eeq{
\vec Y_{\ell,(m)} = \vec Y_\ell^{(m)*}, ~~ \vec Y_\ell^{(m)} = \vec Y_{\ell,(m)}^{*},
~~ \vec Y_{\ell,(m)}= (-1)^{m} \vec Y_\ell^{(-m)}, ~~{\rm and}~~ \vec Y_\ell^{(m)} = (-1)^{m}  \vec Y_{\ell,(-m)}.
}
Symmetric tensors can now be decomposed into the a sum of products of the basis elements and appropriate spherical components 
that do not carry index structure
\eq{
\label{eq:tensor_decomp}
T_{ij} (\vec k) 
&= \sum_{\ell = 0,2} \sum_{m=-\ell}^\ell T^{(m)}_\ell (\vec k) \left(\vec{Y}^{(m)}_\ell(\vec k)\right)_{ij} \\
&= \frac{1}{3}T^{(0)}_0 (\vec k) \dK_{ij}  + \sum_{m=-2}^2T^{(m)}_2 (\vec k) \vec Y^{(m)}_{ij}(\vhat  k),  \non
}
and thus
\eeq{
\tr \big[ T \big] (\vec k) =  T^{(0)}_0(\vec k) \vec Y^{(0)}, ~~
\TF\big[T \big]_{ij} (\vec k) = \sum_{m=-2}^2 T^{(m)}_2 (\vec k)  \vec Y^{(m)}_{ij}.
}
Since $T_{ij}(\vec r)$ is a real tensor field it implies that  $T^*_{ij}(\vec k) = T_{ij}( - \vec k)$.
From Eq. \eqref{eq:basisA} we have a basis corresponding to the $\vec k'$ vector 
$\vhat  k' = - \vhat  k$, $\vec e'^+ = \vec e^-$, $\vec e'^- = \vec e^+$, that corresponds to the rotation $(\psi,\theta,\phi) = (0,\pi,0)$. 
For a general basis vector, we can use the fact that the basis elements are spherical tensors and we have 
\eeq{
\vec Y_\ell^{(m)}(\vhat  k') = \sum_{q= - \ell}^\ell \mathcal D^{\ell}_{mq} (0,\pi,0) \vec Y_\ell^{(q)}(\vhat  k) 
= (-1)^{\ell+m} \vec Y_\ell^{(-m)}(\vhat  k) 
= (-1)^\ell \vec Y_\ell^{(m)*}(\vhat  k),
\label{eq:roted_basis}
}
where $\mathcal D^{\ell}$ is a Wigner rotation matrix, and in particular 
$\mathcal D^{\ell}_{mq} (0,\pi,0) = \dK_{-m,q} (-1)^{\ell+q}$. 
Using this result we have that $T^*_{ij}(\vec k) = T_{ij}( - \vec k)$ implies $T^{(m)*}_{0,2}(\vec k) = T^{(m)}_{0,2}( -\vec k)$.
More generally, we can show (using again the Wigner matrix) that vector components transform as $T^{(m)}_{\ell}(-\vec k) = (-1)^\ell T^{(m)*}_{\ell}(\vec k)$.

Under parity $\mathbb P: k_i \to - k_i$ basis elements transform similar to the ordinary spherical harmonics, i.e. 
\eeq{
\mathbb P \vec Y_{\ell}^{(m)} (\vhat  k) = (-1)^\ell \vec Y^{(m)}_{\ell}( - \vhat  k) 
= (-1)^{m} \vec Y_\ell^{(-m)}(\vhat  k) = \vec Y_\ell^{(m)*}(\vhat  k).
\label{eq:parity}
}
This can be explicitly seen from Eq. \eqref{eq:Y_basisA} above, given that $\mathbb P \vhat  k \to - \vhat  k$ and $\mathbb P  \vec e^{\pm} \to -  \vec e^{\pm}$.
Note that this is different from simply choosing the basis $\vec Y_\ell^{(m)}(\vhat  k')$
around some different vector $\vec k'$ that happens to be $\vec k' = - \vec k$, as we did above.
The difference comes from the fact that parity changes the right-handed basis into a 
left-handed and vice versa (so no rotation can be performed to get from one basis to another).
On the other hand, if we are interested in obtaining, say, a right-handed basis around some 
arbitrary vector $\vec k'$, we can get it by rotating the right-handed basis from the one around the $\vec k$, as we did earlier.
It also follows that tensor components transform as  $\mathbb P T^{(m)}_{\ell} (\vhat  k) \to T_\ell^{(m)*}(\vhat  k)$.

For comparison, consider the SVT decomposition in real space. Any vector can be decomposed into longitudinal (scalar, S) and transverse (proper vector, V) parts through
\eeq{
\xi^i = \xi_S^i + \xi_V^i ,\qquad
\xi_S^i = \partial^i \left[\frac1{\lapl} \partial_j \xi^j \right], \qquad
\xi_V^i = \left[ \d^i_{\  j} - \frac{\partial^i \partial_j}{\lapl}\right] \xi^j\,.
}
Similarly, a symmetric trace-free tensor field $h_{ij}$ is decomposed as
\eeq{
h_{ij} = h_{ij}^S + h_{ij}^V + h_{ij}^T, \qquad
h_{ij}^S = \left[\frac{\partial_i \partial_j}{\lapl} - \frac13 \d_{ij} \right] h^S, \qquad
h_{ij}^V = \partial_{(i} h^V_{j)}\,,
}
where $h^S$ is a scalar field while $h^V_j$ is a transverse vector field. 
We can easily identify the scalar, vector, and tensor contributions with the helicity$-0$, helicity$\pm1$, and helicity$\pm2$ contributions, respectively, in the Fourier-space decomposition described above.

The divergence of the tensor field is then directly related to the longitudinal or scalar part through
\eeq{
h^S = \frac{\partial^i \partial^j}{\lapl} h_{ij}\,,
}
while the curl yields the tensor part:
\eeq{
\eps^{ijk} \partial_i h_{jk} = \eps^{ijk} \partial_i h_{jk}^T\, . 
}

In order not to confuse the helicity decomposition with our classification of
scalar (density) and tensor (shape) observables, we refer to helicity-0,1,2 components throughout this paper, instead of the sometimes used language of ``scalar, vector, tensor'' components.

\subsection{Two-point functions}
\label{App:SVT_PS}

Next we consider the consequences of statistical homogeneity, isotropy, as well as parity invariance on two-point correlators.
First we note the implication of statistical homogeneity for the total two-point function in Fourier space, i.e. power spectrum
\eeq{
\la S_{ij} (\vec k) ~ S_{kl} (\vec k')  \ra = (2\pi)^3 \df^{\rm D}_{\vec k' + \vec k} P_{ij,kl} (\vec k).
}
Notice that in the following we allow for the two fields in this correlator
to be different tracers, although this is not indicated explicitly in the notation.
Moreover we can also define spectra of each field component as well. 
For an arbitrary rotation $ \vec k_i \to \vec k'_i = \vec R. \vec k_i$ we have
\eeq{
\label{eq:ps_isotropy}
\la S_{\ell}^{(m)}(\vec k'_1) ~S_{\ell'}^{(m')}(\vec k'_2)\ra 
= \sum_{q,q'} \mathcal D^{\ell}_{q m}(R)  \mathcal D^{\ell'}_{q' m'}(R) \la S_{\ell}^{(q)}(\vec k_1) ~S_{\ell'}^{(q')}(\vec k_2) \ra.
}
In particular if the rotation is around the $\vec k$ vector, by angle $\phi$, the first $S_\ell^{(m)}$ term gets a phase $e^{im\phi}$
while the second one gets a phase $e^{-im'\phi}$ (since $\vec k' = - \vec k$)\footnote{
Note that we have built our helicity basis out of the Cartesian coordinate system 
given in Eq.~\eqref{eq:basisA} and that $m$ is the eigenvalue of the projected
angular momentum operator on the $\vhat k$ direction, i.e.  $\vec L . \vhat k$. 
Thus, if we change $\vec k \to - \vec k$ the sign of $m$ also changes.
This is why we refer to the $m$ index as the helicity.
}.
From statistical isotropy thus follows
\eq{
\la S_\ell^{(m)}(\vec k) ~S_{\ell'}^{(m')} (\vec k') \ra = (2\pi)^3 \dK_{mm'} \df^{\rm D}_{\vec k + \vec k'} P^{(m)}_{\ell \ell'} (k) .
\label{eq:Pellellm}
}
This shows that by correlating two general tensors $\la S_{ij} (\vec k) S_{lm} (\vec k') \ra$ we 
can expect all helicity contributions, but without mixing of different helicities in the power spectrum. 
Using the facts that $\df^*(\vec k) = \df(-\vec k)$ and $S^*_{ij}(\vec k) = S_{ij}( - \vec k)$ we have shown that 
$S_\ell^{(m)*}(\vec k) = S_\ell^{(m)}(-\vec k)$ (since we are interested in cases $\ell=0$ and 2) and so
\eq{
\la S_\ell^{(m)}(\vec k) ~S_{\ell'}^{(m')} (\vec k') \ra^* = \la S_\ell^{(m)}(-\vec k) ~S_{\ell'}^{(m')} (-\vec k') \ra 
= \la S_\ell^{(-m)}(\vec k) ~S_{\ell'}^{(-m')} (\vec k') \ra,
}
that is to say that the power spectrum behaves as $P^{(m)*}_{\ell \ell'} (k) = P^{(-m)}_{\ell \ell'} (k)$.
Using the expansion in Eq \eqref{eq:tensor_decomp} we have
\eq{
\la  S_{ij} (\vec k) ~ S_{kl} (\vec k')  \ra
&= \frac{1}{9} \dK_{ij} \dK_{kl}  \la S^{(0)}_0 (\vec k) ~ S^{(0)}_0 (\vec k') \ra  \\
&~~~ +  \frac{1}{3}  \dK_{kl} \sum_{m=-2}^2 \vec Y^{(m)}_{ij}(\vhat  k) \la S^{(m)}_2 (\vec k) ~ S^{(0)}_0 (\vec k') \ra  \non\\
&~~~ +  \frac{1}{3}  \dK_{ij} \sum_{m'=-2}^2 \vec Y^{(m')}_{kl}(\vhat  k') \la S^{(0)}_0 (\vec k) ~ S^{(m')}_2 (\vec k') \ra \non\\
&~~~ + \sum_{m, m'=-2}^2 \vec Y^{(m)}_{ij}(\vhat  k)  \vec Y^{(m')}_{kl} (\vhat  k')  \la S^{(m)}_2 (\vec k) ~ S^{(m')}_2 (\vec k') \ra. \non
}
Given that each of the spectra $P^{(m)*}_{\ell \ell'} (k) = P^{(-m)}_{\ell \ell'} (k)$ are real, and we have for $q = \pm1$ or $\pm2$
\eq{
\vec Y^{(q)*}_{ij} \vec Y^{'(q)*}_{kl}  P^{(q)*}_{22} + \vec Y^{(-q)*}_{ij} \vec Y^{'(-q)*}_{kl}  P^{(-q)*}_{22} 
&= \vec Y^{(-q)}_{ij} \vec Y^{'(-q)}_{kl} P^{(-q)}_{22} + \vec Y^{(q)}_{ij} \vec Y^{'(q)}_{kl}  P^{(q)}_{22},
}
it follows that 
\eeq{
P^{*}_{ij,kl} (\vec k) = P_{ij,kl} (\vec k),
}
i.e. the total power spectrum is a real quantity.
This is of course the consequence of the fact that $S^*_{ij}(\vec k) = S_{ij}( - \vec k)$ (i.e. the configuration 
space fields are real).

Simplifying the expression for the power spectrum and using Eq. \eqref{eq:roted_basis} to get the bases elements in terms of the same $\vhat  k$
vector we have
\eeq{
P_{ij,lm} =
 \frac{1}{9} \dK_{ij} \dK_{lm}  P^{(0)}_{00} + \frac{1}{3} \vec Y^{(0)}_{ij}  \dK_{lm}  P^{(0)}_{20}
 +   \frac{1}{3}  \dK_{ij} \vec Y^{(0)}_{lm} P^{(0)}_{02} + \sum_{q=-2}^2 (-1)^q \vec Y^{(q)}_{ij} \vec Y^{(-q)}_{lm}  P^{(q)}_{22},
\label{eq:P_Y_decom} 
}
Note that in general, if different tracers are considered $P^{(0)}_{20}$ and $P^{(0)}_{02}$ give different contributions, 
and become equivalent only when we look at the tracer auto-correlations. For parity-invariant auto-correlations and cross-correlations 
we have $P_{22}^{(q)} = P_{22}^{(|q|)}$ (see below). Thus, for general cross-correlations, assuming parity-invariance, we have 6 distinct 
contributions which reduces to 5 for tracer auto-correlations.  If one does not impose parity-invariance there are two more contributions in both cases. 
For simplicity, from now on we focus only on auto-correlations, which implies $P^{(0)}_{20}=P^{(0)}_{02}$.

It is useful to introduce the following new labels for linear combination of the spectra 
\eeq{
P^{(0)}_{00} = P_a^0, ~~  P^{(0)}_{02} = P^{(0)}_{20} = P_b^0,  ~~ P^{(0)}_{22} = P_c^0,
~~ P^{(\pm 1)}_{22} = - \frac{1}{2} \lb P^{+}_{d} \pm P^{\times}_{d} \rb,
~~ P^{(\pm 2)}_{22} = \frac{1}{2} \lb P^{+}_{e} \pm P^{\times}_{e} \rb.
}
The decomposition above is independent on dynamical equations and thus it is valid also in the fully nonlinear case, as well as at a given PT order.
We see that the power spectrum can be decomposed in seven independent scalar spectra functions $P_n$ depending only on amplitude of each $\vec k$ mode.

Using the new basis we can thus rewrite the power spectrum as
\eeq{
P_{ij,lm}
=  \left( P^{0}_a \vec e^{(0)}_a +  P^{0}_b \vec e^{(0)}_b +  P^{0}_c \vec e^{(0)}_c
+  P^{+}_{d} \vec e^{(+)}_{d} +  P^{\times}_{d} \vec e^{(\times)}_{d}
+  P^{+}_{e} \vec e^{(+)}_{e} +  P^{\times}_{e} \vec e^{(\times)}_{e}
\right)_{ijlm}
}
where we have relations of the two sets of basis vectors 
\eq{
\left(\vec e^{(0)}_a\right)_{ijlm} &= \frac{1}{9} \dK_{ij} \dK_{lm}, \\
\left(\vec e^{(0)}_b\right)_{ijlm} &= \frac{1}{3} \lb \dK_{ij} \vec Y^{(0)}_{lm} +  \vec Y^{(0)}_{ij}  \dK_{lm} \rb, \non\\
\left(\vec e^{(0)}_c\right)_{ijlm} &= \vec Y^{(0)}_{ij} \vec Y^{(0)}_{lm}, \non\\
\left(\vec e^{(+)}_d\right)_{ijlm} &= \frac{1}{2} \lb \vec Y^{(1)}_{ij} \vec Y^{(-1)}_{lm} + \vec Y^{(-1)}_{ij} \vec Y^{(1)}_{lm} \rb \non\\
&= \hat k_i\hat k_j\hat k_l\hat k_m - \frac{1}{4} \lb \dK_{il} \hat k_j \hat k_m + \dK_{im} \hat k_j \hat k_l + \dK_{jl} \hat k_i \hat k_m + \dK_{jm} \hat k_i \hat k_l \rb, \non\\
\left(\vec e^{(+)}_e\right)_{ijlm} &= \frac{1}{2} \lb \vec Y^{(2)}_{ij} \vec Y^{(-2)}_{lm} + \vec Y^{(-2)}_{ij} \vec Y^{(2)}_{lm} \rb \non\\
&= 
\frac{1}{4} \bigg( \dK_{im}\dK_{jl} + \dK_{il}\dK_{jm} - \dK_{ij}\dK_{lm} + \dK_{ij}  \hat k_l \hat k_m + \dK_{lm}  \hat k_i \hat k_j \non\\
&\hspace{0.8cm} - \dK_{im}  \hat k_j \hat k_l  - \dK_{il}  \hat k_j \hat k_m - \dK_{jm}  \hat k_i \hat k_l  - \dK_{jl}  \hat k_i \hat k_m +  \hat k_i \hat k_j  \hat k_l \hat k_m \bigg), \non
}
and for the parity odd part 
\eeq{
\left(\vec e^{(\times)}_d\right)_{ijlm} = \frac{1}{2}  \lb \vec Y^{(1)}_{ij} \vec Y^{(-1)}_{lm} - \vec Y^{(-1)}_{ij} \vec Y^{(1)}_{lm} \rb, ~~
\left(\vec e^{(\times)}_e\right)_{ijlm} = \frac{1}{2}  \lb \vec Y^{(2)}_{ij} \vec Y^{(-2)}_{lm} - \vec Y^{(-2)}_{ij} \vec Y^{(2)}_{lm} \rb.
\non
}
The last symmetry condition we can apply is the parity invariance of the power spectra, i.e.  
\eq{
\la \mathbb P ~ S_{\ell}^{(m)}(\vec k) ~ \mathbb P ~ S_{\ell'}^{(m')}(\vec k')\ra 
= \la S_{\ell}^{(m)}(\vec k) ~S_{\ell'}^{(m')}(\vec k)\ra,
}
from which immediately follows that $P^{(q)}_{22} = P^{(-q)}_{22} = P^{(|q|)}_{22}$. 
Parity invariance thus implies that opposite helicity spectra are equal and 
$P^{\times}_{d} = P^{\times}_{e} = 0$. Our tensor power spectrum is now given by
\eq{
P_{ij,lm} &= \frac{1}{9} \dK_{ij} \dK_{lm}  P^{(0)}_{00} 
 + \frac{1}{3} \lb \dK_{ij} \vec Y^{(0)}_{lm} + \vec Y^{(0)}_{ij}  \dK_{lm} \rb P^{(0)}_{02} 
 +  \vec Y^{(0)}_{ij} \vec Y^{(0)}_{lm}  P^{(0)}_{22} + \sum_{q=1}^2 (-1)^q \vec Y^{(q)}_{\{ ij} \vec Y^{(-q)}_{lm\}}  P^{ab(q)}_{22} \non\\
&=  P^{0}_a \vec e^{(0)}_a +  P^{0}_b \vec e^{(0)}_b +  P^{0}_c \vec e^{(0)}_c +  P^{+}_{d} \vec e^{(+)}_{d} +  P^{+}_{e} \vec e^{(+)}_{e}.
\label{eq:P_decom_last} 
}
Finally, after imposing the statistical homogeneity, isotropy and parity invariance we end up 
with tensor auto-power spectrum described with only five independent scalar spectral functions. 

In section \ref{app:stoch} we give the form of the shot noise contribution to the shape-shape auto-power spectrum on large scales
\eq{
 \left( \dK_{ik}\dK_{jl} + \dK_{il}\dK_{jk} -  \frac{2}{3} \dK_{ij} \dK_{lm}\right) P_{\eps}\, ,
}
which is a result of correlations of the stochastic contributions $\eps_{ij} (\vec k)$. It is quite simple 
to show that this structure arises naturally from our expansion above by requiring that the shot noise spectrum does not depend on
direction nor amplitude of the $\vec k$ modes in the limit $k\to 0$. For example, using this requirement on Eq.~\eqref{eq:P_decom_last} we get
$P_a = P_b = 0$, $P_c = 2 P_\eps$, $P_d = - P_e= -4 P_\eps$.

\subsection{Three-point functions}
\label{app:bispectrum_I}

Here we focus on the consequences of the homogeneity, isotropy, and statistical parity invariance on the form of the bispectra of rank two tensor fields.
Correlating three such fields gives us
\eq{
\la S_{ij} (\vec k_1)  S_{kl} (\vec k_2) S_{rs} (\vec k_3) \ra = (2\pi)^3 \df^{\rm D}_{\vec k_1 + \vec k_2 + \vec k_3} B_{ijklrs} (\vec k_1, \vec k_2, \vec k_3),
}
and using again the expansion given in Eq. \eqref{eq:tensor_decomp} yields
\eq{
\la S_{ij} (\vec k_1)  S_{kl} (\vec k_2) S_{rs} (\vec k_3) \ra
&= \frac{1}{27} \dK_{ij} \dK_{kl}\dK_{rs}  \la S^{(0)}_0 (\vec k_1) S^{(0)}_0 (\vec k_2) S^{(0)}_0 (\vec k_3) \ra  \\
&\hspace{-2.8cm} +   \frac{1}{9}  \dK_{ij}  \dK_{kl} \sum_{m_3=-2}^2 \vec Y^{(m_3)}_{rs} \big( \vhat  k_3 \big)  \la S^{(0)}_0 (\vec k_1) S^{(0)}_0 (\vec k_2) S^{(m_3)}_2 (\vec k_3) \ra
 +  {\rm 2~cycle } \non\\
&\hspace{-2.8cm} +   \frac{1}{3}  \dK_{ij} \sum_{m_2, m_3=-2}^2 \vec Y^{(m_2)}_{kl}\big( \vhat  k_2 \big) \vec Y^{(m_3)}_{rs} \big( \vhat  k_3 \big) \la S^{(0)}_0 (\vec k_1) S^{(m_2)}_2 (\vec k_2) S^{(m_3)}_2 (\vec k_3) \ra 
 +  {\rm 2~cycle } \non\\
&\hspace{-2.8cm} + \sum_{m_i=-2}^2 \vec Y^{(m_1)}_{ij}\big( \vhat  k_1 \big) \vec Y^{(m_2)}_{kl}\big( \vhat  k_2 \big) \vec Y^{(m_3)}_{rs} \big( \vhat  k_3 \big) 
\la S^{(m_1)}_2 (\vec k_1) S^{(m_2)}_2 (\vec k_2) S^{(m_3)}_2 (\vec k_3) \ra. \non
}
Again, while we do not indicate this in the notation here for clarity, our results in the following are valid also for cross-correlations of three different fields.
Note that the statistical rotation isotropy can not be exploited here in the same simple way as it was in the two-point function case 
by just rotating around the $\vhat  k$ axes by angle $\phi$. In the case of the bispectrum, this operation is not very convenient since 
each of the vectors $\vec k_1$, $\vec k_2$ and $\vec k_3$ point in different directions. Nonetheless, statistical isotropy and homogeneity
give further constraints on the bispectrum of spherical tensor components. For an arbitrary rotation $ \vec k_i \to \vec k'_i = \vec R. \vec k_i$ we have
\eq{
\label{eq:bis_isotropy}
\la S_{\ell_1}^{(m'_1)}(\vec k'_1) ~S_{\ell_2}^{(m'_2)}(\vec k'_2)~S_{\ell_3}^{(m'_3)}(\vec k'_3) \ra &= \\
& \hspace{-4.7cm} \sum_{m_1,m_2,m_3} \mathcal D^{\ell_1}_{m_1 m'_1}(R)  \mathcal D^{\ell_2}_{m_2 m'_2}(R) \mathcal D^{\ell_3}_{m_3 m'_3}(R) 
\la S_{\ell_1}^{(m_1)}(\vec k_1) ~S_{\ell_2}^{(m_2)}(\vec k_2)~S_{\ell_3}^{(m_3)}(\vec k_3) \ra, \non
}
given that $S_{\ell}^{(m)}$ transforms as a spherical tensor. 
Due to the statistical homogeneity we can isolate a Dirac delta function that guarantees the 
conservation of the total momenta from the bispectrum of spherical tensor components
\eq{
\la S_{\ell_1}^{(m_1)}(\vec k_1) ~S_{\ell_2}^{(m_2)}(\vec k_2)~S_{\ell_3}^{(m_3)}(\vec k_3) \ra 
= (2\pi)^3 \df^{\rm D}_{\vec k_{123}} B^{(m_1 m_2 m_3)}_{\ell_1 \ell_2 \ell_3} (\vec k_1, \vec k_2, \vec k_3).
}
This allows us to write the tensor bispectrum in terms of bispectra of spherical tensor components
\eq{
\label{eq:bispectrum_decomp}
B_{ijklrs} (\vec k_1, \vec k_2, \vec k_3)
&= \frac{1}{27} \dK_{ij} \dK_{kl}\dK_{rs}  B^{(0,0,0)}_{000} (\vec k_1, \vec k_2, \vec k_3) \hspace{4.5cm} {\rm (\textbf{000})}  \\
&~~~ +   \frac{1}{9}  \dK_{ij}  \dK_{kl} \sum_{m_3=-2}^2 \vec Y^{(m_3)}_{rs}\big( \vhat  k_3 \big)  B^{(0,0,m_3)}_{002} (\vec k_1, \vec k_2, \vec k_3)
 +  {\rm 2~cycle } \hspace{1.9cm} {\rm (\textbf{002})} \non\\
&~~~ +   \frac{1}{3}  \dK_{ij} \sum_{\substack{m_i=-2 \\ i=(2,3)}}^2 \vec Y^{(m_2)}_{kl}\big( \vhat  k_2 \big) \vec Y^{(m_3)}_{rs}\big( \vhat  k_3 \big) 
B^{(0,m_2,m_3)}_{0 2 2} (\vec k_1, \vec k_2, \vec k_3)
 +  {\rm 2~cycle }  \hspace{0.38cm} {\rm (\textbf{022})} \non\\
&~~~ + \sum_{\substack{m_i=-2 \\ i=(1,2,3)}}^2 \vec Y^{(m_1)}_{ij}\big( \vhat  k_1 \big) \vec Y^{(m_2)}_{kl}\big( \vhat  k_2 \big) \vec Y^{(m_3)}_{rs} \big( \vhat  k_3 \big) 
B^{(m_1, m_2, m_3)}_{222} (\vec k_1, \vec k_2, \vec k_3).   \hspace{0.48cm} {\rm (\textbf{222})} \non
}
From the fact that $\df^*_L(\vec k) = \df_L(-\vec k)$ and $S^*_{ij}(\vec k) = S_{ij}( - \vec k)$ it follows that $S_\ell^{(m)*}(\vec k) = (-1)^\ell S_\ell^{(m)}(-\vec k)$,
and thus it follows
\eeq{
\la S_{\ell_1}^{(m_1)*}(\vec k_1) ~S_{\ell_2}^{(m_2)*}(\vec k_2)~S_{\ell_3}^{(m_3)*}(\vec k_3) \ra 
= \la S_{\ell_1}^{(m_1)}(-\vec k_1) ~S_{\ell_2}^{(m_2)}(-\vec k_2)~S_{\ell_3}^{(m_3)}(-\vec k_3) \ra.
}
Using Eq.~\eqref{eq:bis_isotropy} and setting the bispectrum triangle in the plane $\vhat x-\vhat y$ plane and 
rotating around the $\vhat z$ axes for $\pi$ gives
\eeq{
\la S_{\ell_1}^{(m_1)}(-\vec k_1) ~S_{\ell_2}^{(m_2)}(-\vec k_2)~S_{\ell_3}^{(m_3)}(-\vec k_3) \ra
= (-1)^{m_{123}} \la S_{\ell_1}^{(-m_1)}(\vec k_1) ~S_{\ell_2}^{(-m_2)}(\vec k_2)~S_{\ell_3}^{(-m_3)}(\vec k_3) \ra,
}
which thus gives the constraint $B^{(m_1,m_2,m_3)*}_{\ell_1 \ell_2 \ell_3} = (-1)^{m_{123}} B^{(-m_1,-m_2,-m_3)}_{\ell_1 \ell_2 \ell_3}$. 
We can also impose invariance under the parity transformation which simply requires that 
\eeq{
\la  \mathbb P ~ S_{\ell_1}^{(m_1)}(\vec k_1) ~  \mathbb P ~ S_{\ell_2}^{(m_2)}(\vec k_2)~  \mathbb P ~ S_{\ell_3}^{(m_3)}(\vec k_3) \ra 
= \la S_{\ell_1}^{(m_1)}(\vec k_1) ~S_{\ell_2}^{(m_2)}(\vec k_2)~S_{\ell_3}^{(m_3)}(\vec k_3) \ra, 
}
which gives the constraint $B^{(m_1,m_2,m_3)}_{\ell_1 \ell_2 \ell_3} = (-1)^{m_{123}} B^{(-m_1,-m_2,-m_3)}_{\ell_1 \ell_2 \ell_3}$. 

Using these symmetries, we can simplify individual terms in Eq.~\eqref{eq:bispectrum_decomp}. 
Besides the first, trivial, purely scalar, term, the second term has a single nonzero helicity component.
We can simplify it by using 
 \eq{
\vec Y^{(-m)}_{ij}\big( \vhat  k_3 \big)  B^{(0,0,-m)}_{002} + \vec Y^{(m)}_{ij}\big( \vhat  k_3 \big)  B^{(0,0,m)}_{002}
&= \lb (-1)^m \vec Y^{(-m)}_{ij}\big( \vhat  k_3 \big) + \vec Y^{(m)}_{ij}\big( \vhat  k_3 \big) \rb B^{(0,0,m)}_{002} \non\\
&= 2  {\rm Re} \left[ \vec Y^{(m)}_{ij}\big( \vhat  k_3 \big) \right]  B^{(0,0,m)}_{002}.
}
Thus only three components of the $B^{(0,0,m)}_{002}$ bispectrum are independent. 
For the \textbf{(002)} contribution in Eq.~\eqref{eq:bispectrum_decomp}
we thus have
\eq{
\label{eq:bis_sst}
 \sum_{m=-2}^2 \vec Y^{(m)}_{ij}\big( \vhat  k_3 \big)  B^{(0,0,0)}_{002} (\vec k_1, \vec k_2, \vec k_3)
&= \vec Y^{(0)}_{ij}\big( \vhat  k_3 \big)  B^{(0,0,0)}_{002} (\vec k_1, \vec k_2, \vec k_3) \non\\
&\hspace{0.5cm}+ 2 \sum_{m=1}^2  {\rm Re} \left[ \vec Y^{(m)}_{ij}\big( \vhat  k_3 \big) \right]   B^{(0,0,m)}_{002} (\vec k_1, \vec k_2, \vec k_3).
}
For the \textbf{(022)} bispectrum we have more terms contributing. Naively 25 terms would contribute,
but after statistical parity and rotation invariance condition is used only 13 of them are independent. 
This gives us 
\eq{
\label{eq:bis_stt}
\sum_{\substack{m_i=-2 \\ i=(2,3)}}^2 &\vec Y^{(m_2)}_{ij} \big( \vhat  k_2 \big) \vec Y^{(m_3)}_{kl}\big( \vhat  k_3 \big) B^{(0,m_2,m_3)}_{0 2 2} 
= \vec Y^{(0)}_{ij} \big( \vhat  k_2 \big) \vec Y^{(0)}_{kl}\big( \vhat  k_3 \big) B^{(0,0,0)}_{0 2 2} \\
& + 2 \sum_{m=1}^2 \Bigg( {\rm Re} \left[  \vec Y^{(m)}_{ij} \big( \vhat  k_2 \big) \vec Y^{(0)}_{kl}\big( \vhat  k_3 \big) \right]  B^{(0,m,0)}_{0 2 2}
   + {\rm Re} \left[  \vec Y^{(0)}_{ij}\big( \vhat  k_2 \big) \vec Y^{(m)}_{kl} \big( \vhat  k_3 \big) \right] B^{(0,0,m)}_{0 2 2} \non\\
&\hspace{1cm} + {\rm Re} \left[ \vec Y^{(m)}_{ij} \big( \vhat  k_2 \big) \vec Y^{(m)}_{kl}\big( \vhat  k_3 \big) \right] B^{(0,m,m)}_{0 2 2} 
   + {\rm Re} \left[ \vec Y^{(m)}_{ij} \big( \vhat  k_2 \big) \vec Y^{(-m)}_{kl}\big( \vhat  k_3 \big) \right] B^{(0,m,-m)}_{0 2 2} \non\\
&\hspace{1cm} + {\rm Re} \left[  \vec Y^{(m)}_{ij} \big( \vhat  k_2 \big) \vec Y^{(3-m)}_{kl}\big( \vhat  k_3 \big) \right] B^{(0,m,3-m)}_{0 2 2} 
   + {\rm Re} \left[  \vec Y^{(m)}_{ij} \big( \vhat  k_2 \big) \vec Y^{(m-3)}_{kl}\big( \vhat  k_3 \big) \right] B^{(0,m,m-3)}_{0 2 2} \Bigg). \non
}
The final line in Eq.~\eqref{eq:bispectrum_decomp} is the \textbf{(222)} bispectrum. 
Naively 125 terms contribute to these spectra, however, 65 of them are independent, and we can write
\eq{
\label{eq:bis_ttt}
\sum_{\substack{m_i=-2 \\ i=(1,2,3)}}^2 & \vec Y^{(m_1)}_{ij}\big( \vhat  k_1 \big) \vec Y^{(m_2)}_{kl}\big( \vhat  k_2 \big) \vec Y^{(m_3)}_{rs} \big( \vhat  k_3 \big) 
B^{(m_1, m_2, m_3)}_{222} 
 =  \vec Y^{(0)}_{ij}\big( \vhat  k_1 \big) \vec Y^{(0)}_{kl}\big( \vhat  k_2 \big) \vec Y^{(0)}_{rs}  \big( \vhat  k_3 \big)  B^{(0, 0,0)}_{222} \non \\
&+ 2  \sum_{m=1}^2 \Bigg( 
{\rm Re} \left[ \vec Y^{(m)}_{ij}\big( \vhat  k_1 \big) \vec Y^{(m)}_{kl}\big( \vhat  k_2 \big) \vec Y^{(m)}_{rs} \big( \vhat  k_3 \big) \right] B^{(m, m, m)}_{222} \non\\
& +{\rm Re} \left[ \vec Y^{(0)}_{ij}\big( \vhat  k_1 \big) \vec Y^{(0)}_{kl} \big( \vhat  k_2 \big) \vec Y^{(m)}_{rs} \big( \vhat  k_3 \big) \right] B^{(0, 0,m)}_{222} + {\rm 2~cycle} \non\\
& +{\rm Re} \left[ \vec Y^{(m)}_{ij}\big( \vhat  k_1 \big) \vec Y^{(m)}_{kl}\big( \vhat  k_2 \big) \vec Y^{(-m)}_{rs} \big( \vhat  k_3 \big) \right]  B^{(m, m, -m)}_{222} +{\rm 2~cycle} \non\\
&+ {\rm Re} \left[ \vec Y^{(m)}_{ij}\big( \vhat  k_1 \big) \vec Y^{(m)}_{kl}\big( \vhat  k_2 \big) \vec Y^{(0)}_{rs} \big( \vhat  k_3 \big) \right] B^{(m, m, 0)}_{222} +{\rm 2~cycle} \non\\
& +{\rm Re} \left[ \vec Y^{(-m)}_{ij}\big( \vhat  k_1 \big) \vec Y^{(m)}_{kl}\big( \vhat  k_2 \big) \vec Y^{(0)}_{rs} \big( \vhat  k_3 \big) \right] B^{(-m, m, 0)}_{222} + {\rm 2~cycle} \non\\
& +{\rm Re}\left[ \vec Y^{(m)}_{ij}\big( \vhat  k_1 \big) \vec Y^{(3-m)}_{kl}\big( \vhat  k_2 \big) \vec Y^{(0)}_{rs} \big( \vhat  k_3 \big) \right] B^{(m, 3-m, 0)}_{222} + {\rm 2~cycle} \non\\
& +{\rm Re}\left[ \vec Y^{(m)}_{ij}\big( \vhat  k_1 \big) \vec Y^{(m-3)}_{kl}\big( \vhat  k_2 \big) \vec Y^{(0)}_{rs} \big( \vhat  k_3 \big) \right] B^{(m, m-3, 0)}_{222} + {\rm 2~cycle} \non\\
& +{\rm Re}\left[ \vec Y^{(m)}_{ij}\big( \vhat  k_1 \big) \vec Y^{(m)}_{kl}\big( \vhat  k_2 \big) \vec Y^{(3-m)}_{rs} \big( \vhat  k_3 \big) \right]  B^{(m, m, 3-m)}_{222}  + {\rm 2~cycle} \non\\
& +{\rm Re}\left[ \vec Y^{(m)}_{ij}\big( \vhat  k_1 \big) \vec Y^{(m)}_{kl}\big( \vhat  k_2 \big) \vec Y^{(m-3)}_{rs} \big( \vhat  k_3 \big) \right] B^{(m, m, m-3)}_{222}  + {\rm 2~cycle} \non\\
& +{\rm Re}\left[ \vec Y^{(m)}_{ij}\big( \vhat  k_1 \big) \vec Y^{(-m)}_{kl}\big( \vhat  k_2 \big) \vec Y^{(3-m)}_{rs} \big( \vhat  k_3 \big) \right] B^{(m, -m, 3-m)}_{222}  + {\rm 2~cycle} \non\\
& +{\rm Re}\left[ \vec Y^{(m)}_{ij}\big( \vhat  k_1 \big) \vec Y^{(-m)}_{kl}\big( \vhat  k_2 \big) \vec Y^{(m-3)}_{rs} \big( \vhat  k_3 \big) \right] B^{(m, -m, m-3)}_{222}  + {\rm 2~cycle} \Bigg).
}
Note again that these decompositions above are valid for the fully nonlinear case and do not rely on the PT expansion.
It is also clear from these explicit expansion, that all the bispectra are real functions.

\section{Fields in Fourier space}
\label{App:PT_shear_field}

In this appendix, we derive Fourier-space expressions for the operators that appear in the general bias expansion. For brevity, we here employ the summed vector notation
\be
\vec k_{1m} \equiv \vec k_1 + \cdots + \vec k_m.
\ee

We first turn our attention to Eq.~\eqref{eq:Pindef} that recursively defines higher tensor contributions $\vec \Pi^{[n]}$. 
Transforming this relation to Fourier space we have
\eq{
\label{eq:Pindef_Fourer}
\Pi^{[n]}_{ij}(\vec k) = \frac{1}{(n-1)!} \Bigg[ 
\lb \frac{\partial}{\partial \ln D} - (n-1) \rb & \Pi^{[n-1]}_{ij}(\vec k) \\
& - (2\pi)^3 \df^{\rm D}_{\vec k - \vec p_{12}} \frac{\vec p_1 . \vec p_2}{p_1^2} \theta(\vec p_1) \Pi^{[n-1]}_{ij}(\vec p_2)
\Bigg], \non
}
where $\theta$ is the divergence of the peculiar velocity field $\vec v$, i.e. $\theta(\vec k) = - i \vec k \cdot \vec v(\vec k)$ 
and we have 
\eq{
\theta (\vec k)= \sum_{m=1}^\infty (2\pi)^3 \df^{\rm D}_{\vec k - \vec p_{1 m}} G^{(m)} (\vec p_1, \ldots , \vec p_m) \df_L(\vec p_1) \ldots  \df_L(\vec p_m),
}
where $G^{(n)}$ are standard velocity SPT kernels \cite{bernardeau/etal:2001}.
Assuming the ansatz for the $\vec \Pi^{[n]}$ of similar form as for the biased tracers $\df_n$ in Eq.~\eqref{eq:deltag}
we can write
\eq{
\Pi^{[n]}_{ij}(\vec k) = \sum_{m=n}^\infty (2\pi)^3 \df^{\rm D}_{\vec k - \vec p_{1 m}} 
 \pi^{(n, m)}_{ij} (\vec p_1, \ldots , \vec p_m) \df_L(\vec p_1) \ldots  \df_L(\vec p_m).
}
In case of $\Pi^{[1]}_{ij}$ we can immediately deduce  
\eeq{
\pi^{(1, m)}_{ij} (\vec p_1, \ldots , \vec p_m) =  
\frac{(\vec p_{1 m})_i(\vec p_{1 m})_j}{ p^2_{1 m}} F^{(m)} (\vec p_1, \ldots , \vec p_m).
}
Using this result and the recursion relation Eq.~\eqref{eq:Pindef_Fourer} we can obtain the result for the general kernel $\pi^{(n,m)}$.
This gives
\eq{
\pi^{(n,m)}_{ij} (\vec p_1, \ldots , \vec p_m) = \frac{ m - n + 1 }{(n-1)!} & \pi^{(n-1,m)}_{ij} (\vec p_1, \ldots , \vec p_m) \\
&\hspace{-0.3cm} - \sum_{\ell = n-1}^{m-1}
\frac{ \vec p_{1\ell}. \vec p_{\ell+1,m} }{ p_{\ell+1,m}^2}
\pi^{(n-1,\ell)}_{ij} ( \vec p_1, \ldots , \vec p_\ell)  G^{(m - \ell)} (\vec p_{\ell+1}, \ldots , \vec p_m) , \non
\label{eq:pi_rec}
}
where $m \geq n$ and $n\geq2$. 
Note that for the case where $n=m$ only one term $\ell=n-1$ in the last part contributes.
In order to get the symmetric kernels we can symmetrise this solution by summing 
over all the momentum permutations
\eeq{
\pi^{(n,m)}_{{\rm sim}, ij} (\vec p_1, \ldots , \vec p_m) = \frac{1}{m!} \sum_{\pi \rm-all} \pi^{(n,m)}_{ij} (\vec p_{\#_1}, \ldots , \vec p_{\#_m}). 
}
Combining the two relations we have
\eq{
\pi^{(n,m)}_{{\rm sim}, ij} (\vec p_1, \ldots , \vec p_m) &= 
\frac{ m - n + 1 }{(n-1)!} \pi^{(n-1,m)}_{{\rm sim}, ij} (\vec p_1, \ldots , \vec p_m)
 - \sum_{\ell = n-1}^{m-1} \frac{l! (m-l)!}{m!} \\
&\hspace{2.0cm} \times 
 \sum_{\pi \rm-cross} \frac{ \vec p_{1\ell}. \vec p_{\ell+1,m} }{ p_{\ell+1,m}^2}
\pi^{(n-1,\ell)}_{{\rm sim}, ij} ( \vec p_1, \ldots , \vec p_\ell)  G^{(m - \ell)} (\vec p_{\ell+1}, \ldots , \vec p_m), \non
}
where in the sum ``$\pi {\rm-cross}$'' only the cross permutations between the two sets need to be considered.
In the rest of the paper we will drop the label ``sim'' and assume (unless explicitly stated) that it is understood that the kernels are symmetrised.
This gives the $[n]-$th order expression for the $\vec \Pi^{[n]}$ field. 

Let us look at the first few terms coming from the expressions above.
We start by looking at the $\pi^{(2,2)}_{ij} (\vec p_1, \vec p_2)$ term. 
At second order we have $n=m=2$, which gives just one term
\eq{
\pi^{(2,2)}_{ij} (\vec p_1, \vec p_2) &= 
\frac{\vec p_{12,i} \vec p_{12,j}} { \vec p_{12}^2 } F^{(2)} (\vec p_1 , \vec p_2)
- \frac{\vec p_1. \vec p_2}{2p_1^2 p_2^2} \lb p_{1i}p_{1j} + p_{2i}p_{2j} \rb \\
&=  \frac{\vec p_{12,i} \vec p_{12,j}} { \vec p_{12}^2 } \frac{5}{7} \lb 1 - \frac{(\vec p_1 . \vec p_2)^2}{ p_1^2 p_2^2 } \rb
+ \frac{\vec p_1. \vec p_2}{2p_1^2 p_2^2} \lb p_{1i}p_{2j} + p_{2i}p_{1j} \rb . \non
}
At third order, we have two terms contributing for cases $m=2$ and $m=3$. 
The explicit expression can be written in the form
\eq{
\pi^{(2,3)}_{ij} (\vec p_1, \vec p_2, \vec p_3) &=
2 \frac{\vec p_{13,i} \vec p_{13,j}} { p_{13}^2 } F^{(3)} (\vec p_1 , \vec p_2, \vec p_3) \label{eq:pi2333} \\
&\hspace{0.5cm} - \frac{1}{3} \frac{\vec p_1. \vec p_{23} }{p_1^2} \frac{\vec p_{23,i} \vec p_{23,j}} {p_{23}^2 } F^{(2)} (\vec p_2 , \vec p_3)
 + 2~{\rm cycle~perm.} \non\\
&\hspace{0.5cm} - \frac{1}{3} \frac{\vec p_1. \vec p_{23}}{p_{23}^2} \frac{\vec p_{1,i} \vec p_{1,j}} { p_{1}^2 } G^{(2)} (\vec p_2 , \vec p_3)
 + 2~{\rm cycle~perm.}, \non\\
\pi^{(3,3)}_{ij} (\vec p_1, \vec p_2, \vec p_3) &= 
\frac{\vec p_{13,i} \vec p_{13,j}} { p_{13}^2 } F^{(3)} (\vec p_1 , \vec p_2, \vec p_3) \non\\
&\hspace{0.5cm} - \frac{1}{2} \frac{\vec p_1. \vec p_{23}}{p_1^2} \frac{\vec p_{23,i} \vec p_{23,j}} {p_{23}^2 } F^{(2)} (\vec p_2 , \vec p_3)
 + 2~{\rm cycle~perm.} \non\\
&\hspace{0.5cm} - \frac{1}{6} \frac{\vec p_1. \vec p_{23}}{p_{23}^2} \frac{\vec p_{1,i} \vec p_{1,j}} { p_{1}^2 } G^{(2)} (\vec p_2 , \vec p_3)
 + 2~{\rm cycle~perm.} \non\\
&\hspace{0.5cm}
+ \frac{1}{6}  \frac{(\vec p_1.\vec p_2)(\vec p_1.\vec p_3)}
{p_1^2 p_2^2 p_3^2} \lb 2 p_{1i}p_{1j}  + p_{2i}p_{2j} + p_{3i}p_{3j}  \rb
+ 2~{\rm cycle~perm.}~. \non
}

Next we derive explicit expressions for the Fourier space kernels in Eq.~ \eqref{eq:EulBasis_tr}
and Eq.~ \eqref{eq:EulBasis_TF} up to the third order in PT.
These results are summarised in table~\ref{eq:EulBasis_tr}. We start with the first order

\textbf{(1)} For the first order $\vec \Pi^{[1]}$ we have
\eeq{
\Big[\Pi^{[1]}_{ij}\Big]^{(1)}(\vec k) = (2\pi)^3 \df^{\rm D}_{\vec k - \vec p} \pi^{(1,1)}_{ij} \df_L(\vec p),
}
where 
\eeq{
\pi^{(1,1)}_{ij} (\vec p) = \frac{p_ip_j}{p^2} .
}

\textbf{(2)} For the second order we have three new, second generation, operators and the one 
first generation operator which also gives contributions to the second order.
The latter one has the following form 
\eeq{
\Big[\Pi^{[1]}_{ij}\Big]^{(2)}(\vec k) = (2\pi)^3 \df^{\rm D}_{\vec k - \vec p_{12}} \pi^{(1,2)}_{ij} (\vec p_1,\vec p_2) \df_L(\vec p_1)\df_L(\vec p_2),
}
where 
\eeq{
\pi^{(1,2)}_{ij} (\vec p_1,\vec p_2) = \frac{\vec p_{12,i} \vec p_{12,j}} { p_{12}^2 }
F^{(2)} (\vec p_1, \vec p_2).
}
For completeness we write the full form of the $F_2$ kernels 
\eeq{
F^{(2)} (\vec p_1 , \vec p_2) = \frac{5}{7} +   \frac{2}{7} \frac{(\vec p_1.\vec p_2)^2}{p_1^2 p_2^2 } 
+ \frac{\vec p_1.\vec p_2 }{2} \left(\frac{1}{p_1^2}+\frac{1}{p_2^2} \right). 
}
The first of the second generation operators starting from the second order is
\eq{
 \Big[\Pi^{[2]}_{ij}\Big]^{(2)}(\vec k) = (2\pi)^3 \df^{\rm D}_{\vec k - \vec p_{12}} \pi^{(2,2)}_{ij} (\vec p_1,\vec p_2) \df_L(\vec p_1)\df_L(\vec p_2),
}
where 
\eeq{
\pi^{(2,2)}_{ij} (\vec p_1,\vec p_2) = 
\frac{\vec p_{12,i} \vec p_{12,j}} { p_{12}^2 } F^{(2)} (\vec p_1 , \vec p_2)
- \frac{1}{2} \frac{\vec p_1. \vec p_2}{p_1^2 p_2^2} \lb p_{1i}p_{1j} + p_{2i}p_{2j} \rb.
}
The second term gives 
\eq{
\Big[ \big( \Pi^{[1]} \big)^2 \Big]_{ij}^{(2)}(\vec k) &= \int_{\vec r} e^{i \vec k \cdot \vec r} ~ \Pi^{[1]}_{im}(\vec r)\Pi^{[1]}_{m j}(\vec r) \\
&= (2\pi)^3 \df^{\rm D}_{\vec k - \vec p_{12}} ~
   \Big[ \pi^{(1,1)}_{im} (\vec p_1) \pi^{(1,1)}_{mj} (\vec p_2) \Big]   \df_L(\vec p_1) \df_L(\vec p_2) \non\\
&= (2\pi)^3 \df^{\rm D}_{\vec k - \vec p_{12}} ~
   \Big[ \pi^{(1,1)}_{im}\pi^{(1,1)}_{mj } \Big]_{\rm sim} (\vec p_1, \vec p_2)  \df_L(\vec p_1) \df_L(\vec p_2), \non
}
where 
\eeq{
\Big[ \pi^{(1,1)}_{im}\pi^{(1,1)}_{mj} \Big]_ {\rm sim} (\vec p_1, \vec p_2) 
= \frac{1}{2} \frac{\vec p_1 . \vec p_2}{p_1^2 p_2^2} (p_{1i}p_{2j} + p_{2i}p_{1j}).
}
The last of the second generation, second order, terms gives 
\eq{
\Big[ \Pi^{[1]}_{ij} \tr \big[ \Pi^{[1]} \big] \Big]^{(2)}(\vec k)  
&= \int_{\vec r} e^{i \vec k \cdot \vec r} ~ \Pi^{[1]}_{ij}(\vec r) \Pi^{[1]}_{mm}(\vec r) \\
&= (2\pi)^3 \df^{\rm D}_{\vec k - \vec p_{12}}  \Big[ \pi^{(1,1)}_{ij} (\vec p_1) \pi^{(1,1)}_{mm} (\vec p_2) \Big] \df_L(\vec p_1) \df_L(\vec p_2), \non\\
&= (2\pi)^3 \df^{\rm D}_{\vec k - \vec p_{12}}  \Big[ \pi^{(1,1)}_{ij} \Big]_{\rm sim}  (\vec p_1, \vec p_2)  \df_L(\vec p_1) \df_L(\vec p_2), \non
}
where 
\eeq{
 \Big[ \pi^{(1,1)}_{ij}\Big]_{\rm sim} (\vec p_1, \vec p_2) = \frac{1}{2} \lb \frac{p_{1i}p_{1j} }{p_1^2} + \frac{p_{2i}p_{2j} }{p_2^2}  \rb.
}

\textbf{(3)} The third order gives seven operators of (new) third generation, three coming from the second generation, and one originating at the first generation.
We start with the one originating at the first generation that has the following form 
\eeq{
\Big[\Pi^{[1]}_{ij}\Big]^{(3)}(\vec k) = (2\pi)^3 \df^{\rm D}_{\vec k - \vec p_{13}} \pi^{(1,3)}_{ij} (\vec p_1,\vec p_2,\vec p_3) \df_L(\vec p_1)\df_L(\vec p_2)\df_L(\vec p_3),
}
where 
\eeq{
\pi^{(1,3)}_{ij} (\vec p_1,\vec p_2,\vec p_3) = \frac{\vec p_{13,i} \vec p_{13,j}} { p_{13}^2 }
F^{(3)} (\vec p_1, \vec p_2, \vec p_3).
}
The first of the second generation, third order, operators gives
\eq{
 \Big[\Pi^{[2]}_{ij}\Big]^{(3)}(\vec k) = (2\pi)^3 \df^{\rm D}_{\vec k - \vec p_{13}} 
 \pi^{(2,3)}_{ij} (\vec p_1,\vec p_2,\vec p_3) \df_L(\vec p_1)\df_L(\vec p_2)\df_L(\vec p_3),
}
where $\pi^{(2,3)}_{ij}$ is given in \refeq{pi2333}. 
The second term from this group gives 
\eq{
\Big[ \big( \Pi^{[1]} \big)^2 \Big]_{ij}^{(3)}(\vec k) &= \int_{\vec r} e^{i \vec k \cdot \vec r} 
~\bigg( \Big[\Pi^{[1]} \Big]^{(2)}_{im}(\vec r)\Pi^{[1]}_{m j}(\vec r) + \Pi^{[1]}_{i m}(\vec r) \Big[\Pi^{[1]} \Big]^{(2)}_{m j}(\vec r)  \bigg) \\
&=  (2\pi)^3 \df^{\rm D}_{\vec k - \vec p_{13} } \Big[ \pi^{(1,2)}_{im} (\vec p_1, \vec p_2) ~\pi^{(1,1)}_{mj} (\vec p_3) 
     + \pi^{(1,2)}_{mj} (\vec p_1, \vec p_2) ~\pi^{(1,1)}_{im} (\vec p_3)  \Big]
     \df_L(\vec p_1) \df_L(\vec p_2) \df_L(\vec p_3) \non\\
&=  (2\pi)^3 \df^{\rm D}_{\vec k - \vec p_{13} }  \Big[ (\vec p_{12}. \vec p_{3} ) \frac{\vec p_{12,i}  p_{3,j}}{p_{12}^2 p_3^2} 
     + (\vec p_{12}. \vec p_{3} ) \frac{\vec p_{12,j}  p_{3,i}}{p_{12}^2 p_3^2}  \Big] F^{(2)} (\vec p_1, \vec p_2)
     \df_L(\vec p_1) \df_L(\vec p_2) \df_L(\vec p_3) \non\\
&=  (2\pi)^3 \df^{\rm D}_{\vec k - \vec p_{13} } \frac{ (\vec p_{12}. \vec p_{3} ) ~ \vec p_{12,\{ i}  p_{3,j\}}}{p_{12}^2 p_3^2} F^{(2)} (\vec p_1, \vec p_2)
     \df_L(\vec p_1) \df_L(\vec p_2) \df_L(\vec p_3) \non\\          
&=  (2\pi)^3 \df^{\rm D}_{\vec k - \vec p_{13} }
       \Bigg[ \frac{ (\vec p_{12}. \vec p_{3} ) ~ \vec p_{12,\{ i}  p_{3,j\}}}{3 p_{12}^2 p_3^2} F^{(2)} (\vec p_1, \vec p_2) \Bigg]_{\rm sim}
      \df_L(\vec p_1) \df_L(\vec p_2) \df_L(\vec p_3),   \non
}
where 
\eq{
\Bigg[ \frac{ (\vec p_{12}. \vec p_{3} ) ~ \vec p_{12,\{ i}  p_{3,j\}}}{3 p_{12}^2 p_3^2} F^{(2)} (\vec p_1, \vec p_2) \Bigg]_{\rm sim}
= \frac{1}{3} \frac{ (\vec p_{12}. \vec p_{3} ) ~ \vec p_{12,\{ i}  p_{3,j\}}}{p_{12}^2 p_3^2} F^{(2)} (\vec p_1, \vec p_2) + 2~{\rm cycle~perm.} .
}
Note that above both $\pi^{(1,2)}_{ij}$ and $\pi^{(1,1)}_{ij}$ are symmetric in the indices.
The last term from the second generation group is  
\eq{
\Big[ \Pi^{[1]}_{ij} \tr\big[ \Pi^{[1]} \big] \Big]^{(3)} (\vec k) &= \int_{\vec r} e^{i \vec k \cdot \vec r} 
~\bigg( \Big[ \Pi^{[1]}\Big]_{ij}^{(2)}(\vec r) \tr\big[ \Pi^{[1]} \big]^{(1)} (\vec r) + \Big[\Pi^{[1]} \Big]_{ij}^{(1)}(\vec r) \tr\big[ \Pi^{[1]} \big]^{(2)} (\vec r) \bigg) \\
&=   (2\pi)^3 \df^{\rm D}_{\vec k - \vec p_{13}} \bigg[  \pi^{(1,2)}_{ij} (\vec p_1, \vec p_2) 
 + F^{(2)} (\vec p_1, \vec p_2) \pi^{(1,1)}_{ij} (\vec p_3) \bigg] \df_L(\vec p_1) \df_L(\vec p_2) \df_L(\vec p_3)  \non\\
&=   (2\pi)^3 \df^{\rm D}_{\vec k - \vec p_{13}} \bigg( \frac{\vec p_{12,i} \vec p_{12,j}} { p_{12}^2 }
 + \frac{ p_{3,i} p_{3,j}}{p_3^2}  \bigg) F^{(2)} (\vec p_1, \vec p_2) ~ \df_L(\vec p_1) \df_L(\vec p_2) \df_L(\vec p_3)  \non\\
&=   (2\pi)^3 \df^{\rm D}_{\vec k - \vec p_{13}} \Bigg[ \frac{1}{3}\bigg( \frac{\vec p_{12,i} \vec p_{12,j}} { p_{12}^2 }
 + \frac{ p_{3,i} p_{3,j}}{p_3^2}  \bigg) F^{(2)} (\vec p_1, \vec p_2) \Bigg]_{\rm sim} \df_L(\vec p_1) \df_L(\vec p_2) \df_L(\vec p_3), \non
}
where 
\eq{
\Bigg[ \frac{1}{3}\bigg( \frac{\vec p_{12,i} \vec p_{12,j}} { p_{12}^2 }
 + \frac{ p_{3,i} p_{3,j}}{p_3^2}  \bigg) F^{(2)} (\vec p_1, \vec p_2) \Bigg]_{\rm sim}
= \frac{1}{3}\bigg( \frac{\vec p_{12,i} \vec p_{12,j}} { p_{12}^2 }
 + \frac{ p_{3,i}  p_{3,j}}{p_3^2}  \bigg) F^{(2)} (\vec p_1, \vec p_2)  + 2~{\rm cycle~perm.} 
}
The first of the new (third) generation operators starting from the third order is
\eq{
 \Big[\Pi^{[3]}_{ij}\Big]^{(3)}(\vec k) = (2\pi)^3 \df^{\rm D}_{\vec k - \vec p_{13}} 
D^2~ \pi^{(3,3)}_{ij} (\vec p_1,\vec p_2,\vec p_3) \df_L(\vec p_1)\df_L(\vec p_2)\df_L(\vec p_3),
}
where $\pi^{(3,3)}_{ij}$ is given in \refeq{pi2333}. 
The second of the third generation, third order, operators gives
\eq{
\Big[ \big( \Pi^{[1]} \big)^3 \Big]^{(3)}_{ij} & = \int_{\vec r} e^{i \vec k \cdot \vec r} 
\big[ \Pi^{[1]} \big]^{(1)}_{in} (\vec r) \big[ \Pi^{[1]} \big]^{(1)}_{nm} (\vec r) \big[ \Pi^{[1]} \big]^{(1)}_{mj} (\vec r) \non\\
& =  (2\pi)^3 \df^{\rm D}_{\vec k - \vec p_{13} } \frac{ (\vec p_1 . \vec p_2)(\vec p_2.  \vec p_3)}{p_1^2 p_2^2 p_3^2} ( \vec p_{1,i} \vec p_{3,j} ) 
\df_L(\vec p_1) \df_L(\vec p_2) \df_L(\vec p_3) \non\\
& =  (2\pi)^3 \df^{\rm D}_{\vec k - \vec p_{13} } \bigg[ \frac{ (\vec p_1 . \vec p_2)(\vec p_2.  \vec p_3)}{ 6 p_1^2 p_2^2 p_3^2} ( \vec p_{\{1,i} \vec p_{3,j\}} ) \bigg]_{\rm sim}
\df_L(\vec p_1) \df_L(\vec p_2) \df_L(\vec p_3),
} 
where 
\eeq{
\bigg[ \frac{ (\vec p_1 . \vec p_2)(\vec p_2.  \vec p_3)}{ 3 p_1^2 p_2^2 p_3^2} ( \vec p_{1,i} \vec p_{3,j} ) \bigg]_{\rm sim} 
= \frac{ (\vec p_1 . \vec p_2)(\vec p_2.  \vec p_3)}{ 6 p_1^2 p_2^2 p_3^2} ( \vec p_{\{1,i} \vec p_{3,j\}}) + 2~{\rm cycle~perm.}.
}
The third of the third generation, third order, operators gives
\eq{
\Big[ \Pi^{[1]} \Pi^{[2]}  \Big]^{(3)}_{ij} & = \int_{\vec r} e^{i \vec k \cdot \vec r} \big[ \Pi^{[1]} \big]^{(1)}_{\{ im} (\vec r) \big[ \Pi^{[2]} \big]^{(2)}_{mj \}}  (\vec r) \\
& =  (2\pi)^3 \df^{\rm D}_{\vec k - \vec p_{13}}
 \frac{\vec p_{\{1,i} \vec p_{1,m}}{p_1^2} \pi^{(2,2)}_{mj\}} (\vec p_2,\vec p_3)  \df_L(\vec p_1) \df_L(\vec p_2)\df_L(\vec p_3) \non\\
 & =  (2\pi)^3 \df^{\rm D}_{\vec k - \vec p_{13}} \bigg[
 \frac{\vec p_{\{1,i} \vec p_{1,m}}{p_1^2} \pi^{(2,2)}_{mj\} } (\vec p_2,\vec p_3) \bigg]_{\rm s}  \df_L(\vec p_1) \df_L(\vec p_2)\df_L(\vec p_3) \non\\
& =  (2\pi)^3 \df^{\rm D}_{\vec k - \vec p_{13}} \Bigg[
 \frac{ ( \vec p_{1} . \vec p_{23}) \vec p_{\{1,i} \vec p_{23,j\} }} { 3 p_1^2 p_{23}^2 } F^{(2)} (\vec p_2 , \vec p_3)  \non\\
& \hspace{1.5cm} - \frac{1}{6} \frac{\vec p_2. \vec p_3}{p_1^2 p_2^2 p_3^2} \big( (\vec p_1 . \vec p_2) p_{\{1,i} p_{2j\}} + ( \vec p_1 . \vec p_3) p_{\{1,i} p_{3j\}}  \big) 
\Bigg]_{\rm sim}  \df_L(\vec p_1) \df_L(\vec p_2)\df_L(\vec p_3). \non
}
The fourth term gives 
\eq{
\Big[ \Pi^{[2]}_{ij} \tr \big[ \Pi^{[1]} \big] \Big]^{(3)} &=  \int_{\vec r} e^{i \vec k \cdot \vec r}  \big[ \Pi^{[2]} \big]^{(2)}_{ij}  (\vec r) ~ \tr \big[ \Pi^{[1]} \big]^{(1)}  (\vec r) \\
& =(2\pi)^3 \df^{\rm D}_{\vec k - \vec p_{13}} \pi^{(2,2)}_{ij} (\vec p_1,\vec p_2)  \df_L(\vec p_1)\df_L(\vec p_2)  \df_L(\vec p_3) \non\\
& =(2\pi)^3 \df^{\rm D}_{\vec k - \vec p_{13}} \Big[ \tfrac{1}{3} \pi^{(2,2)}_{ij} (\vec p_1,\vec p_2) \Big]_{\rm sim}  \df_L(\vec p_1)\df_L(\vec p_2)  \df_L(\vec p_3) \non\\
& =(2\pi)^3 \df^{\rm D}_{\vec k - \vec p_{13}} \bigg[ \frac{\vec p_{12,i} \vec p_{12,j}} { 3 p_{12}^2 } F^{(2)} (\vec p_1 , \vec p_2) \non\\
& \hspace{3.0cm} - \frac{\vec p_1. \vec p_2}{6p_1^2 p_2^2} \lb p_{1i}p_{1j} + p_{2i}p_{2j} \rb \bigg]_{\rm sim}  \df_L(\vec p_1)\df_L(\vec p_2)  \df_L(\vec p_3). \non
}
The fifth term gives 
\eq{
\Big[ \big[ \big( \Pi^{[1]} \big)^2 \big]_{ij}\tr\big[ \Pi^{[1]} \big] \Big]^{(3)}  
&=  \int_{\vec r} e^{i \vec k \cdot \vec r} ~ \big[ \Pi^{[1]} \big]^{(1)}_{im}  (\vec r) ~  \big[ \Pi^{[1]} \big]^{(1)}_{mj}  (\vec r)  ~ \tr \big[ \Pi^{[1]} \big]^{(1)}  (\vec r) \\
& =(2\pi)^3 \df^{\rm D}_{\vec k - \vec p_{13}} \frac{(\vec p_1.\vec p_2) p_{1,\{ i} p_{2,j\}} }{2 p_1^2 p_2^2}  \df_L(\vec p_1)\df_L(\vec p_2)  \df_L(\vec p_3) \non\\
& =(2\pi)^3 \df^{\rm D}_{\vec k - \vec p_{13}} \bigg[ \frac{(\vec p_1.\vec p_2)  p_{1,\{ i} p_{2,j\}} }{6 p_1^2 p_2^2} \bigg]_{\rm sim}  \df_L(\vec p_1)\df_L(\vec p_2)  \df_L(\vec p_3). \non
}
The sixth term gives 
\eq{
\Big[\Pi^{[1]}_{ij}\tr\big[ \Pi^{[1]} \big]^2 \Big]^{(3)}  
&=  \int_{\vec r} e^{i \vec k \cdot \vec r} ~ \big[ \Pi^{[1]} \big]^{(1)}_{ij}  (\vec r) ~ \Big[ \tr \big[ \Pi^{[1]} \big]^{(1)}  (\vec r) \Big]^2 \\
& =(2\pi)^3 \df^{\rm D}_{\vec k - \vec p_{13}} \frac{ p_{1,i} p_{1,j} }{ p_1^2 }  \df_L(\vec p_1)\df_L(\vec p_2)  \df_L(\vec p_3) \non\\
& =(2\pi)^3 \df^{\rm D}_{\vec k - \vec p_{13}} \bigg[ \frac{ p_{1,i} p_{1,j} }{ 3 p_1^2 }\bigg]_{\rm sim}  \df_L(\vec p_1)\df_L(\vec p_2)  \df_L(\vec p_3).\non
}
Finally, the last of the third order, third generation, terms is
\eq{
\Big[ \Pi^{[1]}_{ij} \tr\big[ \big( \Pi^{[1]} \big)^2 \big] \Big]^{(3)} 
&=  \int_{\vec r} e^{i \vec k \cdot \vec r} ~ \big[ \Pi^{[1]} \big]^{(1)}_{ij}  (\vec r) ~  \big[ \Pi^{[1]} \big]^{(1)}_{nm}  (\vec r) \big[ \Pi^{[1]} \big]^{(1)}_{mn}  (\vec r) \\
& =(2\pi)^3 \df^{\rm D}_{\vec k - \vec p_{13}} \frac{ p_{1,i} p_{1,j} (\vec p_2 . \vec p_3)^2 }{ p_1^2p_2^2p_3^2 }  \df_L(\vec p_1)\df_L(\vec p_2)  \df_L(\vec p_3) \non\\
& =(2\pi)^3 \df^{\rm D}_{\vec k - \vec p_{13}} \bigg[ \frac{ p_{1,i} p_{1,j} (\vec p_2 . \vec p_3)^2 }{3 p_1^2p_2^2p_3^2 } \bigg]_{\rm sim}  \df_L(\vec p_1)\df_L(\vec p_2) 
\df_L(\vec p_3). \non
}

\subsection{Bias expansion and renormalisation of the one-loop power spectrum}
\label{app:renorm}

Naively, if we would count all the operators given in table~\ref{tb:Bias_kernels}, 
we would get eleven bias operators without including the stochastic, or any of the higher 
derivative operators. These operators are independent at the level of the field. However their 
contribution to the two-point (at one-loop level) and three-point (at three-level) statistics generates 
some additional degeneracies. These degeneracies would, of course, be broken if the four-point 
function (at tree-level) would be added into the consideration. In one-loop two-point function, because 
only specific momentum configurations contribute, operator correlations are not all independent, 
and thus the number of independent bias coefficients can be reduced. In this section, we explore 
these degeneracies and derive results described by the full non-degenerate set of operators 
for one-loop power spectrum. 

We first look at the ${(22)}$ term. Operators and bias coefficients appearing as 
$\tr \big[\bm K^{(2)}(\vec p, \vec k - \vec p) \big]$ 
and $\vec Y^{(q)*}_2 . \bm K^{(2)}(\vec p, \vec k - \vec p)$, 
are given by four tensor operators
\eq{
 K_{ij}^{(2)}(\vec p, \vec k - \vec p) = 
    &c_1 \big[ \Pi^{[1]} \big]^{(2)}_{ij} (\vec p, \vec k - \vec p)
+ c_{2,1} \big[ \Pi^{[2]} \big]^{(2)}_{ij}(\vec p, \vec k - \vec p) \non\\
& + c_{2,2} \Big[ \big( \Pi^{[1]} \big)^2 \Big]^{(2)}_{ij}(\vec p, \vec k - \vec p)
 + c_{2,3} \Big[ \Pi^{[1]} \tr \big[ \Pi^{[1]} \big] \Big]^{(2)}_{ij} (\vec p, \vec k - \vec p),
}
where it is understood, as we discussed earlier, that the trace and trace-free part have different bias coeffitients. 
Using the linear dependence of trace terms given in Eq. \eqref{eq:trace_second_order}
we can replace for example the trace of $\vec \Pi^{[2]}$ operator
\eq{
\tr \big[\bm K^{(2)}(\vec p, \vec k - \vec p) \big]
= c^{\rm s}_1 \tr \big[ \Pi^{[1]} \big]^{(2)}  (\vec p, &\vec k - \vec p)
 + \Big( c^{\rm s}_{2,2} + \tfrac{2}{7} c^{\rm s}_{2,1} \Big) \tr \Big[ \big( \Pi^{[1]} \big)^2 \Big]^{(2)} (\vec p, \vec k - \vec p) \non\\
& + \Big( c^{\rm s}_{2,3} + \tfrac{5}{7} c^{\rm s}_{2,1} \Big) \Big[ \big( \tr [\Pi^{[1]} ] \big)^2 \Big]^{(2)} (\vec p, \vec k - \vec p). 
\label{eq:K_tr}
}
In the trace we are sensitive only to three independent operators and as expected have
only two new independent bias combinations: $ c^{\rm s}_{2,2} + \tfrac{2}{7} c^{\rm s}_{2,1}$ and $c^{\rm s}_{2,3} + \tfrac{5}{7} c^{\rm s}_{2,1}$.
A similar reduction can be obtained using the projection with $\vec Y^{(0)}_2(\vec k)$ basis function and 
the resulting operator dependence given in Eq.~\eqref{eq:Y0_projection_lin_dep}. 
Using this, we get the expression
\eq{
\vec Y^{(0)*}_2(\vec k) . \bm K^{(2)}(\vec p, \vec k - \vec p) 
=&c^{\rm g}_1 \vec Y^{(0)}_2(\vec k) . \big[ \Pi^{[1]} \big]^{(2)} (\vec p, \vec k - \vec p) \\
&+ \Big( c^{\rm g}_{2,2} - \tfrac{13}{7} c^{{\rm g}}_{2,1} \Big) \vec Y^{(0)*}_2(\vec k) .\Big[ \big( \Pi^{[1]} \big)^2 \Big]^{(2)} (\vec p, \vec k - \vec p) \non\\
&+ \Big( c^{\rm g}_{2,3} + \tfrac{20}{7} c^{{\rm g}}_{2,1} \Big) \vec Y^{(0)*}_2(\vec k) . \Big[ \Pi^{[1]}  \tr \big[ \Pi^{[1]} \big] \Big]^{(2)} (\vec p, \vec k - \vec p). \non
}
Thus both the trace and $\vec Y^{(0)}_2(\vec k)$ projection for the $(22)$ contribution have only three independent
second order tensor operators: $\Pi^{[1]}$, $\big( \Pi^{[1]} \big)^2$ and $\Pi^{[1]}  \tr \big[ \Pi^{[1]} \big]$. 
However, after projections, these operators can be further reduced using the relations given in Eq.~\eqref{eq:tr&Y0_projection_lin_dep}, 
and we can replace the $\vec Y^{(0)}_2. \big[ \Pi^{[1]} \big] $ and $\vec Y^{(0)}_2. \big[  \Pi^{[1]} \tr \big[ \Pi^{[1]}\big] \big]$ operators to get
\eq{
\vec Y^{(0)*}_2(\vec k) . \bm K^{(2)}(\vec p, \vec k - \vec p) 
 = &c^{{\rm g}}_1  \tr \big[ \Pi^{[1]} \big]^{(2)}(\vec p, \vec k - \vec p) \\
&+ \Big( c^{{\rm g}}_{2,3} + \tfrac{20}{7} c^{{\rm g}}_{2,1} \Big) \tr \left[ \Pi^{[1]} \tr \big[ \Pi^{[1]}\big] - \big( \Pi^{[1]} \big)^2 \right]^{(2)}(\vec p, \vec k - \vec p) \non\\
&+ \Big( c^{{\rm g}}_{2,1} + c^{{\rm g}}_{2,2} + c^{{\rm g}}_{2,3}  \Big) \vec Y^{(0)*}_2(\vec k) .\Big[ \big( \Pi^{[1]} \big)^2 \Big]^{(2)} (\vec p, \vec k - \vec p). \non
\label{eq:Y0K2}
}
Next we look at the projections using the higher helicity basis vectors $\vec Y^{(1)*}_2(\vec k)$ and $\vec Y^{(2)*}_2(\vec k)$. 
Using the relations in Eq.~\eqref{eq:Y12_projection_lin_dep} it follows
\eq{
\vec Y^{(1)*}_2(\vec k) . \bm K^{(2)}(\vec p, \vec k - \vec p) 
&= \Big( c^{{\rm g}}_{2,1} + c^{{\rm g}}_{2,2} + c^{{\rm g}}_{2,3} \Big) \vec Y^{(1)*}_2(\vec k) .\Big[ \big( \Pi^{[1]} \big)^2 \Big]^{(2)} (\vec p, \vec k - \vec p), \\
\vec Y^{(2)*}_2(\vec k) . \bm K^{(2)}(\vec p, \vec k - \vec p) 
&= \Big(c^{{\rm g}}_{2,1} + c^{{\rm g}}_{2,2} \Big) \vec Y^{(2)*}_2(\vec k) .\Big[ \big( \Pi^{[1]} \big)^2 \Big]^{(2)} (\vec p, \vec k - \vec p) \non\\
&\hspace{3cm} + c^{{\rm g}}_{2,3} \vec Y^{(2)*}_2(\vec k) .\Big[ \Pi^{[1]} \tr \big[ \Pi^{[1]} \big] \Big]^{(2)} (\vec p, \vec k - \vec p) \non\\
 &=
 \Big(c^{{\rm g}}_{2,1} + c^{{\rm g}}_{2,2} + c^{{\rm g}}_{2,3} \Big) \vec Y^{(2)*}_2(\vec k) .\Big[ \big( \Pi^{[1]} \big)^2 \Big]^{(2)} (\vec p, \vec k - \vec p) \non\\
 &\hspace{3cm} + c^{{\rm g}}_{2,3} \vec Y^{(2)*}_2(\vec k) . \left[  \Pi^{[1]}\tr \big[ \Pi^{[1]} \big] - \big( \Pi^{[1]} \big)^2 \right]^{(2)} (\vec p, \vec k - \vec p). \non
}
Combining all these results leads to seven independent scalar operators at second order:
\eq{
\bm{ \mathcal K}^{(2)}(\vec{k}_1,\vec{k}_2) \equiv \Big\{  &
\tr \big[ \Pi^{[1]} \big], \tr \left[\big( \Pi^{[1]} \big)^2 \right], \tr \left[ \Pi^{[1]} \tr \big[ \Pi^{[1]} \big] \right], \vec Y^{(0)*}_2(\vec k_{12}) . \big( \Pi^{[1]} \big)^2, \\
&\vec Y^{(1)*}_2(\vec k_{12}) . \big( \Pi^{[1]} \big)^2, \vec Y^{(2)*}_2(\vec k_{12}) . \big( \Pi^{[1]} \big)^2, \vec Y^{(2)*}_2(\vec k_{12}) . \left[ \Pi^{[1]} \tr \big[ \Pi^{[1]} \big] \right] \Big\}(\vec{k}_1,\vec{k}_2). \non
}
In terms of bias coefficients we would have two independent second order trace coefficients, and three trace-free coefficients. 
Note that in the basis we have chosen here most of the operators depend only on $c^{\rm g}_{2,2} - \tfrac{13}{7} c^{{\rm g}}_{2,1}$ 
and $c^{{\rm g}}_{2,3} + \tfrac{20}{7} c^{{\rm g}}_{2,1}$ coefficients since 
\eeq{
\lb c^{\rm g}_{2,2} - \tfrac{13}{7} c^{{\rm g}}_{2,1} \rb + \lb c^{g}_{2,3} + \tfrac{20}{7} c^{{\rm g}}_{2,1} \rb =  c^{{\rm g}}_{2,1} + c^{{\rm g}}_{2,2} + c^{{\rm g}}_{2,3}, \non
}
and so the only operator that remains multiplied with the new coefficients is  
\eeq{
\vec Y^{(2)*}_2(\vec k) . \left[ \big[ \Pi^{[1]}\big] - \big( \Pi^{[1]} \big)^2 \right]^{(2)} (\vec p, \vec k - \vec p) \subset \vec Y^{(2)*}_2(\vec k) . \bm K^{(2)}(\vec p, \vec k - \vec p), 
}
carrying the bare $c^{{\rm g}}_{2,3}$ coefficient.

\begin{figure}[t!]
\centering
\includegraphics[width=0.7\linewidth]{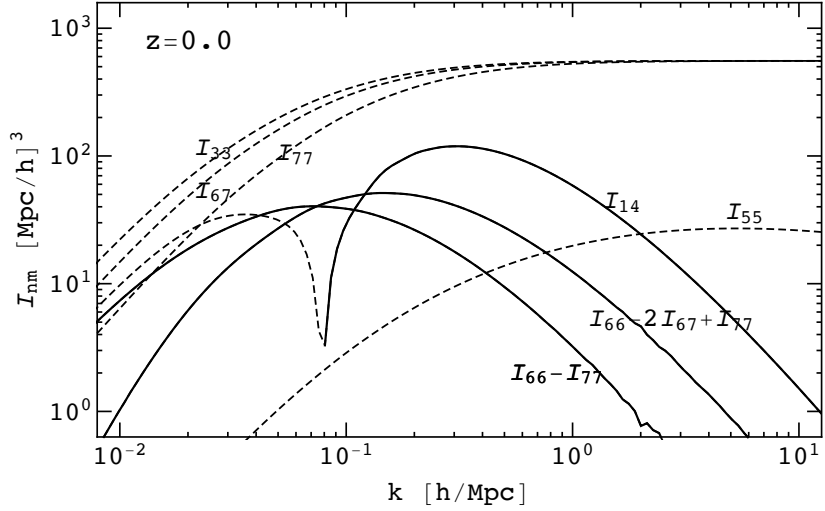}
\caption{Linearly dependent mode coupling terms $I_{14}$, $I_{66}$, $I_{67}$ and $I_{66}$ shown  at $z=0$. 
Dashed lines represent negative contributions while the solid ones are positive.}
\label{fig:asymptotic_I6677}
\end{figure}

Next we look at the ${(22)}$ loop integrals given in table~\ref{tb:ps_loops}. 
For $q=1$ and $2$, $P_{22}^{ab(q)}$ integrand terms also contain integration 
over the azimuthal angle. However these integration can be performed exactly, given that we can choose the coordinate frame in 
which none of the $P_L$ arguments depend on the azimuthal angle. For trace terms and terms with zero helicity $q=0$, the azimuthal angle integration is trivial. 
For this reason all the contributions to the $P_{22}$ components in table~\ref{tb:ps_loops} can be obtained from the
list of $\bm{\mathcal K}^{(2)}$ operators in the form of cross-correlation integrals
\eeq{
I_{nm}(k) = 
\big[ \bm{\mathcal K} \big]_n \lb\vec p, \vec k - \vec p\rb \big[ \bm{\mathcal K} \big]_m\lb\vec p, \vec k - \vec p\rb 
P_L (\vec p)P_L (\vec k - \vec p),
\label{eq:I_master_22_II}
}
which are also explicitly are given in table~\ref{tb:ps_scalar_loops_22}.. 
Note that the integral components are symmetric, i.e. $I_{ij}(k)=I_{ji}(k)$.
\begin{table}
\captionof{table}{Relevant one-loop auto- and cross-correlation contributions to $P_{22}$.}
\resizebox{\textwidth}{!}{
\begin{tabular}{ c| c c c c c c c } 
 \hline
 Op. & $\tr \big[ \Pi^{[1]} \big] $
& $\tr \left[\big( \Pi^{[1]} \big)^2 \right] $
& $\tr \left[ \Pi^{[1]} \tr \big[ \Pi^{[1]} \big] \right] $
& $\vec Y^{(0)}_2(\vec k) . \big( \Pi^{[1]} \big)^2$
& $\vec Y^{(1)}_2(\vec k) . \big( \Pi^{[1]} \big)^2$
& $\vec Y^{(2)}_2(\vec k) . \big( \Pi^{[1]} \big)^2$
& $\vec Y^{(2)}_2(\vec k) . \left[ \Pi^{[1]} \tr \big[ \Pi^{[1]} \big] \right]$ \\[0.3cm] \hline 
$\tr \big[ \Pi^{[1]} \big]$ & $I_{11}$ & $I_{12}$ & $I_{13}$ & $I_{14}$ & 0 & 0 & 0 \\[0.3cm] \hline
$\tr \left[\big( \Pi^{[1]} \big)^2 \right] $ &   & $I_{22}$ & $I_{23}$ & $I_{24}$ & 0 & 0 & 0 \\[0.3cm] \hline
$\tr \left[ \Pi^{[1]} \tr \big[ \Pi^{[1]} \big] \right] $ &  &  & $I_{33}$ & $I_{34}$ & 0 & 0 & 0 \\[0.3cm] \hline
$\vec Y^{(0)}_2 . \big( \Pi^{[1]} \big)^2$ &   &  &  & $I_{44}$ & 0 & 0 & 0 \\[0.3cm] \hline
$\vec Y^{(1)}_2 . \big( \Pi^{[1]} \big)^2$ &   &  &  &  & $I_{55}$ & 0 & 0 \\[0.3cm] \hline
$\vec Y^{(2)}_2 . \big( \Pi^{[1]} \big)^2$ &  &  & &  &  & $I_{66}$ & $I_{67}$ \\[0.3cm] \hline
$\vec Y^{(2)}_2 . \left[ \Pi^{[1]} \tr \big[ \Pi^{[1]} \big] \right]$ &  &  &  & &  &  & $I_{77}$ \\[0.3cm] \hline
\end{tabular}}
\label{tb:ps_scalar_loops_22}
\end{table}
After we have exhausted the linear dependencies at the level of the field operators, 
we explore the ones at the level of correlators.
These are obtained, once the integration over some of the variables is performed.
First we list the dependencies in the $P_{22}$ type of integrals. These are structured into the 
contributions to the integrals $I_{nm}(k)$ given in Eq. \eqref{eq:I_master_22_II}.
Considering first the trace and helicity-0 contributions we can find one dependence relation
\eeq{
28 I_{12} - I_{22}+I_{23} - 2 \sqrt{6} \lb 7 I_{14}  + 5 I_{24} - 5  I_{34} \rb = 0.
\label{eq:I_dep_scalar}
}
Higher helicity correlators exhibit similar dependencies after the azimuthal integration is performed. 
Specifically, for helicity $q=1$ the integration gives $(\vec e^+ . \vec p)(\vec e^- . \vec p) \to -\frac{1}{2} p^2(1-\mu^2)$
and similarly for $q=2$ we have $(\vec e^+ . \vec p)^2(\vec e^- . \vec p)^2 \to \frac{1}{4} p^4(1-\mu^2)^2$.
We get the dependencies:
\eq{
2 I_{22} - 2 \sqrt 6 I_{24}+3 I_{44} - 18 I_{66} &= 0, \non\\
2 I_{22} + 6 I_{23} - 5 \sqrt 6 I_{24} - 3 \sqrt 6 I_{34} + 12 I_{44} - 72 I_{67} &=0, \non\\
I_{22} + 6 I_{23} + 9 I_{33} - 4 \sqrt 6 I_{24} - 12 \sqrt 6  I_{34} +24 I_{44} - 144 I_{77} &= 0 .
\label{eq:I_dep_higher_hel}
}
Once we have explored all the dependencies we can list explicit expressions 
of all the remaining, independent, one-loop mode coupling correlators:
\eq{
\label{eq:I_xy_contributions}
I_{11}(k) &= 
\frac{\left( k^2 (7 \vec k\cdot \vec p+3 p^2)-10 (\vec k\cdot \vec p)^2\right)^2}{196 p^4 (\vec k- \vec p)^4} P_L (\vec p) P_L (\vec k - \vec p), \\
I_{12}(k) &= 
\frac{(\vec k\cdot \vec p-p^2)^2 \left( k^2 (7 \vec k\cdot \vec p+3 p^2)-10 (\vec k \cdot \vec p)^2\right)}{14 p^4 (\vec k- \vec p)^4} P_L (\vec p) P_L (\vec k - \vec p), \non\\
I_{13}(k) &= 
\frac{ k^2 (7 \vec k\cdot \vec p+3 p^2)-10 (\vec k\cdot \vec p)^2}{14 p^2 (\vec k- \vec p)^2} P_L (\vec p) P_L (\vec k - \vec p), \non\\
I_{22}(k) &= 
\frac{(\vec k\cdot \vec p-p^2)^4}{p^4 (\vec k- \vec p)^4} P_L (\vec p) P_L (\vec k - \vec p), \non\\
I_{23}(k) &= 
\frac{(\vec k\cdot \vec p-p^2)^2}{p^2 (\vec k- \vec p)^2} P_L (\vec p) P_L (\vec k - \vec p), \non\\
I_{24}(k) &= 
\frac{(\vec k\cdot \vec p-p^2)^3 \lb k^2 (2 \vec k\cdot \vec p+ p^2 )-3 (\vec k\cdot \vec p)^2 \rb }{\sqrt 6 k^2 p^4 (\vec k- \vec p)^4} P_L (\vec p)P_L (\vec k - \vec p), \non\\
I_{33}(k) &= 
P_L (\vec p)P_L (\vec k - \vec p), \non\\
I_{34}(k) &= 
\frac{(\vec k\cdot \vec p - p^2) \left(k^2 (2 \vec k\cdot \vec p + p^2)-3 (\vec k\cdot \vec p)^2\right)}{\sqrt 6 k^2 p^2 (\vec k- \vec p)^2} P_L (\vec p)P_L (\vec k - \vec p), \non\\
I_{44}(k) &= 
\frac{(\vec k\cdot \vec p - p^2)^2 \left(k^2 (2 \vec k\cdot \vec p + p^2)-3 (\vec k\cdot \vec p)^2\right)^2}{ 6 k^4 p^4 (\vec k- \vec p)^4} P_L (\vec p)P_L (\vec k - \vec p), \non\\
I_{55}(k) &= 
\frac{(k^2 - 2 \vec k\cdot \vec p)^2 (\vec k\cdot \vec p - p^2)^2 \left((\vec k\cdot \vec p)^2 - k^2 p^2 \right)}{ 4 k^2 p^4 (\vec k- \vec p)^4} P_L (\vec p)P_L (\vec k - \vec p). \non
}
These contributions are shown in the left panel of figure~\ref{fig:asymptotic_IandJ} at $z=0$. 
All $I_{nm}$ terms, except  $I_{11}$, have $\sim k^2$ asymptotic functional dependence on large scales once the constant, UV sensitive, part has been subtracted.
The UV sensitive constant piece can always be subtracted the large scale since this contributions can always be taken into account by the renormalised stochastic operators. 
The matter $I_{11}$ term starts with the asymptotic form $\sim k^4$, which is a manifestation of the mass and momentum conservation. 
The dependent terms $I_{14}$, $I_{66}$, $I_{67}$ and $I_{66}$ are given in figure~\ref{fig:asymptotic_I6677}.
For comparison we also show the $I_{55}$ on the same figure, showing that it is relatively suppressed on the scales of interest $k<1$Mpc/$h$.

All this contributions to the ${(22)}$ correlators are the combined to terms that are controlled by two trace, 
and three trace-free second order bias coefficients, that can be redefined as 
\eq{
\label{eq:bias_b_second}
a \in \{ {\rm n},\, {\rm s} \}: ~~ \Big\{ b^{a}_{2,1}, b^{a}_{2,2} \Big\}&, ~~ {\rm where:} ~~  b^{a}_{2,1} = c^a_{2,2} + \tfrac{2}{7} c^a_{2,1}, 
~~~~~~~~  b^{a}_{2,2} =  c^a_{2,3} + \tfrac{5}{7} c^a_{2,1},  \\
{\rm g}: ~ \Big\{ b^{\rm g}_{2,1}, b^{\rm g}_{2,2}, b^{\rm g}_{2,3} \Big\}&, ~~ {\rm where:} ~~b^{\rm g}_{2,1} = c^{\rm g}_{2,1} + c^{\rm g}_{2,2} +c^{\rm g}_{2,3}, 
~~b^{\rm g}_{2,2} = c^{\rm g}_{2,3} + \tfrac{20}{7} c^{\rm g}_{2,1},~~ b^{\rm g}_{2,3} = c^{\rm g}_{2,3}.  \non
}

We next turn to ${(13)}$ terms in table~\ref{tb:ps_loops}.  Only helicity-0 terms ($q=0$) are contributing to this correlator. 
These are $\tr \big[\bm K^{(3)} \big]$ terms and $\vec Y^{(0)*}_2 . \bm K^{(3)}$ terms. 
Given the specific form of the input vectors $\vec k$, $\vec p$ and $- \vec p$ we have additional constraints and linear dependencies among the contributing operators. 
For the trace term, we have five relations given in Eq.~\eqref{eq:tr_third_operators_II}, which leaves six independent terms in the $\tr \big[\bm K^{(3)} \big]$ term. 
The second helicity-0 contribution $\vec Y^{(0)*}_2 . \bm K^{(3)}$ gives similar linear dependencies listed in Eq.~\eqref{eq:Y0_third_operators_II}. 
However, here eight terms survive. These dependence relations can be used to reshuffle the bias coefficients leaving only these independent $k-$dependent terms.
Required bias combinations, multiplying these surviving terms, are given in table~\ref{tb:ps_scalar_loops_13}.

All of these (13) contributions can exhibit UV sensitive and potentially divergent behavior, and thus they need to be regularised and renormalised. 
We can isolate this UV sensitive contributions by considering the leading terms when expanding in low $k$ expansion. 
Looking at the set of independent operators from table~\ref{tb:ps_scalar_loops_13} from which we want to isolate the leading UV part  
\eeq{
\lb 3 ~ \int_{\vp}~ \bm K^{(3)}(\vec k, \vec p, - \vec p) P_L (q)  \rb_{\rm UV}, \non
}
gives us 
\eq{
\label{eq:UV_tr}
\bigg( 3 \int_{\vp} \tr &
\bigg\{  \Pi^{[1]}, \big( \Pi^{[1]} \big)^2, \Pi^{[1]} \tr \big[ \Pi^{[1]} \big], \Pi^{[1]} \Pi^{[2]},  \Pi^{[1]} \Big(\tr\big[ \Pi^{[1]} \big] \Big)^2, \\
&\hspace{3cm}  \Pi^{[1]} \tr\big[ \big( \Pi^{[1]} \big)^2 \big] \bigg\} (\vec k, \vec p, - \vec p) P_L (p)  \bigg)_{\rm UV}
\underset{k \to 0}{\implies} \left\{0,\frac{68}{63},\frac{68}{63},\frac{82}{63},1,\frac{5}{9}\right\} \sigma^2 \non,
}
where $\sigma^2 = \int_{\vp} P_L (p)$. We can now renormalise the linear bias coefficient $c^s_1$, in order to absorb these UV contributions. 
Note that the first $\vec \Pi^{[1]}$ operator does not have UV divergencies of this kind, as expected given that it is precisely the dark matter contribution 
given by the standard $F^{(3)}$ kernel. In order to perform this procedure we proclaim a linear bias $c^s_1$ parameter to be a `bare' parameter, and to be in fact a 
sum of a finite part and a UV counter term part:
\eeq{
c^{\rm s}_1= \big(c^{\rm s}_1\big)_{\rm fin.} + \big(c^{\rm s}_1\big)_{\rm UV}.
}
The UV counter term can be set to precisely cancel the contributions obtained in Eq.~\eqref{eq:UV_tr} above.
After this renormalisation procedure, only three operators contributing to the $\tr \big[\bm K^{(3)} \big]$ kernel survive 
\eeq{
\tr \Big[ \Big\{ \Pi^{[1]} ,  ~ \big( \Pi^{[1]} \big)^2,  ~ \Pi^{[1]} \Pi^{[2]} \Big\} \Big]_{\rm fin.}. 
}
However, after the renormalisation we have a further degeneracy in last two operators 
\eeq{
\tr  \Big[ \big( \Pi^{[1]} \big)^2 - \Pi^{[1]} \Pi^{[2]} \Big]_{\rm fin.} =0.
}
Using this degeneracy to eliminate one of them, 
we end up with two independent trace operators
\eeq{
\tr \Big[ \Big\{ \Pi^{[1]},~ \big( \Pi^{[1]} \big)^2  \Big\} \Big]_{\rm fin.}.
\label{eq:tr13_indepe_fncA}
}
Since $\vec \Pi^{[1]}$ operator multiplies the renormalised linear bias $\big(c^s_1\big)_{\rm fin.}$,
this means that we have only one independent third order bias parameter at one-loop order. 
This result agrees with one-loop power spectrum result obtained for the number density 
of bias tracers (see e.g. \cite{angulo/etal:2015, Fujita2016}.

As we did for the mode coupling ${(22)}$ integral $I_{nm}$, it is also useful to introduce 
integrals for the ${(13)}$ contribution
\eq{
J_{n}(k) = 3 
\big[ K^{(3)} \big]_n \lb \vec k, \vec p, - \vec p\rb P_L (\vec p).
\label{eq:I_master_13_II}
}
In total $J_{n}$ has three independent components, only two of which contribute 
to the trace $\tr \big[\bm K^{(3)}\big]$ part, as we noted above.
The third one is contributing only to the trace-free helicity-0 part $\vec Y^{(0)*}_2 . \bm K^{(3)}$, 
which will be shown shortly.
For the first two components we can write explicitly 
\eq{
\label{eq:J1J2_integral}
J_1(k)&= 
\frac{ f_1(k,p) + g_1(k,p) (\vec k\cdot \vec p)^2 + h_1(k,p) (\vec k\cdot \vec p)^4} {42 k^2 p^4 \left(\left(k^2+p^2\right)^2-4 (\vec k\cdot \vec p)^2\right)} P_L (\vec k) P_L (\vec p), \non\\
J_2(k)&= 
\frac{ f_2(k,p) + g_2(k,p) (\vec k\cdot \vec p)^2 + h_2(k,p)  (\vec k\cdot \vec p)^4} {21 k^2 p^4 \left(\left(k^2+p^2\right)^2-4 (\vec k\cdot \vec p)^2\right)} P_L (\vec k) P_L (\vec p), 
}
where we introduced functions
\eq{
f_1(k,p) &= 10 k^4 p^4 \left(k^2+p^2\right),
&&f_2(k,p) = - 4 k^2 p^4 \left(k^2+p^2\right) \left(17 k^2+2 p^2\right) , \non\\
g_1(k,p) & = - \left(21 k^6+44 k^4 p^2+59 k^2 p^4\right),
&&g_2(k,p) = 8 \left(18 k^4 p^2+25 k^2 p^4+3 p^6\right) , \non\\
h_1(k,p) & = 4 \left(19 k^2+7 p^2\right), 
&&h_2(k,p) = - 12 \left(5 k^2+13 p^2\right) . \non
}
These components thus contribute to both $\tr \big[\bm K^{(3)}\big]$ and $\vec Y^{(0)*}_2 . \bm K^{(3)}$ terms.

\begin{table}[t!]
\captionof{table}{One-loop helicity-0 contributions to $\tr \big[\bm K_s^{(3)} \big]$ and $\vec Y^{(0)*}_2 . \bm K_s^{(3)}$.}
\resizebox{\textwidth}{!}{
\begin{tabular}{ c | c | c } 
 \hline
 Op. & $\tr \big[\bm { K}^{(3)} \big]$
&  $\vec Y^{(0)*}_2 . \bm{ K}^{(3)}$ \\[0.3cm] \hline 
$ \Pi^{[1]}$ & $c^s_{1}$  & $c^g_{1}$  \\[0.3cm] \hline
$\big( \Pi^{[1]} \big)^2 $ & $c^s_{2,2}$ & $c^g_{2,2}$  \\[0.3cm] \hline
$\Pi^{[1]} \tr \big[ \Pi^{[1]} \big] $ & $c^s_{2,3}$ & $c^g_{2,3}$  \\[0.3cm] \hline
$\Pi^{[1]} \Pi^{[2]}  $ 
& $c^s_{3,3} + \frac{2}{15} c^s_{2,1} + \frac{1}{15} c^s_{3,1}$ 
& $c^g_{3,3} - \frac{28}{15} c^g_{2,1} - \frac{73}{30} c^g_{3,1}$  \\[0.3cm] \hline
$ \Pi^{[2]} \tr \big[ \Pi^{[1]} \big]$ & 0 & $c^g_{3,4} + \frac{8}{3} c^g_{2,1} + \frac{10}{3} c^g_{3,1}$  \\[0.3cm] \hline
$ \big( \Pi^{[1]} \big)^2 \tr\big[ \Pi^{[1]} \big] $  & 0 
& $c^g_{3,5} + c^g_{3,2}  +\frac{46}{15} c^g_{2,1} +\frac{38}{15} c^g_{3,1} $  \\[0.3cm] \hline
$ \Pi^{[1]} \Big(\tr\big[ \Pi^{[1]} \big] \Big)^2$ & 
$c^s_{3,6} - \frac{1}{2} c^s_{3,2} + \frac{5}{7} c^s_{3,4} +\frac{44}{105} c^s_{2,1} + \frac{62}{105} c^s_{3,1} $ 
& $c^g_{3,6} + \frac{1}{2} c^g_{3,2} + \frac{101}{105} c^g_{2,1} + \frac{58}{105} c^g_{3,1} $  \\[0.3cm] \hline  
$ \Pi^{[1]} \tr\big[ \big( \Pi^{[1]} \big)^2 \big] $  
&$c^s_{3,7} + c^s_{3,5} + \frac{3}{2} c^s_{3,2} + \frac{2}{7} c^s_{3,4} + \frac{68}{105} c^s_{2,1}+\frac{64}{105} c^s_{3,1}$ 
&$c^g_{3,7} - \frac{1}{2} c^g_{3,2} - \frac{57}{35} c^g_{2,1} - \frac{23}{105} c^g_{3,1} $   \\[0.3cm] \hline
\end{tabular}}
\label{tb:ps_scalar_loops_13}
\end{table}

Next we look at the $\vec Y^{(0)*}_2 . \bm K^{(3)}$ term. Following a similar procedure as earlier we have
\eq{
\label{eq:renorm_13_Y0K3}
\bigg( 3 ~ \int_{\vp}~ \vec Y^{(0)*}_2 .
\bigg\{ & \Pi^{[1]}, \big( \Pi^{[1]} \big)^2, \Pi^{[1]} \tr \big[ \Pi^{[1]} \big], \Pi^{[1]} \Pi^{[2]}, 
\Pi^{[2]} \tr \big[ \Pi^{[1]} \big],  \big( \Pi^{[1]} \big)^2 \tr\big[ \Pi^{[1]} \big], \\
& \Pi^{[1]} \Big(\tr\big[ \Pi^{[1]} \big] \Big)^2, \Pi^{[1]} \tr\big[ \big( \Pi^{[1]} \big)^2 \big] \bigg\}^{(3)}
(\vec k, \vec p, - \vec p) P_L (p)  \bigg)_{\rm UV} \non\\
&\hspace{4.5cm} \underset{k \to 0}{\implies} \left\{0,\frac{58}{105},\frac{58}{105},\frac{128}{105},\frac{10}{21},\frac{2}{3},1,\frac{19}{15}\right\} \sqrt{\frac{2}{3}} \sigma^2, \non
}
which after the renormalisation yields five surviving terms 
\eeq{
 \vec Y^{(0)*}_2 . \Big[  \Big\{ 
 \Pi^{[1]}, ~\big( \Pi^{[1]} \big)^2, ~\Pi^{[1]} \tr \big[ \Pi^{[1]} \big],
 ~\Pi^{[1]} \Pi^{[2]}, ~\Pi^{[2]} \tr \big[ \Pi^{[1]} \big] \Big\} \Big]_{\rm fin.}. 
}
These, however, are again not independent and we have additional redundancies
\eq{
\vec Y^{(0)*}_2 . \Big[ \big( \Pi^{[1]} \big)^2 - \Pi^{[1]} \Pi^{[2]} \Big]_{\rm fin.} &=0, \non\\
\vec Y^{(0)*}_2 . \Big[ \Pi^{[1]} \tr \big[ \Pi^{[1]} \big] - \Pi^{[2]} \tr \big[ \Pi^{[1]} \big] \Big]_{\rm fin.} &=0.
}
Moreover, we can also explore the possible dependencies among residual terms in
the trace and trace-free basis for the $\bm K^{(3)}$ kernel. 
We find the following additional dependencies
\eq{
N_0~ \vec Y^{(0)*}_2 . \Big[ \Pi^{[1]} \Big]_{\rm fin.} - \tr  \Big[ \Pi^{[1]} \Big]_{\rm fin.}  &= 0 \non\\
 4 N_0~\vec Y^{(0)*}_2 . \Big[ \Pi^{[1]} \tr \big[ \Pi^{[1]} \big]  -  \big( \Pi^{[1]} \big)^2\Big]_{\rm fin.} + \tr  \Big[ \big( \Pi^{[1]} \big)^2 \Big]_{\rm fin.} &= 0,
 \label{eq:tr_Y0_degen}
}
where $N_0=\sqrt{3/2}$ is the normalization of the $\vec Y^{(0)}_2$ tensor basis vector. 
Note that when considering the constraints above we take into account the integration over $\vhat  k \cdot \vhat  q$ angle, 
otherwise trivial angle dependent remainders might make integrand functions not strictly degenerate. 
For the $\vec Y^{(0)*}_2 . \bm K^{(3)}$ we obtain three independent contributions
\eeq{
\bigg\{  \tr  \Big[ \Pi^{[1]} \Big]_{\rm fin.},  ~~~  \tr  \Big[ \big( \Pi^{[1]} \big)^2 \Big]_{\rm fin.},
 ~~N_0 \vec Y^{(0)*}_2 . \Big[ \big( \Pi^{[1]} \big)^2 \Big]_{\rm fin.} \bigg\}.
}
In contrast to the case of trace above, no further degeneracies appear here and thus all of these operators are indeed independent.  
Thus, besides the two independent terms in Eq.~\eqref{eq:J1J2_integral}, we have an additional independent term, 
and thus one additional bias parameter.
This third component is
\eq{
J_3(k)&= 
\frac{ f_3( k, p) +g_3(k,p) (\vec k\cdot \vec p)^2  +h_3(k,p) (\vec k\cdot \vec p)^4 + w_3(k,p) (\vec k\cdot \vec p)^6 }
{105 k^2 p^4 \left(\left(k^2+p^2\right)^2-4 (\vec k\cdot \vec p)^2\right)} P_L (\vec k) P_L (\vec p),
\label{eq:J3_integral}
}
where 
\eq{
f_3(k,p) &= -2 k^4 p^4 \left(k^2+p^2\right) \left(29 k^2+104 p^2\right), \\
g_3(k,p) & = 405 k^6 p^2+1042 k^4 p^4+705 k^2 p^6, \non\\
h_3(k,p) & =15 \left(k^4-76 k^2 p^2-9 p^4\right), \non\\
w_3(k,p) & = -360. \non
}
The third $J_3$ term contributes only to the $\vec Y^{(0)*}_2 . \bm K^{(3)}$ integral.

\subsection{Degeneracy of the bias operators}
\label{App:degeneracy}

By construction all the operators summarised in table~\ref{tb:Bias_kernels} are independent, 
but the trace components are not. We give the linear dependence relations here for these 
trace components.
For any pair of vectors $\vec p_1$, $\vec p_2$, at second order we have
\eeq{
\tr \left[ 7~ \Pi^{[2]}_{ij} - 2  ~\big( \Pi^{[1]} \big)^2_{ij} - 5~ \Pi^{[1]}_{ij} \tr \big[ \Pi^{[1]} \right]^{(2)} = 0,
\label{eq:trace_second_order}
}
and at third order, for any set of $\vec p_1$, $\vec p_2$ and $\vec p_3$ we have
\eq{
 \label{eq:tr_third_operators}
\tr \left[   \big( \Pi^{[1]} \big)^2_{ij}\tr\big[ \Pi^{[1]} \big] -   \Pi^{[1]}_{ij} \tr\Big[ \big( \Pi^{[1]} \big)^2 \Big]  \right]^{(3)} & = 0, \\
\tr \left[ 7~ \Pi^{[2]}_{ij} \tr \big[ \Pi^{[1]} \big] - 2  ~ \Pi^{[1]}_{ij} \tr\Big[ \big( \Pi^{[1]} \big)^2 \Big]   
- 5~ \Pi^{[1]}_{ij} \Big(\tr\big[ \Pi^{[1]} \big] \Big)^2 \right]^{(3)}&=0, \non\\
\tr \bigg[ 90~ \Pi^{[3]}_{ij} + 45~ \Pi^{[2]}_{ij} - 12~ \Big[ \Pi^{[1]} \Pi^{[2]} \Big]_{ij} 
-16~ \big( \Pi^{[1]} \big)^3_{ij} - 60~ \Pi^{[1]}_{ij} \tr\Big[ \big( \Pi^{[1]} \big)^2 \Big] & \non\\
- 80 ~ \Pi^{[1]}_{ij} \Big(\tr\big[ \Pi^{[1]} \big] \Big)^2 \bigg]^{(3)}&=0, \non\\
\tr \bigg[ 2205~ \Pi^{[2]}_{ij} + 336 ~ \Big[ \Pi^{[1]} \Pi^{[2]} \Big]_{ij} 
-392~ \big( \Pi^{[1]} \big)^3_{ij} - 630~ \big( \Pi^{[1]} \big)^2_{ij} - 1575 ~ \Pi^{[1]}_{ij}\tr\big[ \Pi^{[1]}& \big]  \non\\
- 570 ~  \Pi^{[1]}_{ij}\tr\Big[ \big( \Pi^{[1]} \big)^2 \Big]
+ 290 ~  \Pi^{[1]}_{ij} \Big( \tr\big[ \Pi^{[1]} \big] \Big)^2
 \bigg]^{(3)}& =0. \non
}
Note that the second dependence relation above (in the third order) is the direct consequence 
of the trace linear dependence at the second order.

If we pick the pair of vectors to be $\vec p$, and $\vec k - \vec p$ we can 
get another operator relation by projecting onto $\vec Y^{(0)*}_2(\vec k)$ 
basis function
\eeq{
\vec Y^{(0)*}_2(\vec k) . \left[ 7~ \Pi^{[2]} (\vec p, \vec k - \vec p) +13  ~\big( \Pi^{[1]} \big)^2(\vec p, \vec k - \vec p)  - 20~ \Pi^{[1]} \tr \big[ \Pi^{[1]}\big] (\vec p, \vec k - \vec p) \right]^{(2)} = 0,
\label{eq:Y0_projection_lin_dep}
}
as well as mixed linear dependence relations between the two projections
\eq{
\label{eq:tr&Y0_projection_lin_dep}
\tr \big[ \Pi^{[1]} \big]^{(2)}(\vec p, \vec k - \vec p) - N_0 \lb \vec Y^{(0)*}_2(\vec k) . \big[ \Pi^{[1]} \big]^{(2)} \rb(\vec p, \vec k - \vec p) &= 0, \\
\tr \left[ \big( \Pi^{[1]} \big)^2 - \Pi^{[1]} \tr \big[ \Pi^{[1]}\big] \right]^{(2)}(\vec p, \vec k - \vec p) \hspace{4.7cm} & \non\\
- 4N_0 \lb \vec Y^{(0)*}_2(\vec k) . \left[ \big( \Pi^{[1]} \big)^2 - \Pi^{[1]} \tr \big[ \Pi^{[1]}\big] \right]^{(2)} \rb (\vec p, \vec k - \vec p) &= 0. \non
} 
Next we turn to the higher helicity projections $\vec Y^{(1)*}_2(\vec k)$ and $\vec Y^{(2)*}_2(\vec k)$. 
Since the $\vec \Pi^{[1]}$ operator is orthogonal to both of these higher helicity basis projections 
we have
\eq{
\label{eq:Y12_projection_lin_dep}
 \vec Y^{(1)*}_2(\vec k) .\big[ \Pi^{[2]}\big]^{(2)} (\vec p, \vec k - \vec p)  
 & = \vec Y^{(1)*}_2(\vec k) .\big[ \big( \Pi^{[1]} \big)^2 \big]^{(2)} (\vec p, \vec k - \vec p) \\
 &= \vec Y^{(1)*}_2(\vec k) . \big[ \Pi^{[1]}  \tr \big[ \Pi^{[1]} \big] \big]^{(2)} (\vec p, \vec k - \vec p), \non\\
 \vec Y^{(2)*}_2(\vec k) .\big[ \Pi^{[2]}\big]^{(2)} (\vec p, \vec k - \vec p) 
& = \vec Y^{(2)*}_2(\vec k) .\big[ \big( \Pi^{[1]} \big)^2 \big]^{(2)} (\vec p, \vec k - \vec p). \non
}
Next we list similar constraints for the third order kernels, with kernel input vectors $\vec k$, $\vec p$, $- \vec p$. 
We need to consider only the helicity-0 contributions given that at one-loop level only 
$\tr \big[\bm K^{(3)} \big]$ and $\vec Y^{(0)*}_2 . \bm K^{(3)}$ terms contribute.
Trace gives five such operator dependence relations 
\eq{
\label{eq:tr_third_operators_II}
0 &= \tr \left[   \big( \Pi^{[1]} \big)^2 \tr\big[ \Pi^{[1]} \big] -   \Pi^{[1]} \tr\Big[ \big( \Pi^{[1]} \big)^2 \Big]  \right]^{(3)}( \vec k, \vec p, - \vec p), \\
0 &= \tr \left[ 7~ \Pi^{[2]} \tr \big[ \Pi^{[1]} \big] - 2  ~ \Pi^{[1]} \tr\Big[ \big( \Pi^{[1]} \big)^2 \Big]  - 5~ \Pi^{[1]}\Big(\tr\big[ \Pi^{[1]} \big] \Big)^2 \right]^{(3)}( \vec k, \vec p, - \vec p), \non\\
0 &= \tr \left[ 105~ \Pi^{[3]} -  7~ \Big( \Pi^{[1]} \Pi^{[2]} \Big) - 64~ \Pi^{[1]} \tr\Big[ \big( \Pi^{[1]} \big)^2 \Big]  - 62 ~ \Pi^{[1]} \Big(\tr\big[ \Pi^{[1]} \big] \Big)^2 \right]^{(3)}( \vec k, \vec p, - \vec p), \non\\
0 &= \tr \left[ 105~ \Pi^{[2]} -14~ \Big( \Pi^{[1]} \Pi^{[2]} \Big) - 68~ \Pi^{[1]} \tr\Big[ \big( \Pi^{[1]} \big)^2 \Big]  - 44 ~ \Pi^{[1]} \Big(\tr\big[ \Pi^{[1]} \big] \Big)^2 \right]^{(3)}( \vec k, \vec p, - \vec p), \non\\
0 &= \tr \left[     2~ \big( \Pi^{[1]} \big)^3 - 3~ \Pi^{[1]} \tr\Big[ \big( \Pi^{[1]} \big)^2 \Big] + \Pi^{[1]} \Big(\tr\big[ \Pi^{[1]} \big] \Big)^2 \right]^{(3)}( \vec k, \vec p, - \vec p). \non
}
This thus leaves six independent operators in the $\tr \big[\bm K^{(3)}\big]$ term.
When contracting with $\vec Y^{(0)*}_2$ there are three similar constraints 
\eq{
\label{eq:Y0_third_operators_II}
0&= \vec Y^{(0)*}_2 . \Big[ 210~ \Pi^{[3]} + 511~ \Big( \Pi^{[1]} \Pi^{[2]} \Big) - 532 \big( \Pi^{[1]} \big)^2 \tr\big[ \Pi^{[1]} \big] - 700~ \Pi^{[2]} \tr \big[ \Pi^{[1]} \big] \\
&\hspace{3.1cm}  - 116~ \Pi^{[1]} \tr\Big[ \big( \Pi^{[1]} \big)^2 \Big] + 46 ~ \Pi^{[1]} \Big(\tr\big[ \Pi^{[1]} \big] \Big)^2 \Big]^{(3)} ( \vec k, \vec p, - \vec p), \non\\
0&= \vec Y^{(0)*}_2 . \Big[ 105~ \Pi^{[2]} + 196~ \Big( \Pi^{[1]} \Pi^{[2]} \Big) - 322 \big( \Pi^{[1]} \big)^2 \tr\big[ \Pi^{[1]} \big] - 280~ \Pi^{[2]} \tr \big[ \Pi^{[1]} \big] \non\\
&\hspace{3.1cm} - 101~ \Pi^{[1]} \tr\Big[ \big( \Pi^{[1]} \big)^2 \Big] + 171 ~ \Pi^{[1]} \Big(\tr\big[ \Pi^{[1]} \big] \Big)^2 \Big]^{(3)}( \vec k, \vec p, - \vec p), \non\\
0&= \vec Y^{(0)*}_2 . \Big[ 2~ \big( \Pi^{[1]} \big)^3 - 2 \big( \Pi^{[1]} \big)^2 \tr\big[ \Pi^{[1]} \big] -  \Pi^{[1]} \tr\Big[ \big( \Pi^{[1]} \big)^2 \Big] 
+ \Pi^{[1]} \Big(\tr\big[ \Pi^{[1]} \big] \Big)^2 \Big]^{(3)}( \vec k, \vec p, - \vec p), \non
}
leaving eight operators surviving in the $\vec Y^{(0)*}_2 . \bm K^{(3)}$ term. 

\subsection{Bias expansion of the tree-level bispectrum}
\label{app:bispectrum_II}

We apply the decomposition to the PT tree-level results. In order to do this we have to expand one of the fields to the second order.
This gives the general expression  
\eeq{
B^{\alpha\beta\gamma, {\rm tree}}_{ijklrs} (\vec k_1,\vec k_2,\vec k_3) 
= 2 K_{ij,\alpha}^{(1)}(\vec k_1) K_{kl,\beta}^{(1)}(\vec k_2) K_{rs,\gamma}^{(2)}(\vec k_1,\vec k_2)P_L (\vec k_1) P_L (\vec k_2) +  {\rm 2~cycle },
}
where the $\alpha,~\beta$ and $\gamma$ label three, in principle different, generic biased tracers. 
Since the linear kernel $K_{ij}^{(1)}$ has only helicity-0 components, and given that at tree-level at least two linear kernels contribute to each correlator, 
the decomposition in Eq. \eqref{eq:bis_decomp_0} reduces to a much less general form, and nonzero helicity contributions only appear in terms with 
the second order shear fields.  
Projecting out kernels for each of the rotational symmetry components we get  
\eq{
B^{\alpha\beta\gamma,(0,0,0)}_{000} (\vec k_1, \vec k_2, \vec k_3) &= 2 b^{(\alpha)}_1 b^{(\beta)}_1 \tr\big[ \bm{ K}_{\gamma}^{(2)}(\vec k_1,\vec k_2) \big] P_L (\vec k_1) P_L (\vec k_2) +  {\rm 2~cycle }, \\
B^{\alpha\beta\gamma,(0,0,m)}_{002} (\vec k_1, \vec k_2, \vec k_3) 
&= 2 b^{(\alpha)}_1 b^{(\beta)}_1 \vec Y^{(m)*}_{2} \big( \vhat  k_3 \big) . \bm{ K}_{\gamma}^{(2)}(\vec k_1,\vec k_2) P_L (\vec k_1) P_L (\vec k_2) \non\\
                    &~~~~ +  2 \tilde \df^{\rm K}_{0,m} b^{(\gamma)}_1 \Big( b^{(\alpha)}_1 \tr\big[ \bm{ K}_{\beta}^{(2)}(\vec k_1,\vec k_3) \big] P_L (\vec k_1) \non\\
                    &~~~~~~~~~~~~~~~~~~~~~   +     b^{(\beta)}_1 \tr\big[ \bm{K}_{\alpha}^{(2)}(\vec k_2,\vec k_3) \big] P_L (\vec k_2)  \Big) P_L (\vec k_3), \non\\
B^{\alpha\beta\gamma,(0,m_2,m_3)}_{022} (\vec k_1, \vec k_2, \vec k_3) 
&= 2 \tilde \df^{\rm K}_{0,m_2}  b^{(\alpha)}_1 b^{(\beta)}_1 \vec Y^{(m_3)*}_{2} \big(\vhat  k_3 \big) . \bm{K}_{\gamma}^{(2)}(\vec k_1,\vec k_2) P_ L (\vec k_1) P_L (\vec k_2) \non\\
                    &~~~~ +  2 \tilde \df^{\rm K}_{0,m_3} b^{(\alpha)}_1 b^{(\gamma)}_1 \vec Y^{(m_2)*}_{2} \big( \vhat  k_2 \big) . \bm{ K}_{\beta}^{(2)}(\vec k_1,\vec k_3) P_L (\vec k_1) P_L (\vec k_3) \non\\
                    &~~~~ +  2 \tilde \df^{\rm K}_{0,m_2} \tilde \df^{\rm K}_{0,m_3} b^{(\beta)}_1 b^{(\gamma)}_1 \tr \big[ \bm{ K}_{\alpha}^{(2)}(\vec k_2,\vec k_3) \big] P_L (\vec k_2) P_ L (\vec k_3), \non\\
B^{\alpha\beta\gamma,(m_1,m_2,m_3)}_{222} (\vec k_1, \vec k_2, \vec k_3) 
&= 2 \tilde \df^{\rm K}_{0,m_1} \tilde \df^{\rm K}_{0,m_2}  b^{(\alpha)}_1b^{(\beta)}_1 \vec Y^{(m_3)*}_{2} \big( \vhat  k_3 \big) . \bm{ K}_{\gamma}^{(2)}(\vec k_1,\vec k_2)  P_ L (\vec k_1) P_L (\vec k_2) +  {\rm 2~cycle }, \non
} 
where we used our normalised Kronecker delta $\tilde \df^{\rm K}_{n,m} =N_0^{-1}\dK_{n,m}$. Notice that there is no integration over the $\vec k_i$, as these are the external wavenumbers. 
For the (022) and (222) bispectra, higher helicity modes, where at least two $m_i$
are different from zero can not be obtained at tree-level PT and arise only at higher order contributions.
Indeed, at the tree-level only four distinct terms are present, and these can be constructed using two basic functional forms.
We can thus introduce the shorthand for these components 
\eq{
\mathcal F^{\alpha\beta\gamma,(0)}_{0} (\vec k_1, \vec k_2) &\equiv 2 b^{(\alpha)}_1 b^{(\beta)}_1 \tr\big[ \bm K_{\gamma}^{(2)}(\vec k_1,\vec k_2) \big] P_L (\vec k_1) P_L (\vec k_2), \non\\
\mathcal F^{\alpha\beta\gamma,(m)}_{2} (\vec k_1, \vec k_2) &\equiv 2 b^{(\alpha)}_1 b^{(\beta)}_1 \vec Y^{(m)*}_{2} \big( \vhat  k_3 \big) . \bm K_{\gamma}^{(2)}(\vec k_1,\vec k_2) P_L (\vec k_1) P_L (\vec k_2),
\label{eq:Fbis_kernel}
}
where the kernels for the $\mathcal F^{\alpha\beta\gamma,(0)}_{0}$ are readily given as in Eq.~\eqref{eq:K_tr} 
and we can write 
\eeq{
\tr \big[\bm K^{(2)}(\vec k_1, \vec k_2) \big]
= b^{\rm s}_1 \tr \big[ \Pi^{[1]} \big]^{(2)} (\vec k_1, \vec k_2)
 + b^{\rm s}_{2,1} \tr \Big[ \big( \Pi^{[1]} \big)^2 \Big]^{(2)} (\vec k_1, \vec k_2)
 + b^{\rm s}_{2,2} \Big[ \big( \tr [\Pi^{[1]} ] \big)^2 \Big]^{(2)} (\vec k_1, \vec k_2).  
 \label{eq:trace_K2}
}
However, for the $\mathcal F^{\alpha\beta\gamma,(m)}_{2}$ we do not use the  
Eq.~\eqref{eq:Y0K2}, since the projection is now relative to the $\vec Y^{(0)}_{2} $ of the $\vec k_3$ mode, 
not present in the kernels. We have
\eq{
\vec Y^{(m)*}_2(\vec k_3) . \bm K^{(2)}(\vec k_1, \vec k_2) 
=\:& c^{\rm g}_1 \vec Y^{(m)*}_2(\vec k_3) .\big[ \Pi^{[1]} \big]^{(2)} (\vec k_1, \vec k_2)
+ c^{\rm g}_{2,1} \vec Y^{(m)*}_2(\vec k_3) .\big[ \Pi^{[2]} \big]^{(2)}(\vec k_1, \vec k_2) \non\\
& + c^{\rm g}_{2,3} \vec Y^{(m)*}_2(\vec k_3) .\Big[ \Pi^{[1]} \tr \big[ \Pi^{[1]} \big] \Big]^{(2)} (\vec k_1, \vec k_2) \non\\
& + c^{\rm g}_{2,2} \vec Y^{(m)*}_2(\vec k_3) .\Big[ \big( \Pi^{[1]} \big)^2 \Big]^{(2)}(\vec k_1, \vec k_2),
\label{eq:Y0_K2}
}
where the bias relation in two basis are given in Eq.~\eqref{eq:bias_b_second}.
However note that since we do not expect any a priori relation between shear and size bias coefficients 
we can use the notation $c^{\rm g} \to b^{\rm g}$ for any of the above biases, as long as this is consistently 
performed also in the one-loop power spectrum. 
All the bispectra can now be built by cycling over variables of these kernels to obtain 
the results  
\eq{
B^{\alpha\beta\gamma,(0)}_{000} (\vec k_1, \vec k_2, \vec k_3) 
&= \mathcal F^{\alpha\beta\gamma,(0)}_{0} (\vec k_1, \vec k_2) + \mathcal F^{\beta\gamma\alpha,(0)}_{0} (\vec k_2, \vec k_3) + \mathcal F^{\gamma\alpha\beta,(0)}_{0} (\vec k_3, \vec k_1), \\
B^{\alpha\beta\gamma,(m)}_{002} (\vec k_1, \vec k_2, \vec k_3) 
&= \mathcal F^{\alpha\beta\gamma,(m)}_{2} (\vec k_1, \vec k_2) + \tilde \df^{\rm K}_{0,m} \mathcal F^{\beta\gamma\alpha,(0)}_{0} (\vec k_2, \vec k_3) 
+ \tilde \df^{\rm K}_{0,m} \mathcal F^{\gamma\alpha\beta,(0)}_{0} (\vec k_3, \vec k_1), \non\\
B^{\alpha\beta\gamma,(m_2,m_3)}_{022} (\vec k_1, \vec k_2, \vec k_3) 
&= \tilde \df^{\rm K}_{0,m_2} \mathcal F^{\alpha\beta\gamma,(m_3)}_{2} (\vec k_1, \vec k_2) 
+ \tilde \df^{\rm K}_{0,m_2} \tilde \df^{\rm K}_{0,m_3}  \mathcal F^{\beta\gamma\alpha,(0)}_{0} (\vec k_2, \vec k_3) \non\\
&\hspace{4.2cm}+ \tilde \df^{\rm K}_{0,m_3} \mathcal F^{\gamma\alpha\beta,(m_2)}_{2} (\vec k_3, \vec k_1), \non\\
B^{\alpha\beta\gamma,(m_1,m_2,m_3)}_{222} (\vec k_1, \vec k_2, \vec k_3) 
&= \tilde \df^{\rm K}_{0,m_1} \tilde \df^{\rm K}_{0,m_2}  \mathcal F^{\alpha\beta\gamma,(m_3)}_{2} (\vec k_1, \vec k_2) 
+ \tilde \df^{\rm K}_{0,m_1} \tilde \df^{\rm K}_{0,m_3}  \mathcal F^{\beta\gamma\alpha,(m_2)}_{2} (\vec k_2, \vec k_3) \non\\
&\hspace{5.05cm}+ \tilde \df^{\rm K}_{0,m_2} \tilde \df^{\rm K}_{0,m_3}  \mathcal F^{\gamma\alpha\beta,(m_1)}_{2} (\vec k_3, \vec k_1).  \non
}

\section{Bias renormalisation for tensor fields}
\label{app:reno}

In section \ref{subsec:PT_Fourer} we introduced the perturbative expansion of the
tensorial fields and we asserted that trace and trace-free parts of the tensor field 
require separate counterterms (i.e. bias coefficients). In this section, we show how this 
property naturally follows from the requirement  that our PT expansion should be closed under 
the renormalisation. First, we decompose the biased tracer field into trace and trace-free parts:
\eeq{
S_{ij}(\vec x) = \dK_{ij} S(\vec x) + g_{ij}(\vec x).
}
Considering that the trace $S$ part is a scalar under 3D rotations, its general bias expansion is given 
by the standard set of terms that appear in the expansion of the galaxy density (see Eq.~\eqref{eq:EulBasis_tr}). 
This expansion is closed under renormalisation. That is, if $g_{ij}$ vanishes then the statistics of the tracer 
$S$ are all consistently described by the correlators (with UV-sensitive parts removed) involving the operators 
appearing in the scalar bias expansion, multiplied by renormalised bias parameters.

When we consider the trace-free part $g_{ij}$, the corresponding bias expansion is given in 
Eq.~\eqref{eq:EulBasis_TF}. The conjecture is that the latter bias expansion is also closed under renormalisation. 
This follows from the fact that all local gravitational observables can (at fixed order in perturbation theory) be written 
as combinations of the $\vec \Pi^{[n]}$ operators. Eq.~\eqref{eq:EulBasis_TF} contains all these combinations, 
with their trace subtracted. As an example, showing that counter-terms again come from the same list of bias operators, 
consider the following one-loop contribution:
\eq{
\left\langle \d(\vk') \text{TF}\left[\left(\Pi^{[1]}\Pi^{[1]}\Pi^{[1]}\right)_{ij}(\vk)\right]\right\rangle' &= 
\d^{\rm D}_{\vk-\vp_{123}} \langle \d(\vk') \d(\vp_1)\d(\vp_2)\d(\vp_3) \rangle' \\
   &\hspace{-0.5cm}\times \left[\frac{p_1^i p_3^j (\vp_1\cdot\vp_2)(\vp_3\cdot\vp_2)}{p_1^2p_2^2p_3^2} - \frac13 \d^{ij}\frac{(\vp_1\cdot\vp_2)(\vp_1\cdot\vp_3)(\vp_3\cdot\vp_2)}{p_1^2p_2^2p_3^2}\right]\,. \non
}
One sees that this loop integral becomes
\eeq{
\left\langle\d(\vk') \text{TF}\left[\left(\Pi^{[1]}\Pi^{[1]}\Pi^{[1]}\right)_{ij}(\vk)\right]\right\rangle' \propto \left[\frac{k_i k_j}{k^2} - \frac13 \dK_{ij}\right] P_L(k) \sigma^2\,.
}
This is absorbed by a counter term to the cubic operator given by $\propto \sigma^2\, \text{TF}[\Pi^{[1]}_{ij}]$, 
or equivalently by a contribution $\propto c_{\text{TF}[\Pi^3]} \sigma^2$ to the renormalised bias coefficient 
of $\text{TF}[\Pi^{[1]}_{ij}]$, the first operator in Eq.~\eqref{eq:EulBasis_TF}.

Thus, trace and trace-free part can be consistently treated within their respective complete bias expansions. 
However, this does not imply that the bias coefficients of the trace-free parts are the same as the corresponding 
trace part. Indeed, the fact that the UV parts of the trace and trace-free components of the correlators involve two 
different eigenvalues implies that we have to allow for different bias parameters multiplying the trace and trace-free 
parts of a symmetric tensor tracer $\vec \Pi$. 
If we do, then the respective bias parameters can consistently absorb the UV-dependent pieces. 
The fact that we need these different bias parameters was perhaps not immediately obvious. 
However, the trace and trace-free parts clearly transform differently under rotations, which leads 
to different structures in the correlation functions, e.g. in the power spectrum
\eeq{
\dK_{ij} \quad\mbox{vs}\quad \frac{k_i k_j}{k^2}-\frac13 \dK_{ij}\,.
}
The preferred direction provided by $\vec k$ then requires one to allow for separate bias parameters 
for trace and trace-free parts.

As proof of the fact that different counter terms are needed for trace and trace-free part, let us consider 
a toy quadratic biasing model for the rank two tensor field. We start by assuming the opposite claim, 
i.e. that bias expansion of the tensor field is given by the single parameter for both the trace and trace-free parts. 
This gives us the expansion
\eeq{
S_{ij} (\vec r) = c_1  \Pi_{ij}^{[1]} (\vec r)  + c_2 \big( \Pi^{[1]} (\vec r) \big)_{ij}^2.
\label{eq:S_toy}
}
The simplest two-point function we can consider is the correlation with the linear density 
\eeq{
\la \df_L(\vec k_1) S_{ij} (\vec k_2) \ra  = c_1 \la  \df_L(\vec k_1) \Pi_{ij}^{[1]} (\vec k_2) \ra + c_2 \la \df_L(\vec k_1) \big( \Pi^{[1]}\big)_{ij}^2(\vec k_2) \ra.
}
If we look just at the second contribution above, we get  
\eq{
2 P_L (k)  \frac{\vec p . (\vec k - \vec p)}{p^2 (\vec k - \vec p)^2} p_{\{i} (\vec k - \vec p)_{j\}} & F_2 (\vec k , - \vec p) P_L(p) \\
&\sim 
\frac{1}{105}  \begin{pmatrix}
    94  & 0 & 0 \\
    0  & 94 & 0  \\ 
    0  & 0 & 152
\end{pmatrix} \sigma^2 P_L(k), ~~{\rm as} ~~ k\to0, \non
}
where we have set the coordinate frame so that ${\vec k}$ is along the $z$ axis and we introduced $\sigma^2 = \int_{\vec p} P_L(p)$. 
It is evident from the structure of the matrix that the trace and trace-free components require different renormalisation contributions.
If we are interested just in the trace we get the usual $68/21 \sigma^2 P_L(k)$ contribution which simply renormalises the linear bias. 
Indeed we get the standard renormalisation scheme (with several standard terms omitted, due to the fact that we study the simplified 
toy bias model from Eq.~\eqref{eq:S_toy})
\eeq{
c^{\rm s}_1 \to c^{\rm s}_1 + c^{\rm s}_2 \frac{68}{21}  \sigma^2.
}
However, we can see that the trace-free component requires a different scheme implying that 
our initial assumption, used in Eq.~\eqref{eq:S_toy}, of the sufficiency of a single bias parameter set for 
trace and trace-free components was incorrect. 
This thus proves that we need two different bias parameter sets, as was initially claimed.

\bibliography{submit}
\end{document}